\documentclass{article}

\usepackage{PRIMEarxiv}

\usepackage[utf8]{inputenc} 
\usepackage[T1]{fontenc}    
\usepackage{hyperref}       
\usepackage{url}            
\usepackage{booktabs}       
\usepackage{amsfonts}       
\usepackage{nicefrac}       
\usepackage{microtype}      
\usepackage{lipsum}
\usepackage{fancyhdr}       
\usepackage{graphicx}       
\graphicspath{{media/}}     
\usepackage{xcolor}
\usepackage{amsmath}
\usepackage{amsfonts}
\usepackage{amssymb}
\usepackage{amsmath,tabu,booktabs}
\usepackage{stfloats}
\usepackage{caption}
\usepackage{subcaption}
\newcommand{\overbar}[1]{\mkern 1.5mu\overline{\mkern-1.5mu#1\mkern-1.5mu}\mkern 1.5mu}

\title{Bayesian weighted time-lapse full-waveform inversion using a receiver-extension strategy
\thanks{Submitted as a Journal Paper to IEEE Transactions on Geoscience and Remote Sensing} 
}

\author{
  Sérgio Luiz E. F. da Silva$^{1,2,3}$, Ammir Karsou$^{3}$, Roger M. Moreira$^{3}$,  Marco Cetale$^{3}$ \\ \\
  $^{1}$ Dipartimento di Scienza Applicata e Tecnologia (DISAT), Politecnico di Torino, Turin, Italy\\
  $^{2}$Institute for Complex Systems of the National Research Council  c/o  Politecnico di Torino, Turin, TO, Italy \\
  $^{3}$  Grupo de Imageamento Sísmico e Inversão Sísmica (GISIS), Fluminense Federal University, Niter\'oi, RJ, Brazil \\
  Corresponding author email: \texttt{sergio.dasilva@polito.it; sergioluizsilva@id.uff.br} 
}

\begin{document}
\maketitle

\begin{abstract}
Time-lapse full-waveform inversion (FWI) has become a powerful tool for characterizing and monitoring subsurface changes in various geophysical applications. However, non-repeatability (NR) issues caused, for instance, by GPS inaccuracies, often make it difficult to obtain unbiased time-lapse models. In this work we explore the portability of combining a receiver-extension FWI approach and Bayesian analysis to mitigate time-lapse noises arising from NR issues. The receiver-extension scheme introduces an artificial degree of freedom in positioning receivers, intending to minimize kinematic mismatches between modeled and observed data. Bayesian analysis systematically explores several potential solutions to mitigate time-lapse changes not associated with reservoir responses, assigning probabilities to each scenario based on prior information and available evidence. We consider two different subsurface models to demonstrate the potential of proposed approaches. First, using the Marmousi model, we investigate two NR scenarios associated with background noise in seismic data. Second, using a challenging deep-water Brazilian pre-salt setting, we investigate several NR scenarios to simulate real-world challenges. Our results demonstrate that combining Bayesian analysis with the receiver-extension FWI strategy can mitigate adverse NR effects successfully, producing cleaner and more reliable time-lapse models than conventional approaches. The results also reveal that the proposed Bayesian weighted procedure is a valuable tool for determining time-lapse estimates through statistical analysis of pre-existing models, allowing its application in ongoing time-lapse (4D) projects.
\end{abstract}


\section{Introduction}

Time-lapse (4D) seismic processing plays a crucial role in accurately characterizing and monitoring oil and gas reservoirs, providing valuable insights into their temporal evolution \cite{Sambo_et_al_2020__TimeLapseReview}. This procedure allows the determination of subtle and complex changes that occur in the subsurface over time by directly comparing geophysical data collected at different surveys. Such information is essential for adequate reservoir management and planning, which are vital, for instance, in $\textrm{CO}_2$ sequestration projects \cite{Ivandic_et_al_2015__TimeLapse_CO2} as well as for a better understanding of production-related changes in reservoirs \cite{Cruz_et_al_2021__TimeLapse_TupiNodesPilot}. Time-lapse 4D processing is also essential in several other analyses, such as in geodynamics and volcanic activity monitoring \cite{Pivetta_et_al_2023__TimeLapse_GeodinamicsVolcano__GJI}, in groundwater reservoir managements \cite{Bai_et_al_2021__TimeLapse_Groundwater_monitoring}, and in geotechnical risk assessments \cite{Pringle_et_al_2012__TimeLapse_Geotecnical_monitoring}. 

The term 4D refers to the repetition of 3D seismic surveys (spatial dimension) over the same geographic area, with the fourth dimension representing the times of these surveys. In this context, a 4D dataset consists of seismic measures taken in sequential geophysical surveys at identical exploration regions. The initial survey (typically completed before production begins) is the \textit{baseline survey}, and the following assessments are the \textit{monitor surveys}. The baseline and monitor datasets are analyzed to retrieve the so-called baseline and monitor models. Then, time-lapse models are produced by matching the retrieved monitor and baseline models. Due to the nature of time-lapse seismic projects, some challenges arise since 4D seismic comprises acquisitions carried out in different periods. In this regard, non-repeatability (NR) issues make the time-lapse analysis susceptible to 4D noises appearing in the time-lapse estimates. NR effects stem from many factors, including the difficulty of perfectly repeating the same seismic acquisition parameters, seismic noise (from natural or anthropogenic origins) unrepeatable from one acquisition to another, and variable weather conditions and seasons.

Full-waveform inversion (FWI) \cite{Fichtner_2010_book_FWI} has recently been assessed as a robust reservoir monitoring tool in time-lapse (4D) studies, generating less noisy time-lapse models than other procedures \cite{Hicks_et_al_2016_TIMELAPSEFEI_LeadingEdge}. Thus, several successful 4D FWI applications have been reported in the literature. For example, Ref. \cite{Hicks_et_al_2016_TIMELAPSEFEI_LeadingEdge} demonstrated, using a permanent reservoir monitoring system installed at the Grane oil field in the North Sea, that current 4D FWI implementations could reliably detect changes in P-wave velocities related to oil-to-gas substitution in hydrocarbon production. Ref. \cite{Maharramov_et_al_2016_TimeLapse_FieldApp} appraised the potentialities of obtaining 4D changes due to overburden induced by reservoir compaction and overburden dilation above the Gulf of Mexico Genesis field by applying a regularized 4D FWI strategy. Ref. \cite{Bortoni_et_al_2021_SEG_4DFWI} presents a case study in the deep-water Campos Basin field that demonstrates the potential of FWI applications to obtain a good time-lapse model of post-salt reservoirs using a narrow-azimuth towed streamer.

In this way, several 4D FWI strategies have been proposed to obtain time-varying models from time-lapse seismic data and make it an accurate and practical approach. The procedure known as parallel 4D FWI is the most straightforward time-lapse strategy, which involves conducting independent FWIs of the baseline and monitor data using the same initial model \cite{Lumley_2001_Geophysics}. Then, the retrieved models are subtracted from each other to generate the final time-lapse model. However, artifacts in the individual retrieved FWI models can accumulate and negatively impact the quality of the time-lapse estimates \cite{Zhou_Lumley_2021_Geophysics}. Another approach is the sequential 4D FWI strategy (also known as the cascaded 4D FWI), in which the retrieved baseline model serves as the initial model for inverting the monitor data \cite{Routh_et_al_2012_SEG_sequential_4DFWI}; finally, these models are subtracted to generate the associated time-lapse model. The difficulty with this strategy lies in its dependence on a good baseline estimate since artifacts generated in the retrieved baseline and monitor models are unlikely to be eliminated by direct subtraction between them \cite{Zhou_Lumley_2021_Geophysics}. 

Although parallel and sequential 4D FWI are considered primary techniques in 4D seismic processing, they may not yield the most accurate time-lapse estimate compared to more advanced methods. The double-difference 4D FWI strategy is an advanced method that directly inverts the difference between the monitor and baseline data (time-lapse data) using the retrieved baseline estimate as an initial model \cite{Zhang_Huang_2013_Geophysics,Zhang_Huang_2015_Geophysics}. The associated time-lapse model is the difference between the resulting model from the inversion of the time-lapse data and the firstly retrieved baseline model. This strategy holds significant promise but requires substantial seismic pre-processing. From a practical point of view, data subtraction is inherently dangerous whenever variations between datasets originate from sources other than reservoir responses \cite{Zhou_Lumley_2021_Geophysics}. 
 Nevertheless, in the data inversion sense, the parallel, sequential and double-difference strategies require a lower computational cost than other techniques (only two FWI runs are demanded). 

Several robust strategies have been suggested, albeit requiring additional inversions, in an effort to suppress time-lapse (4D) noises \cite{daSilva_et_al_2024_EAGE_Baye_4D_FWI}. For example, Ref. \cite{Maharramov_Biondi_2014__arxiv__cross_updating_time_lapse} proposed a cross-updating strategy consisting of four sequential inversions. This strategy comprises two main sequential branches. The first one consists of a standard execution of the sequential 4D FWI strategy. Then the resulting monitor model is used as the initial model to invert the baseline data, obtaining a second baseline model estimate. This resulting model is then used as the initial model to invert the monitor data again, obtaining a second monitor model. Finally, the resulting time-lapse model is obtained from the subtraction between the second retrieved monitor and baseline models. Ref. \cite{Hicks_et_al_2016_TIMELAPSEFEI_LeadingEdge} introduced the common-model 4D FWI approach, which comprises two main sequential branches, each following the parallel strategy. The first consists of a standard execution of the parallel 4D FWI strategy. The retrieved monitor and baseline models are combined by arithmetically averaging them, generating the so-called common model. Then, the common model is employed as the initial model for a new round of independent FWIs of the monitor and baseline data. These last two FWI models are subtracted to generate the final time-lapse estimate. Ref. \cite{Zhou_Lumley_2021_Geophysics_central_difference_4DFWI} recently proposed the central-difference 4D FWI, in which two independent sequential inversions are performed (forward and reverse bootstraps) followed by a model combination. The forward bootstrap FWI consists in performing a standard sequential 4D FWI \cite{Brenno_RBGF_2024}. Then, the final time-lapse model is derived by combining the two FWI models from the forward bootstrap with the two retrieved models from the reverse bootstrap FWI. 

Naturally, all time-lapse approaches mentioned above have advantages and disadvantages. A common difficulty factor lies in the non-linearity of the FWI problem. So, improving the quality of FWI estimates and mitigating 4D noise using robust statistical tools are very important. In this way, we propose a robust methodology for time-lapse processing with two sequential main steps. First, we consider (i) a new FWI procedure to construct high-resolution P-wave velocity models and, secondly, (ii) a Bayesian analysis approach to alleviate existing biases and artifacts in the final time-lapse models. 
Regarding point (i), we consider an FWI approach with one more degree of freedom associated with the receiver positions to generate the time-lapse models, mitigating the NR effects by decreasing the FWI non-linearity. Then, with the time-lapse models in hand, we consider point (ii), which consists of applying Bayesian analysis to reduce 4D noises by statistically determining mathematical weights. 

In particular, we consider three forms of model weighting; the first one consists of linearly combining the forward and reverse bootstraps resulting models. The second consists of a linear combination between the time-lapse models generated from the sequential and parallel approaches, while the third is based on redefining the central difference approach as the difference between weighted means rather than arithmetic ones. The coefficients of the linear combinations and the weighted average are determined by applying the Bayes' theorem, which provides a solid statistical framework based on evidence and incorporation of prior information \cite{Bayes___Nature___Review____2021}.

We opt for Bayesian analysis due to its powerful and flexible statistical approach to obtain more accurate and informed estimates. Indeed, Bayesian analysis has played an important role in geosciences, allowing the effective integration of prior information and direct observations to obtain a more accurate and complete understanding of subsurface models in different contexts, such as time-lapse problems involving post-stack data inversion \cite{Buland_Ouair_2006_Geophysics} and stochastic frameworks  \cite{Zhang_Curtis_2023_arXiv}, seismic tomography issues using machine learning methodologies \cite{Agata_et_al_2023_IEEE_Trans_Geoscien}, as well as in seismic facies classification \cite{Feng_et_al_2021_IEEE_Trans_Geoscien}. 

This work is organized as follows. In Section \ref{sec:conventional_FWI}, we present the basic concepts and theoretical foundations of the conventional FWI formulation. In Section \ref{sec:receiver_extension_FWI}, we present the receiver-extension FWI approach by paralleling the concepts presented in the previous section. In Section \ref{sec:time_lapse_FWI}, we briefly review the three time-lapse strategies employed in this work, including the parallel, sequential and central-difference 4D FWI. Then, in Section \ref{sec:bayesian_analysis}, we introduce new time-lapse model settings in a Bayesian analysis context. In Section \ref{sec:numerical_experiments}, we present numerical simulations and results of several time-lapse experiments by considering a wide variety of NR issues. We consider two different subsurface models, the first being the Marmousi model and the second a realistic subsurface model representing a deep-water Brazilian pre-salt oil region and an ocean bottom node (OBN) geometry acquisition. Finally, in Section \ref{sec:conclusion}, we give the concluding remarks and future perspectives.







\section{Conventional FWI formulation \label{sec:conventional_FWI}}

FWI is a wave-equation-based procedure that aims to estimate physical parameters that cannot be observed directly by exploiting the full waveforms recorded during a seismic survey \cite{Fichtner_2010_book_FWI}. It is solved by iteratively reconstructing a subsurface model by matching modeled data to observed data using an optimization technique. Classically, FWI in the time domain is formulated as the following constrained least-squares problem \cite{Virieux_Operto_2009_Geophysics}:
\begin{subequations}
    \begin{equation}
        \underset{m,\psi}{min} \quad
        \frac{1}{2}\sum_{s,r} \int_{0}^{T}
        \Big(\Gamma_{s,r} \psi_s(\textbf{x},t) - d_{s,r}(\textbf{x}_{s,r},t)\Big)^2 dt,
        \label{eq:fwi_least_squares}
    \end{equation}
    subject to
    \begin{equation}
        \mathcal{A}(m)\psi_s(\textbf{x},t) = g_s(\textbf{x}_s,t),
        \label{eq:fwi_constrain}
    \end{equation}
    \label{eq:fwi_constrain_leastsquares}
\end{subequations}    
where $m$ represents the subsurface model parameter, $d_{s,r}^{mod}=\Gamma_{s,r} \psi_s$ and $d_{s,r}$ denote, respectively, the modeled and observed data, in which $\Gamma_{s,r}$ is an extraction operator that take out the wavefields $\psi_s$ at the receiver $r$ associated with the source $s$. The acquisition time is represented by $t \in [0,T]$, while the spatial coordinates are denoted by $\textbf{x}$. The constraint in \eqref{eq:fwi_constrain} represents a wave equation reported in a compact form, in which $\mathcal{A}$ symbolizes the d’Alembert wave operator, while $g_s(\textbf{x}_s,t)$ means the seismic source $s$ at the position $\textbf{x} = \textbf{x}_s$.

An elegant technique to solve the constrained problem formulated in \eqref{eq:fwi_constrain_leastsquares} consists of determining the stationary point of the following Lagrangian functional:
\begin{multline}
\mathcal{L}(m,\psi,\lambda) = \frac{1}{2}\sum_{s,r} \int_{0}^{T} \Big(\Gamma_{s,r} \psi_s(\textbf{x},t) - d_{s,r}(\textbf{x}_{s,r},t)\Big)^2 dt  
- \sum_{s} \int_{0}^{T} \Big\langle \lambda_s(x,t) ,   \mathcal{A}(m)\psi_s(\textbf{x},t) - g_s(\textbf{x}_s,t)\Big\rangle_\textbf{x} dt,
\label{eq:lagrangian}
\end{multline}
where $\lambda$ are Lagrange multipliers (namely the adjoint wavefields in FWI problems), and $\langle a , b \rangle_x = \int_\mathcal{X} a(x) b(x) \,dx$ represents the inner product in the subsurface geological domain $\mathcal{X}$. The stationary point of the Lagrangian \eqref{eq:lagrangian} is determined by computing $\nabla \mathcal{L}(m,\psi_s,\lambda_s)$ $ = 0$, where the gradient of the
Lagrangian is given by $\nabla \mathcal{L}(m,\psi_s,\lambda_s) = \left(\frac{\partial \mathcal{L}}{\partial m}, \frac{\partial \mathcal{L}}{\partial \psi_s}, \frac{\partial \mathcal{L}}{\partial \lambda_s} \right)$, which reads:
\begin{subequations}
\begin{equation}
\frac{\partial \mathcal{L}}{\partial m} = -\sum_{s}\int_{0}^{T}\Big\langle \lambda_s(t)   ,   \frac{\partial \mathcal{A}(m)}{\partial m}\psi_s(t)\Big\rangle_x dt ,
\label{eq:lagrangian_gradient_m}
\end{equation}
\begin{equation}
    \frac{\partial \mathcal{L}}{\partial \psi_s} = \sum_{s,r} \int_{0}^{T}
    \Gamma_{s,r}^\dagger \, \left(\Gamma_{s,r} \psi_s(t) - d_{s,r}(t)\right) dt - \sum_{s} \int_{0}^{T} \mathcal{A}^\dagger(m) \lambda_s(t) dt,
    \label{eq:lagrangian_gradient_psi}
\end{equation}
\begin{equation}
\frac{\partial \mathcal{L}}{\partial \lambda_s} = - \sum_{s} \int_{0}^{T} \Big(\mathcal{A}(m)\psi_s(t) - g_s(t)\Big) dt,
\label{eq:lagrangian_gradient_lambda}
\end{equation}
\end{subequations}
where the spatial coordinates $x$ are implicit for the sake of a simplified notation.

Note that at the stationary point of the Lagrangian, Eq. \eqref{eq:lagrangian_gradient_lambda} becomes, for each source $s$, the wave equation \eqref{eq:fwi_constrain}. This result means that the wave equation must always be satisfied in FWI problems, resulting in $\psi_s(x,t) = \mathcal{A}^{-1}(m)\,g_s(\textbf{x}_s,t)$. Analyzing Eq. \eqref{eq:lagrangian_gradient_psi} at the stationary point, we obtain a new wave equation given by
\begin{equation}
    \mathcal{A}^\dagger(m) \lambda_s(x,t) = \Gamma_{s,r}^\dagger \, \big(\Gamma_{s,r} \psi_s(x,t) - d_{s,r}(x,t)\big)
    \label{eq:lagrangian_gradient_psi_adjointequation}
\end{equation}
for each seismic source $s$. Note that the latter wave equation solves for the Lagrange multiplier $\lambda_s$, where the associated wave operator $\mathcal{A}^\dagger$ is the adjoint operator of $\mathcal{A}$ and the seismic source is the difference between the modeled and observed data (residual data). Equation \eqref{eq:lagrangian_gradient_psi_adjointequation} is known as the adjoint wave equation \cite{Plessix_2006_GJI}.

In other words, considering that the wave equation \eqref{eq:fwi_constrain_leastsquares} is always satisfied and that the Lagrange multipliers $\lambda_s$ associated with the Lagrangian function \eqref{eq:lagrangian} are computed through Eq. \eqref{eq:lagrangian_gradient_psi_adjointequation}, we can rewrite the constrained optimization problem formulated in \eqref{eq:fwi_constrain_leastsquares} as follows:
\begin{subequations}
\begin{equation}
    \underset{m}{min} \,\,\,\, \phi(m) = 
    \frac{1}{2}\sum_{s,r} \int_{0}^{T}
        \Delta d_{s,r}^2(m, \textbf{x}, t) dt
    \label{eq:fwi}
\end{equation}
with
\begin{equation}
\Delta d_{s,r}(m, \textbf{x}, t) = \Gamma_{s,r}\, \mathcal{A}^{-1}(m)\,g_s(\textbf{x}_s,t)  - d_{s,r}(\textbf{x}_{s,r},t),
    \label{eq:residual_data}
\end{equation}
\end{subequations}
where $\phi$ represents the objective function and $\Delta d_{s,r}$ is the residual data. In this work, we consider the acoustic wave equation, and therefore, the d’Alembert wave operator is given by $\mathcal{A}(m) = \nabla^2 - m\,\frac{\partial^2}{\partial t^2}$, in which $\nabla^2$ is the Laplace operator and $m = 1/c^2$ represents the slowness squared; $c$ is the P-wave velocity. In this way, the modeled data for each model $m$ are obtained after solving the following acoustic wave equation: 
\begin{equation}
\nabla^2 \psi_s (\textbf{x},t) - m(\textbf{x})\,\frac{\partial^2 \psi_s (\textbf{x},t)}{\partial  t^2 }   
= g_s(\textbf{x}_s,t).
    \label{eq:wave_eq_time}
\end{equation}

Due to the high computational effort demanded by FWI algorithms, it is typically solved using a local optimization scheme by employing gradient-based optimizers. Starting from an initial model $m_0$, the FWI process iteratively retrieves a subsurface model in according to:
\begin{equation}
    m_{j+1}=m_j-\xi_{j} H_j^{-1} \nabla_{m} \mathbf{\phi}({m}_j),
    \label{eq:metodoquasiNewton}
\end{equation}
where $j = 0, 1, 2, \cdots, N_{iter}$ represents the FWI iteration with $N_{iter}$ the maximum number, $\xi_j > 0$ represents a step-length \cite{nocedal_book_2006}, $H$ represents the Hessian (second-order partial derivatives of the objective function $\phi$), and $\nabla_{m}\phi({m}_j) = \frac{\partial \phi({m}_j)}{\partial {m}}$ is the first-order partial derivatives of the objective function with respect to model parameters, i.e., the gradient of the objective function.

It is worth noting that the Lagrangian combines the objective function with the problem's constraint. Therefore, the gradient of the objective function $\mathbf{\phi}(m)$ can be calculated considering the result in Eq. \eqref{eq:lagrangian_gradient_m}, i.e., $\nabla_m \phi(m) = \partial \mathcal{L} / \partial m$. In addition, let us define a new adjoint-state variable as $\mu_s(x,t) = \lambda_s(x,T-t)$ to give a physical sense, as suggested by Ref. \cite{lailly1983}. In this regard, the gradient of the objective function can be expressed as:
\begin{equation}
    \nabla_m \phi(m) =
- \sum_{s} \int_{0}^{T} \Big\langle \mu_s(\textbf{x},T-t) ,   \frac{\partial \mathcal{A}(m)}{\partial m}\psi_s(\textbf{x},t)\Big\rangle_\textbf{x} dt.
\label{eq:gradient_adjoint_state}
\end{equation}
The new adjoint-state variable $\mu$ is known as the backpropagated wavefield and satisfies 
\begin{equation}
    \mathcal{A}^\dagger(m) \mu_s(t) = \sum_{r} \Gamma_{s,r}^\dagger \, \Big(\Gamma_{s,r} \psi_s(T-t) - d_{s,r}(T-t)\Big).
\end{equation}
Thus, the gradient of the objective function is, in fact, the result of the cross-correlation between  the second-order time derivative of the wavefield obtained in the forward modeling (since $\partial \mathcal{A}(m) / \partial m = -\partial^2 / \partial t^2$ in the acoustic case) with the backpropagated wavefield $\mu$.

\section{Receiver-extension FWI \label{sec:receiver_extension_FWI}}

The receiver-extension strategy adopted in this work consists of adding an artificial degree of freedom in the positioning of the receivers. This task is carried out only when sampling the modeled data using the $\Gamma_{s,r}$ operator and for each source-receiver pair \cite{Metivier_Brossier_2022_Geophysics_receiverextension}. In particular, we consider that the extraction operator $\Gamma_{s,r}$ now depends on a receiver position correction parameter $\Delta \textbf{x}_{s,r}$, i.e., $\Gamma_{s,r}(\Delta x_{s,r})$ \cite{daSilva_et_all_2023_EAGE_receiverextension_4D_FWI}. In other words, instead of sampling the modeled wavefields only at the nominal positions $x_r$ of the receivers, the $\Gamma_{s,r}$ operator will sample at the position $x_r^* = x_r + \Delta x_{s,r}$, where $\Delta x_{s,r}$ is a real number. In this way, we can formulate the receiver-extension FWI as the following joint optimization problem:
\begin{subequations}
\begin{equation}
\underset{m,\Delta x}{min} \quad \phi_{re}(m,\Delta x) = \phi_0(m,\Delta x) + \sum_{s,r} \gamma_{s,r}^2\,\Delta x_{s,r}^2,    \label{eq:receiver_extension_objective_function}
\end{equation}
with
\begin{equation}
    \phi_0 = \frac{1}{2}\sum_{s,r} \int_{0}^{T} \Big(\Gamma_{s,r}(\Delta x_{s,r})\, \psi_s(t)  - d_{s,r}(t)\Big)^2 dt,
\end{equation}
\label{eq:receiver_extension_FWI_formulation}
\end{subequations}
where $\phi_0 = \phi_0(m, \Delta x)$, $d_{s,r}(t) = d_{s,r}(x_{s,r},t)$, $\psi_s(t) = \psi_s(m,t) = \mathcal{A}^{-1}(m)\,g_s(\textbf{x}_s,t)$, $\phi_{re}$ represents the receiver-extension objective function, and $\gamma_{s,r}$ are weight parameters that balance the contributions between the first term (associated with the data misfit) and the second term (associated with receiver position corrections). The receiver-extension FWI determines optimal values for the model parameters $m$ and the (artificial) reallocation of the receivers from $x_r$ to $x_r + \Delta x_{s,r}$.  It is worth noting that this approach does not impact the wave equation solution in the forward modeling process. Indeed, the new extraction operator only samples the modeled wavefields in relocated positions for the receivers. 

The receiver-extension FWI can be formulated as a bilevel optimization problem, where one is embedded within the other. Assuming that the receiver-extension objective function $\phi_{re}$ is a twice differentiable function and that its second-order partial derivatives are continuous, the joint problem formulated in Eq. \eqref{eq:receiver_extension_FWI_formulation} is equivalently solved in two sequential steps (levels), where the first one is the so-called inner loop (lower-level optimization task) and the second is the outer loop (higher-level optimization task); each level may adopt its own strategy to optimize its payoffs \cite{Colson_et_al_book_bilevelOptimization__2005}. 

In this work, we consider that the inner loop consists of selecting the receiver position corrections given a model $m$, solving the following single optimization problem:
\begin{equation}
    \Delta\hat{x}_{s,r}(m) =  \underset{\Delta x}{\textrm{argmin}} \quad  \phi_{re}(m,\Delta x_{s,r}),
\end{equation}
where $\Delta\hat{x}_{s,r}(m)$ is the optimal receiver position correction for each source-receiver pair. Although there is no analytical solution with a closed-form expression for $\Delta\hat{x}_{s,r}(m)$, the inner loop solution can be quickly obtained through, for example, heuristic optimization techniques \cite{Rardin_Uzsoy_2001_Heuristic_Optimization}. In this sense, the receiver-extension strategy entails minimal additional computational cost and does not require significant additional computational resources. This is because the solution of the inner loop can be easily determined by, for example, sampling several values of $\Delta \hat{x}$ and determining which one gives the smallest value of the objective function.

The outer loop can be performed once the optimal receiver position corrections are obtained, which consists of performing a conventional FWI iteration considering the relocated receivers at $x_r + \Delta\hat{x}_{s,r}$, i.e.,
\begin{equation}
    \hat{m} =  \underset{m}{\textrm{argmin}} \quad  \phi_{re}(m,\Delta \hat{x}),
\end{equation}
where $\hat{m}$ is the optimal model given the correction $\Delta\hat{x}$. The gradient of the receiver-extension objective function $\phi_{re}$, given a collection of receiver position corrections $\Delta\hat{x}_{s,r}$, via adjoint-state method is given by Eq. \eqref{eq:gradient_adjoint_state}, in which the adjoint wavefield is calculated through: 
\begin{subequations}
    \begin{equation}
    \mathcal{A}^\dagger(m) \mu_s(t) = \sum_{r} \Upsilon_{s,r}
    \end{equation}
where
    \begin{equation}
        \Upsilon_{s,r} = \Gamma_{s,r}^\dagger(x_r^*) \, \Big(\Gamma_{s,r}(x_r^*) \psi_s(T-t) - d_{s,r}(T-t)\Big)
    \end{equation}
\end{subequations}
represents the adjoint source and $x_r^* = x_r + \Delta\hat{x}_{s,r}$.

\newpage

\section{Time-lapse FWI strategies \label{sec:time_lapse_FWI}}

\begin{figure}[!htb]
    \centering
    \includegraphics[width=.5\columnwidth]{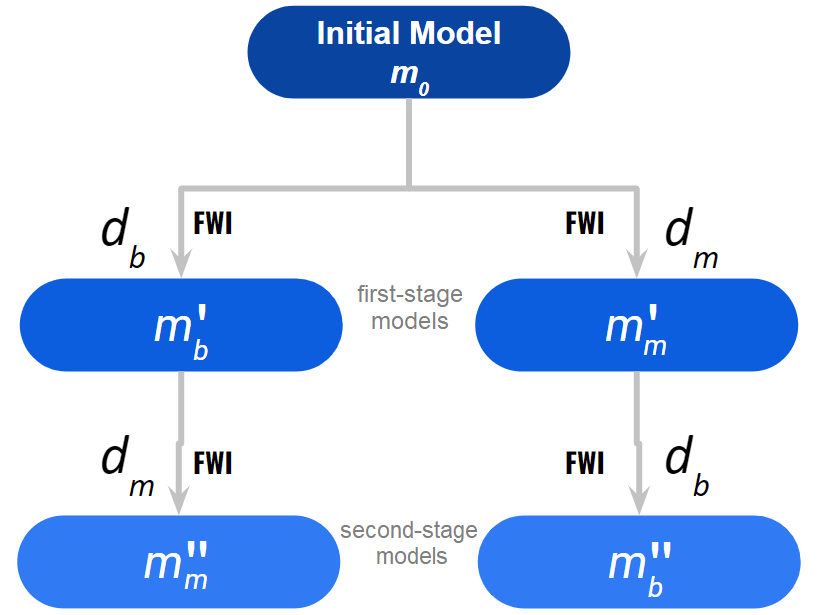}
    \caption{Flowchart illustrating obtaining the first- and second-stage models.}
    \label{fig:time_lapse_approaches}
\end{figure}

In this section, we review three different strategies for 4D FWI, including parallel 4D FWI, sequential 4D FWI, and central-difference 4D FWI. In these approaches, time-lapse models are obtained by combining monitor and baseline models. The first step consists of independently executing parallel FWIs using the baseline data $d_b$ and the monitor data $d_m$ from the same initial model $m_0$, obtaining a baseline model $m_b'$ and a monitor model $m_m'$, respectively. Such models independently retrieved from $m_0$ are named first-stage retrieved models (or first-stage models). The second-stage retrieved models are obtained by conducting FWIs starting from the first-stage retrieved models. The second-stage monitor model $m_m''$ is recovered by inverting the monitor data $d_m$ using the first-stage recovered baseline model $m_b'$ as the initial model. The second-stage baseline model $m_b''$ is recovered by inverting the baseline data $d_b$ from the first-stage recovered monitor model $m_m'$. The flowchart in Fig. \ref{fig:time_lapse_approaches} summarizes how to obtain first- and second-stage models. 

In the following, we describe these three classical time-lapse strategies: \\ \\

\begin{enumerate}
    \item[(i)] The \textit{parallel 4D FWI strategy} \cite{Lumley_2001_Geophysics} considers only the first-stage retrieved models. The associated time-lapse model $\Delta m_{par}$ is obtained by subtracting the first-stage monitor model $m_m'$ from the first-stage baseline model $m_b'$ following:
    \begin{equation}
        \Delta m_{par} = m_m' - m_b'.    
        \label{eq:delta_m__parallel}
    \end{equation}
    \item[(ii)] The \textit{sequential 4D FWI} \cite{Routh_et_al_2012_SEG_sequential_4DFWI} considers only the left arm of the flowchart depicted in Fig. \ref{fig:time_lapse_approaches}. The associated time-lapse model $\Delta m_{seq}$ is obtained by subtracting the second-stage monitor model $m_m''$ from the first-stage baseline model $m_b'$ following:
    \begin{equation}
        \Delta m_{seq} = m_m'' - m_b'.    
        \label{eq:delta_m__sequential}
    \end{equation}
    \item[(iii)] The \textit{central-difference 4D FWI} \cite{Zhou_Lumley_2021_Geophysics_central_difference_4DFWI} comprises three key steps: forward bootstrap, reverse bootstrap, and model averaging. The forward and reverse bootstraps correspond, respectively, to the left and right arms of the flowchart depicted in Fig. \ref{fig:time_lapse_approaches}. The model averaging step involves calculating the difference between two averages, one related to the retrieved baseline models and the other to the retrieved monitor models. The associated time-lapse model $\Delta m_{cd}$ is obtained through:
    \begin{equation}
        \Delta m_{cd} = \frac{m_m' + m_m''}{2} - \frac{m_b' + m_b''}{2}.
        \label{eq:delta_m_cd}
    \end{equation}
\end{enumerate}

\newpage
\section{Time-lapse models using Bayesian analysis\label{sec:bayesian_analysis}}
In this work we propose three distinct Bayesian weighted time-lapse 4D FWI strategies, namely \textit{$\text{BW}_1$}, \textit{$\text{BW}_2$} and \textit{$\text{BW}_3$} in reference to Bayesian weighted, as follows:

\begin{itemize}
    \item[\textit{$\text{BW}_1$}:] The authors in Ref. \cite{Zhou_Lumley_2021_Geophysics_central_difference_4DFWI} state that existing artifacts in the forward bootstrap should appear in the reverse bootstrap with a phase reversal. Thus, a linear combination between bootstraps can succeed in canceling out several artifacts if weighted appropriately. Defining the bootstrap-related time-lapse models as 
    \begin{subequations}
    \begin{equation}
        \Delta m_{seq}^+ = m_m^{''} - m_b^{'}
        \label{eq:forward_bootstrap}
    \end{equation}
    and
    \begin{equation}
        \Delta m_{seq}^- = m_m^{'} - m_b^{''},
        \label{eq:reverse_bootstrap}
    \end{equation}
    \end{subequations}
    
    we propose a new time-lapse model based on the weighted average:
    \begin{equation}
        \Delta m_{{}_{{BW}_1}} = \frac{\alpha\,\Delta m_{seq}^+ + \beta\,\Delta m_{seq}^-}{\alpha + \beta},
        \label{eq:delta_m__bw1}
    \end{equation}
    where $\alpha$ and $\beta$ are (non-negative) weights.  
    \item[\textit{$\text{BW}_2$}:] The second proposed model involves combining the parallel and sequential strategies. This procedure offers the advantage of reducing the computational efforts, as it requires performing only three FWIs instead of four, as is the case with the central-difference approach and the \textit{$\text{BW}_1$} model. We define the \textit{$\text{BW}_2$} model as the following linear combination: 
    \begin{equation}
        \Delta m_{{}_{{BW}_2}} = \frac{\alpha\,\Delta m_{par} + \beta\,\Delta m_{seq}}{\alpha + \beta},
        \label{eq:delta_m__bw2}
    \end{equation}
    where $\Delta m_{par}$ and $\Delta m_{seq}$ are defined by Eqs. \eqref{eq:delta_m__parallel} and \eqref{eq:delta_m__sequential}, while $\alpha$ and $\beta$ are weights.
    It is worth noting that the \textit{$\text{BW}_2$} model actually consists of  obtaining a weighted-average monitor model to compare with the baseline model. To see this, note that Eq. \eqref{eq:delta_m__bw2} can be rewritten as:
    \begin{equation}
    \Delta m_{{}_{{BW}_2}} = \frac{\alpha\,m_m^{'} + \beta\,m_m^{''}}{\alpha + \beta} - m_b^{'}.
    \end{equation}
    \item[\textit{$\text{BW}_3$}:] The third strategy involves a new strategy to redefine the central difference 4D FWI in \eqref{eq:delta_m_cd}. Instead of estimating it as the difference between simple arithmetic means, we propose computing it as the difference between weighted means. The weighted average allows us to assign different weights to model elements. It is useful when some model is more important than another in calculating the average model, being flexible for several different time-lapse projects. We define the \textit{$\text{BW}_3$} model as follows:
    \begin{equation}
        \Delta m_{{}_{{BW}_3}} =  \frac{\alpha\,m_m' + \beta\,m_m''}{\alpha + \beta} - \frac{\vartheta\,m_b' + \varepsilon\,m_b''}{\vartheta + \varepsilon},
        \label{eq:delta_m__bw3}
\end{equation}
in which the parameters $\alpha$, $\beta$, $\vartheta$, and $\varepsilon$ are (non-negative) weights.
\end{itemize}

We emphasize that although the three models (Eqs. \eqref{eq:delta_m__bw1}-\eqref{eq:delta_m__bw3}) use the same notation for the constants $\alpha$ and $\beta$, their values in each equation differ from those in the others. To simplify the notation, we use the same letters.

Fundamentally, our proposal models are based on weighted averages between baseline and monitor models. Such a formulation is interesting from a statistical point of view since the weighted mean is more flexible than the arithmetic mean, as it allows the adjustment of the importance of each model in the average depending on the situation. This feature makes it suitable for situations where some models are more relevant than others, and therefore, the weighted mean can provide a more accurate estimate of the average measure. 

Although the weighted mean is a valuable measure of central tendency in statistical analyses it has some limitations, such as the need to assign appropriate weights. In this work we propose to determine the weights using Bayesian analysis. In this context, such an approach has several advantages since we have previous information (a priori information) about the distribution of weights, and we want to update this information based on observations. Bayesian analysis is crucial for parameter estimation tasks because it offers a flexible and illuminating method for drawing conclusions about unknown parameter values in statistical models \cite{Bayes___Nature___Review____2021}. This statistical technique differs from traditional frequentist approaches, where probabilities are treated as relative frequencies in numerous repetitions of the same experiment. The Bayesian analysis considers hypotheses or models by incorporating prior information and updating it as new data is observed. 

The foundation of Bayesian analysis is the Bayes' theorem, which is stated as:
\begin{equation}
    P(\boldsymbol{\Theta} | \mathbf{D}, M) =
\frac{P(\mathbf{D} | \boldsymbol{\Theta}, M)
P(\boldsymbol{\Theta} | M)}{P(\mathbf{D} | M)},
\end{equation}
where $P$ represents a probability function and $\boldsymbol{\Theta}$ stands for a set of parameters for a model $M$ given some data $\mathbf{D}$. In the latter equation, $P(\boldsymbol{\Theta} | \mathbf{D}, M)$ denotes the posterior probability, which represents the updated belief or probability after considering new evidences, i.e., the probability of a hypothesis in light of the observed evidence. The maximum value of such a probability function is what we are interested in determining through the Bayesian analysis. $P(\mathbf{D} | \boldsymbol{\Theta}, M)$ and $P(\boldsymbol{\Theta} | M)$ represent the likelihood function and the prior probability, respectively, while    
\begin{equation}
    P(\mathbf{D} | M) = \int
P(\mathbf{D} | \boldsymbol{\Theta}, M) P(\boldsymbol{\Theta}| M) \,
d\boldsymbol{\Theta}
\end{equation}
represents the evidence (or marginal likelihood), with the integral encompassing all possible values of $\boldsymbol{\Theta}$.

In our case, $\boldsymbol{\Theta}$ represents the set of parameters $\boldsymbol{\Theta} = \{\alpha, \beta\}$ for the \textit{$\text{BW}_1$} and \textit{$\text{BW}_2$} models, while $\boldsymbol{\Theta} = \{\alpha, \beta, \vartheta, \varepsilon\}$ is for the \textit{$\text{BW}_3$} model, and $M$ correspond to the proposed \textit{$\text{BW}$} models. Suppose a given model $M$ (\textit{$\text{BW}_j$}, with $j = 1, 2, 3$) can be represented by a vector $\Delta\mathbf{m} = \{\Delta\,m_i\}_{i= 1}^{n} = \{\Delta\,m_1, \Delta\,m_2, ..., \Delta\,m_n\}$. This means that $\Delta \mathbf{m}$ represents a discretized time-lapse model with $n$ real elements. In this case, $\mathbf{D}$ represents the elements of the time-lapse model ($\mathbf{D} = \Delta \mathbf{m} = \{D_i = \Delta\,m_i \mid i = 1, 2, \ldots, n \}$), associated with the retrieved baseline and monitor models.

In this work, we consider a Gaussian likelihood function with zero mean and standard deviation $\sigma$, defined, for a single model point $i$, as
\begin{equation}
    P(D_{i} | \boldsymbol{\Theta}, M) = \frac{1}{\sigma\sqrt{2\,\pi}}\exp\left(-\frac{1}{2\,\sigma^2}\chi^{2}(D_{i} | \boldsymbol{\Theta}, M)\right),
\end{equation}
where $\chi$ is given by
\begin{subequations}
\begin{equation}
\chi(D_{i} | \boldsymbol{\Theta}, M \!=\! \textit{$\text{BW}_1$}) = \frac{\alpha\,\Delta m_{\text{seq}, i}^{+} + \beta\,\Delta m_{\text{seq}, i}^{-}}{\alpha + \beta},
\end{equation}
\begin{equation}
\chi(D_{i} | \boldsymbol{\Theta}, M \!=\! \textit{$\text{BW}_2$}) = \frac{\alpha\,\Delta m_{\text{par}, i}  + \beta\,\Delta m_{\text{seq}, i}}{\alpha + \beta},
\end{equation}
and
\begin{equation}
   \chi(D_{i} | \boldsymbol{\Theta}, M \!=\! \textit{$\text{BW}_3$}) = \frac{\alpha\,m_{m,i}' + \beta\,m_{m,i}''}{\alpha + \beta} - \frac{\vartheta\,m_{b,i}' + \varepsilon\,m_{b,i}''}{\vartheta + \varepsilon}
\end{equation}
\label{eq:receiver_extension_FWI_formulation}
\end{subequations}
for the $\text{BW}_1$, $\text{BW}_2$ and $\text{BW}_3$, respectively. Then, the Gaussian likelihood for all model points is $P(\mathbf{D} | \boldsymbol{\Theta}, M) =  \prod_{i=1}^n \,P(D_i | \boldsymbol{\Theta}, M)$.

We notice that in all proposed models, Bayesian analysis attempts to maximize the likelihood function by determining weighting parameters that minimize the dissimilarities between monitor and baseline models while maximizing the similarities between monitor models and between baseline models. Since we assume a likelihood function with zero mean in our analyses, the weighting parameters are determined in such a way that the differences between the resulting monitor and baseline models are minimal. Moreover, we conducted several simulations varying only the value of $\sigma$ and found that its variation had no significant effect on determining the parameters for weighting the time-lapse models. Therefore, we fixed $\sigma = 1$ for all numerical experiments.

We consider uniform distributions as prior probabilities because they ensure the objectivity and transparency of the analysis and establish that all values within a given range are equally likely \cite{Bayes___Nature___Review____2021}. In particular, we restrict that the values of $\alpha$, $\beta$, $\vartheta$ and $\varepsilon$ lie within the range of $(0,1]$.

Bayesian analysis is based on the estimation of posterior probabilities and the comparison of statistical evidence between competing models \cite{Bayes___Nature___Review____2021}. In this work, we use a dynamic nested (dynesty) sampler \cite{Speagle_DYNESTY_2020}, a sampling algorithm that is very effective in exploring parameter spaces and computing Bayesian posteriors and evidences accurately. We invite readers who are particularly interested in this sampling technique's theory and implementation to consult Refs. \cite{Speagle_DYNESTY_2020,Higson_et_al_DYNESTY_2019,Skilling_DYNESTY_2004}. In this work, we use the version v2.1.3 \cite{Koposov_et_al_DYNESTY_zenodo_2023}, which is written in Python language. Finally, the weighting parameters that generate a model with the highest posterior probability are selected.

By tweaking parameters $\alpha$, $\beta$, $\vartheta$ and $\varepsilon$, we can significantly mitigate time-lapse noises, making artifacts more or less pronounced. Consequently, as we experiment with these parameters and analyze the time-lapse estimates, we disclose that while parameter adjustments should alleviate artifacts, true anomalies should exhibit only subtle value changes, as will be demonstrated later.  In this regard, Bayesian analysis allows the incorporation of uncertainties in determining these weighting parameters, exploring several potential solutions.

\section{Numerical Experiments\label{sec:numerical_experiments}}

In this section we present two case studies. In the first, we consider the Marmousi model as the baseline model. The Marmousi case study aims to show how Bayesian analysis in the context of the proposed weighted time-lapse models effectively mitigates time-lapse noises after analyzing noisy data. In the second case, we investigate the capabilities of the receiver-extension strategy and Bayesian analysis by examining a realistic subsurface P-wave velocity model representing a deep-water Brazilian pre-salt region. In the Brazilian pre-salt case study, the challenge lies in characterizing and monitoring deep oil and gas reservoirs.

In all numerical tests, we employ the limited-memory Broyden-Fletcher-Goldfarb-Shanno (\textit{l}-BFGS) method \cite{nocedal_book_2006} to solve the FWI problem, and set the stopping criterion at 100 iterations. The \textit{l}-BFGS method involves computing an approximation of the inverse of the Hessian matrix based on previous gradients. We employ a line search procedure that satisfies the Wolfe conditions \cite{nocedal_book_2006} to determine the step length, ensuring the algorithm's convergence and efficiency.

\subsection{Marmousi case study}\label{sec:numerical_experiments_marmousi}

\begin{figure}[!b]
    \centering
    \subfloat[\label{fig:marmousi_true_velocity_model}]{%
       \includegraphics[width=.35\linewidth]{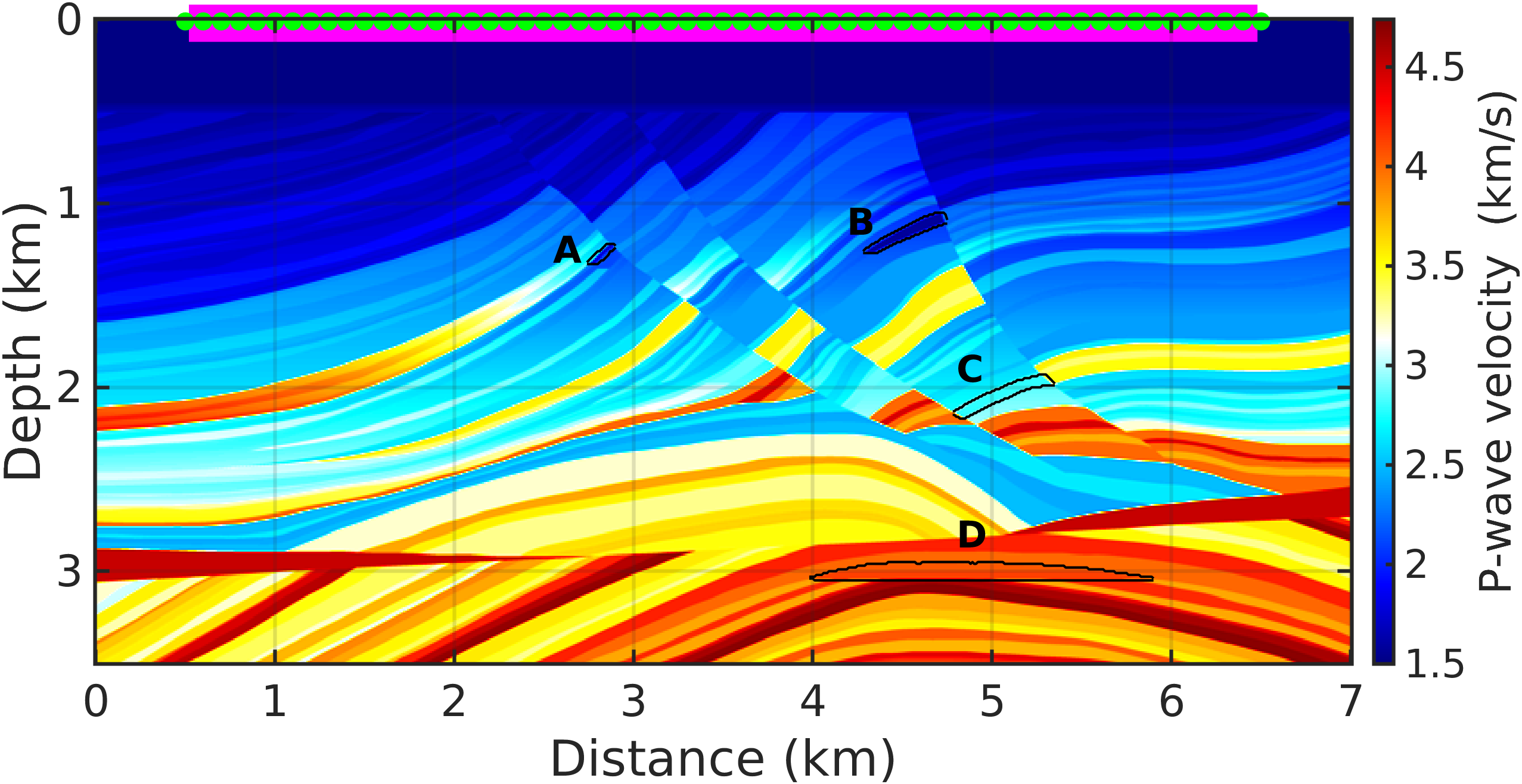}} 
    \\
  \subfloat[\label{fig:marmousi_true_4D_model}]{%
       \includegraphics[width=.35\linewidth]{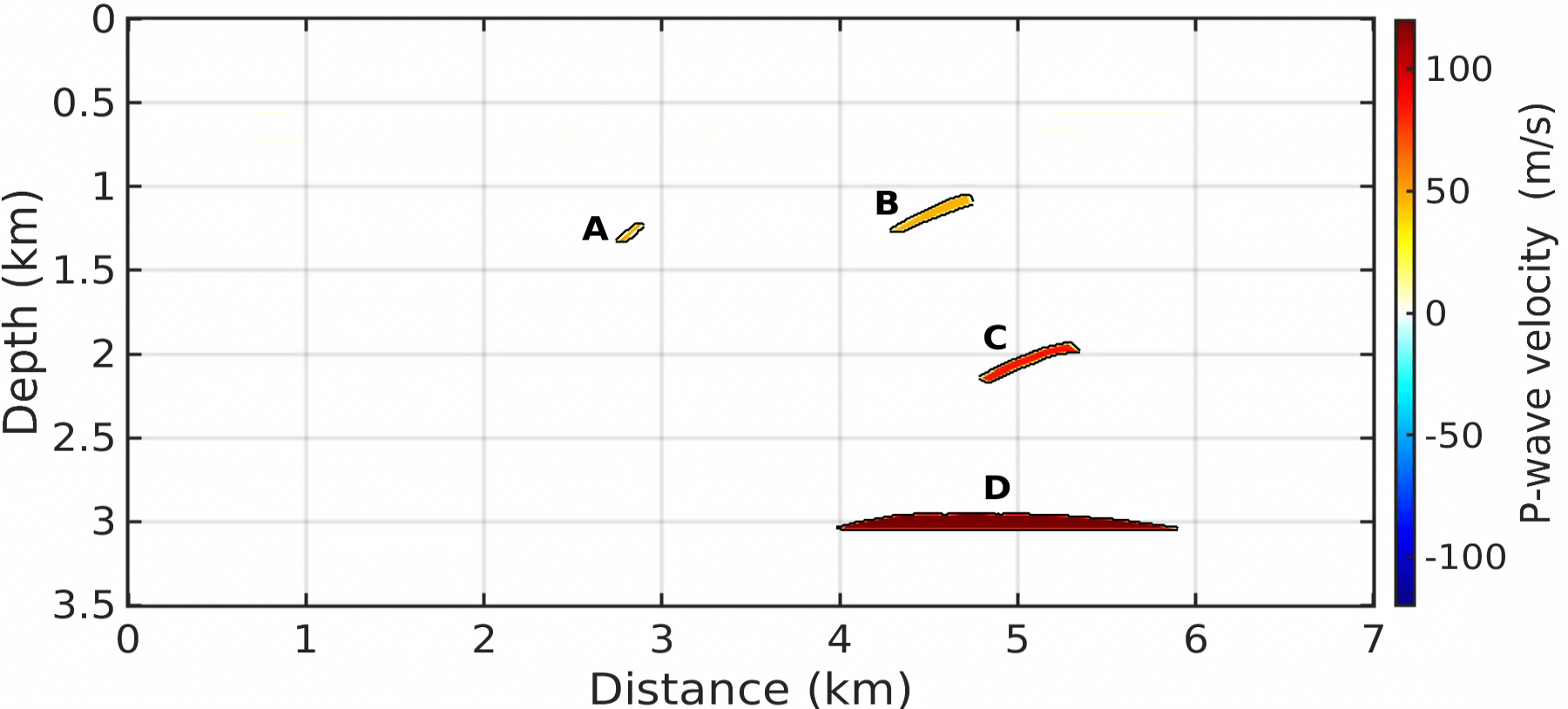}}
       \\
  \subfloat[\label{fig:marmousi_initial_model}]{%
       \includegraphics[width=.35\linewidth]{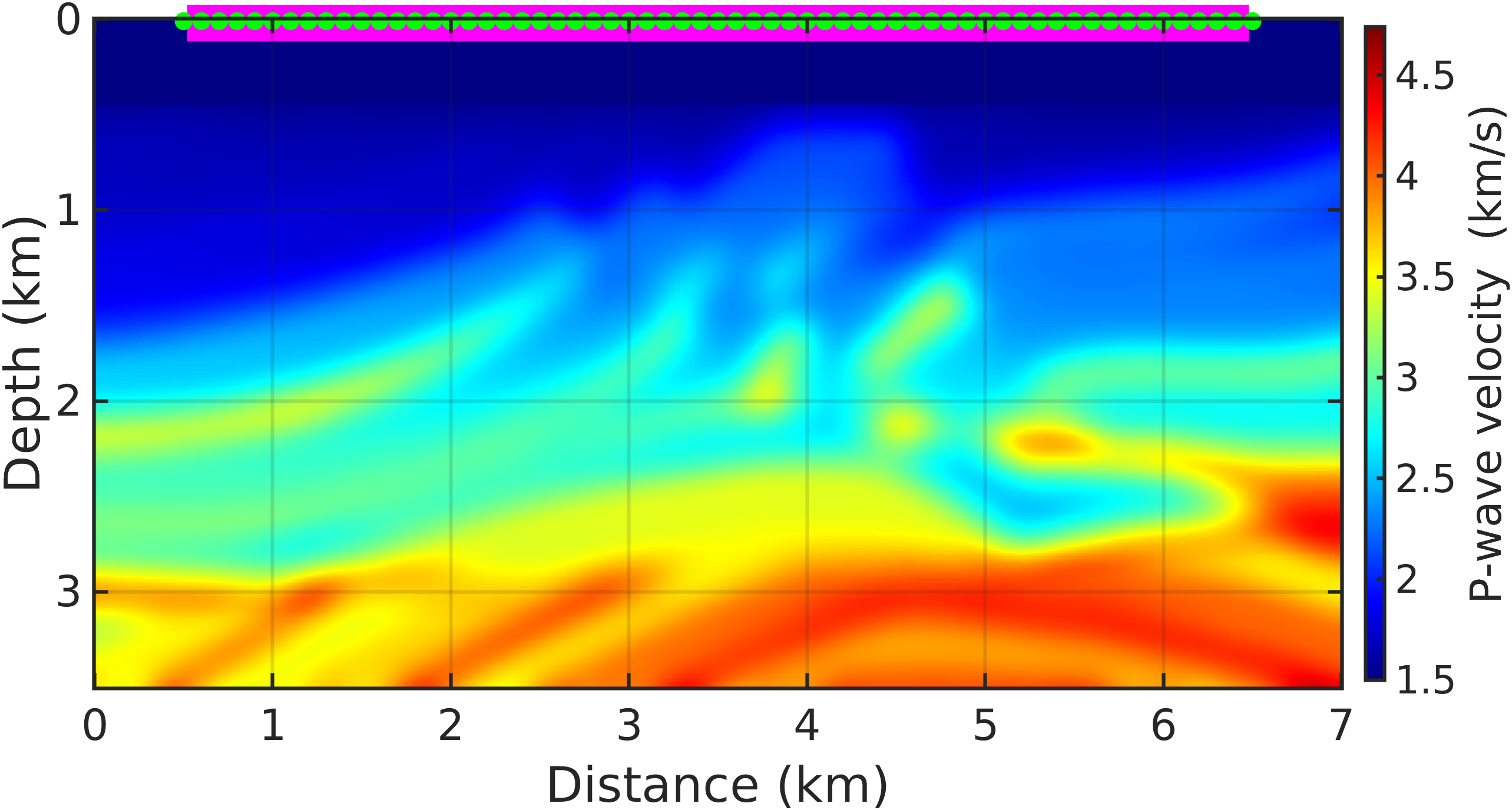}}
    \caption{Velocity models: \textbf{(a)} Baseline model considered in the Marmousi case study. The magenta line represents the receivers' positions, while the green dots represent the locations of the seismic sources. \textbf{(b)} The true time-lapse model (the difference between monitor and baseline models). \textbf{(c)} Initial model $m_0$ used in the forward and reverse bootstrap steps' first stages.}
    \label{fig:marmousi__input_models}
\end{figure}

To demonstrate how the Bayesian analysis in the proposed weighted time-lapse models can mitigate time-lapse noises associated with noisy data analysis, we consider the Marmousi acoustic velocity model \cite{Versteeg_1994_Marmousi} as a reference (true baseline model). This P-wave velocity model is a subsurface representation created based on the geology of the Kwanza Basin region (Angola) \cite{Martin_et_al_2006__Marmousi2}. It includes a variety of complex geologic structures covering an area 7.0 km long and 3.5 km deep, as depicted in Figure \ref{fig:marmousi_true_velocity_model}. We consider a fixed-spread acquisition comprising 597 receivers at regular intervals of 10 m and 61 isotropic explosive sources at regular intervals of 100 m. All sources and receivers are positioned at 10 m depth, as depicted in Figs. \ref{fig:marmousi_true_velocity_model} and \ref{fig:marmousi_initial_model} by the green dots and the magenta line, respectively. Each seismic source is a Ricker wavelet with a cutoff frequency of 20 Hz. The recording time is 4.5 s.

We generate the true monitor model by perturbing the true baseline model. In particular, we changed four regions of the Marmousi model associated with gas sand and oil gaps, increasing P-wave velocities by $3\%$. Figure \ref{fig:marmousi_true_4D_model} shows the true time-lapse model, where the black curves enclose the time-lapse changes for a better visualization. It is important to emphasize that the true monitor model is formed from the sum of the true baseline model with the velocity anomaly. Using the true baseline and monitor models and the acquisition geometry described earlier, we generate 'observed' data sets using the 2D time-domain acoustic wave equation. In particular, we apply the finite difference method with approximations of order $2nd$ and $8th$ for time and space, respectively, in a grid with regular discretization of 10 m.

In this case study, we consider the FWI process under two different scenarios in terms of signal-to-noise ratio (SNR). Both scenarios include seismic data contaminated by Gaussian noise, one with an SNR of 8 dB and the other with an SNR of 15 dB. Therefore, the only NR parameter is related to the seismic noise, which is different between the baseline data and the monitor data.

\begin{figure*}[!htb]
    \centering
    \subfloat[\label{fig:model1_snr15_corner}]
        {%
        \includegraphics[width=0.351\linewidth]{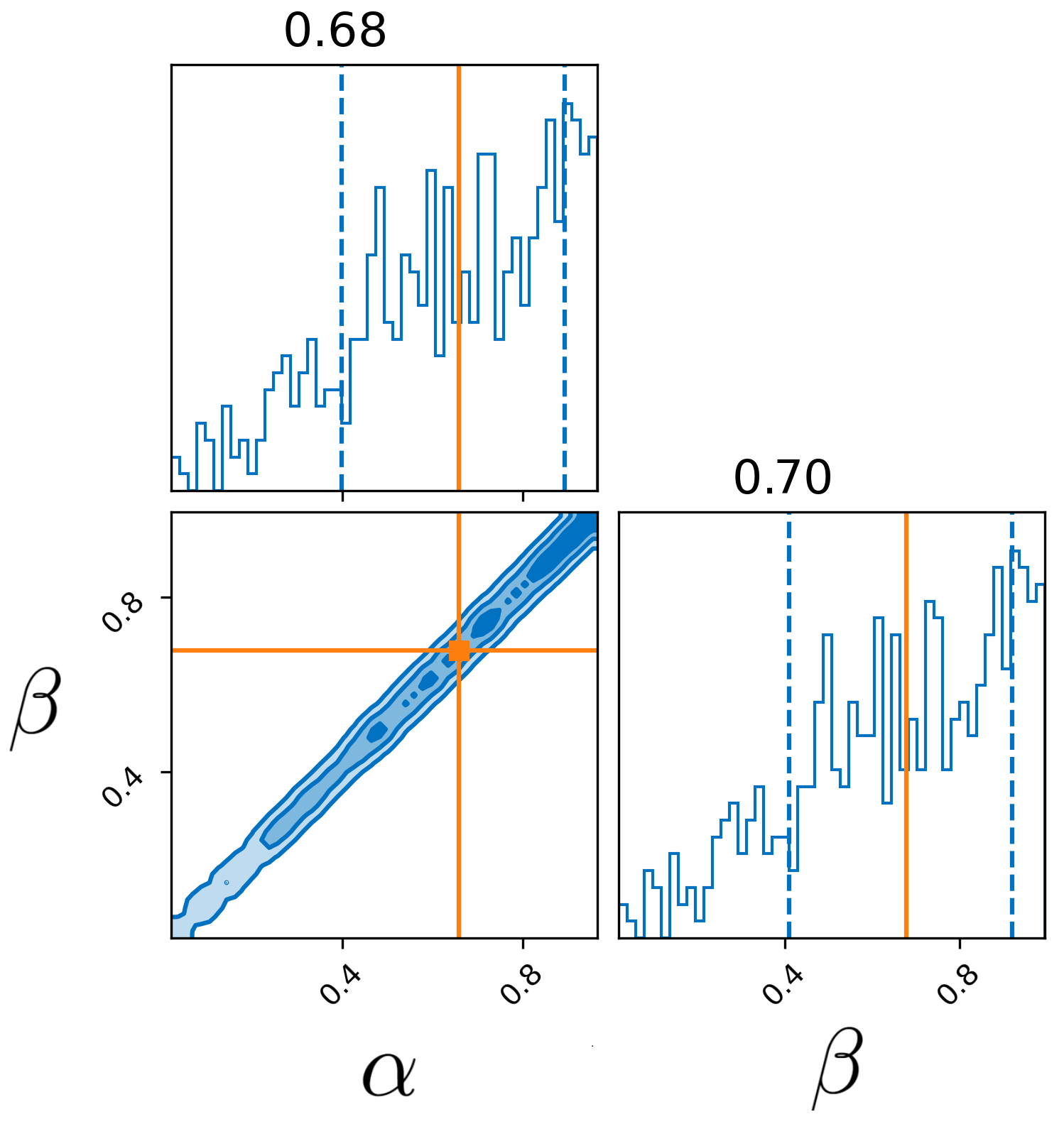}
        }
    \subfloat[\label{fig:model2_snr15_corner}]
        {%
        \includegraphics[width=0.351\linewidth]{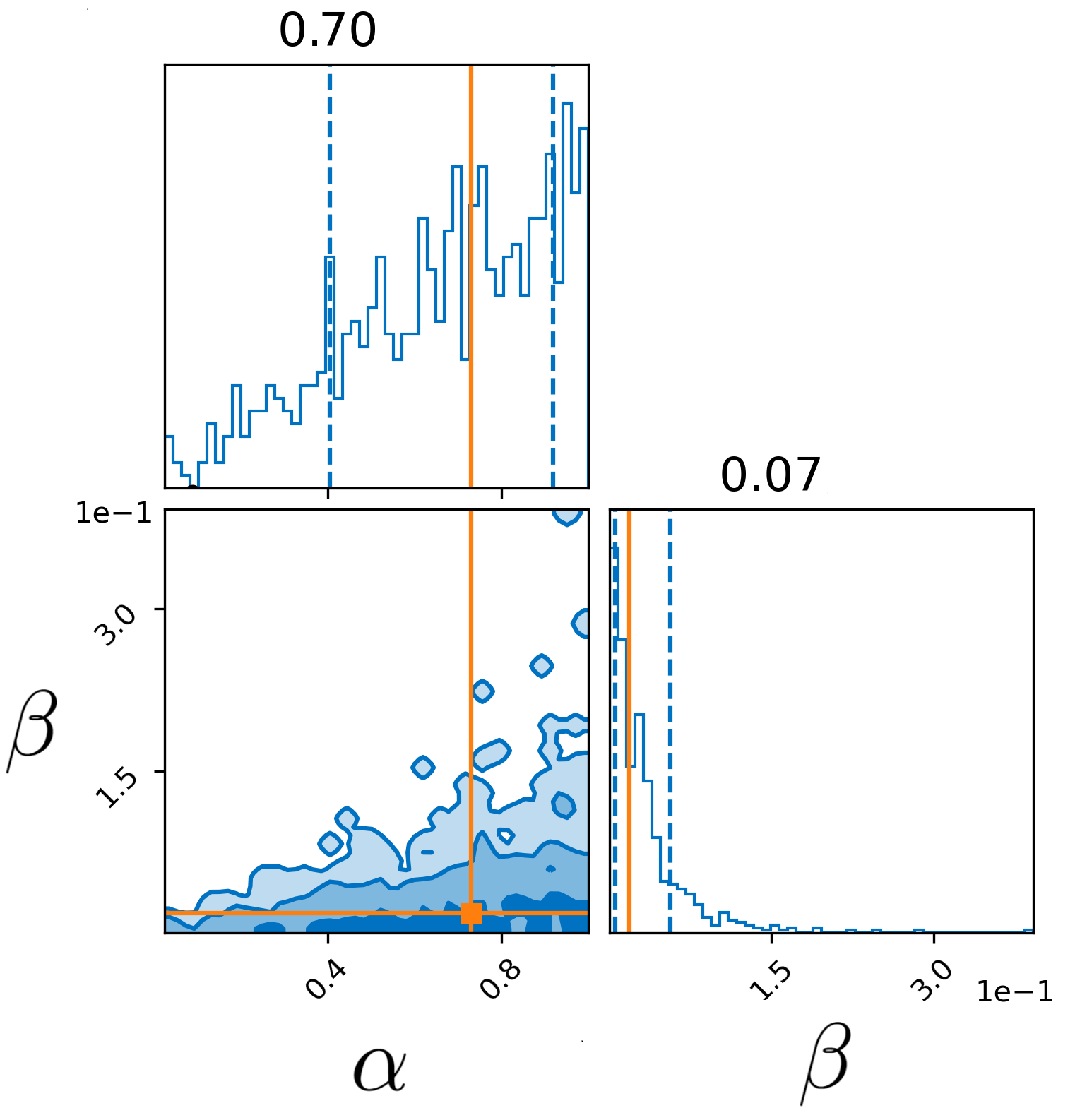}
        }  
    \subfloat[\label{fig:model3_snr15_corner}]
        {%
        \includegraphics[width=0.351\linewidth]{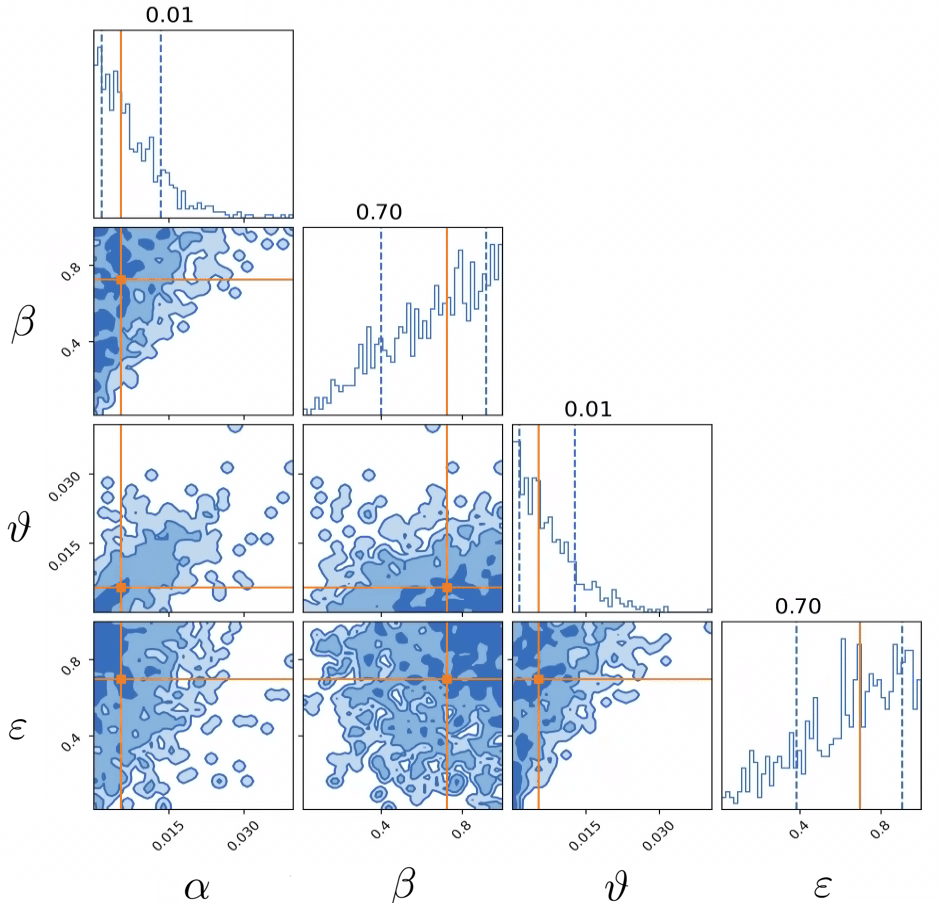}
        }  
        \\
    \subfloat[\label{fig:model1_snr8_corner}]
        {%
        \includegraphics[width=0.351\linewidth]{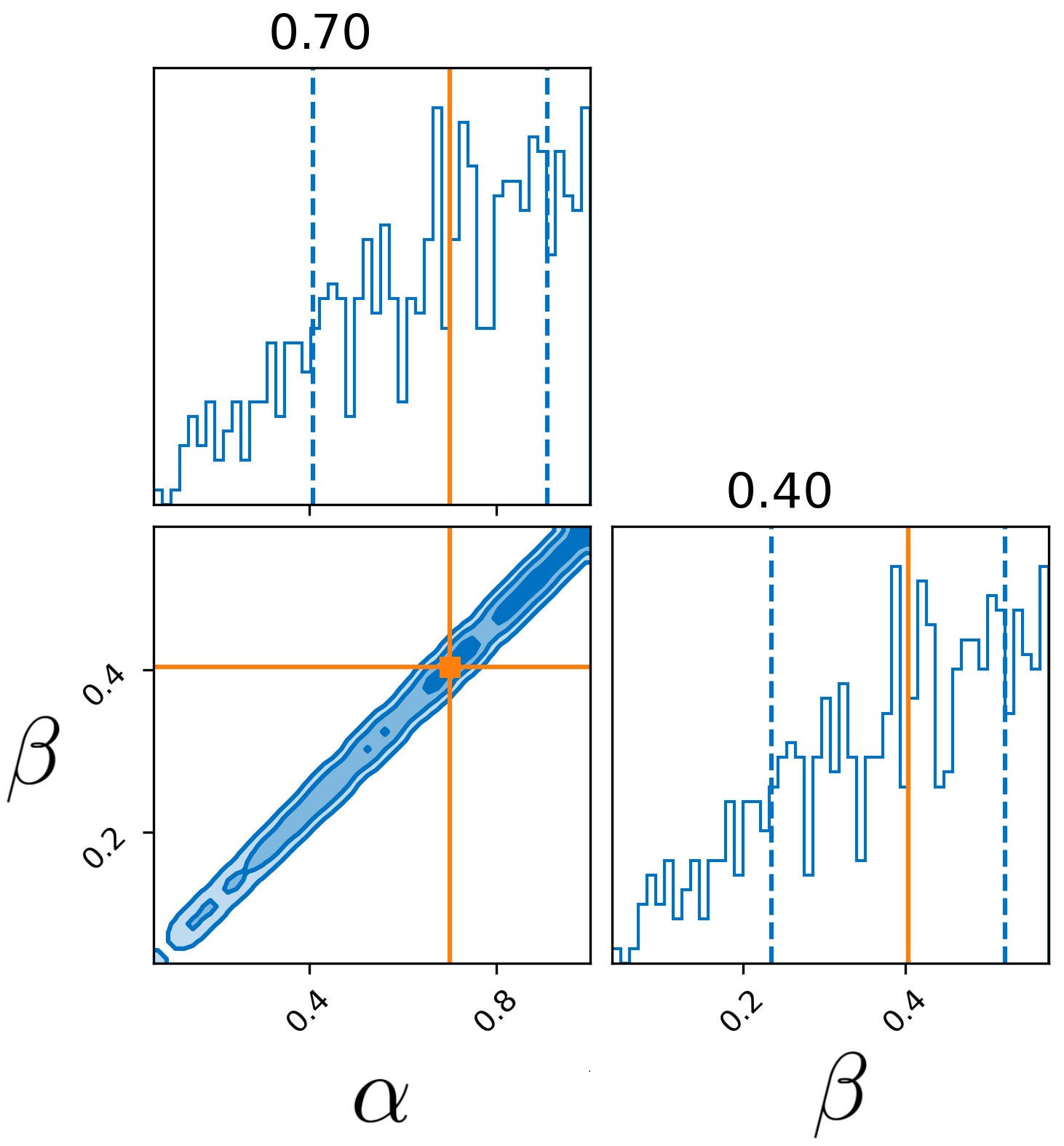}
        }
    \subfloat[\label{fig:model2_snr8_corner}]
        {%
        \includegraphics[width=0.351\linewidth]{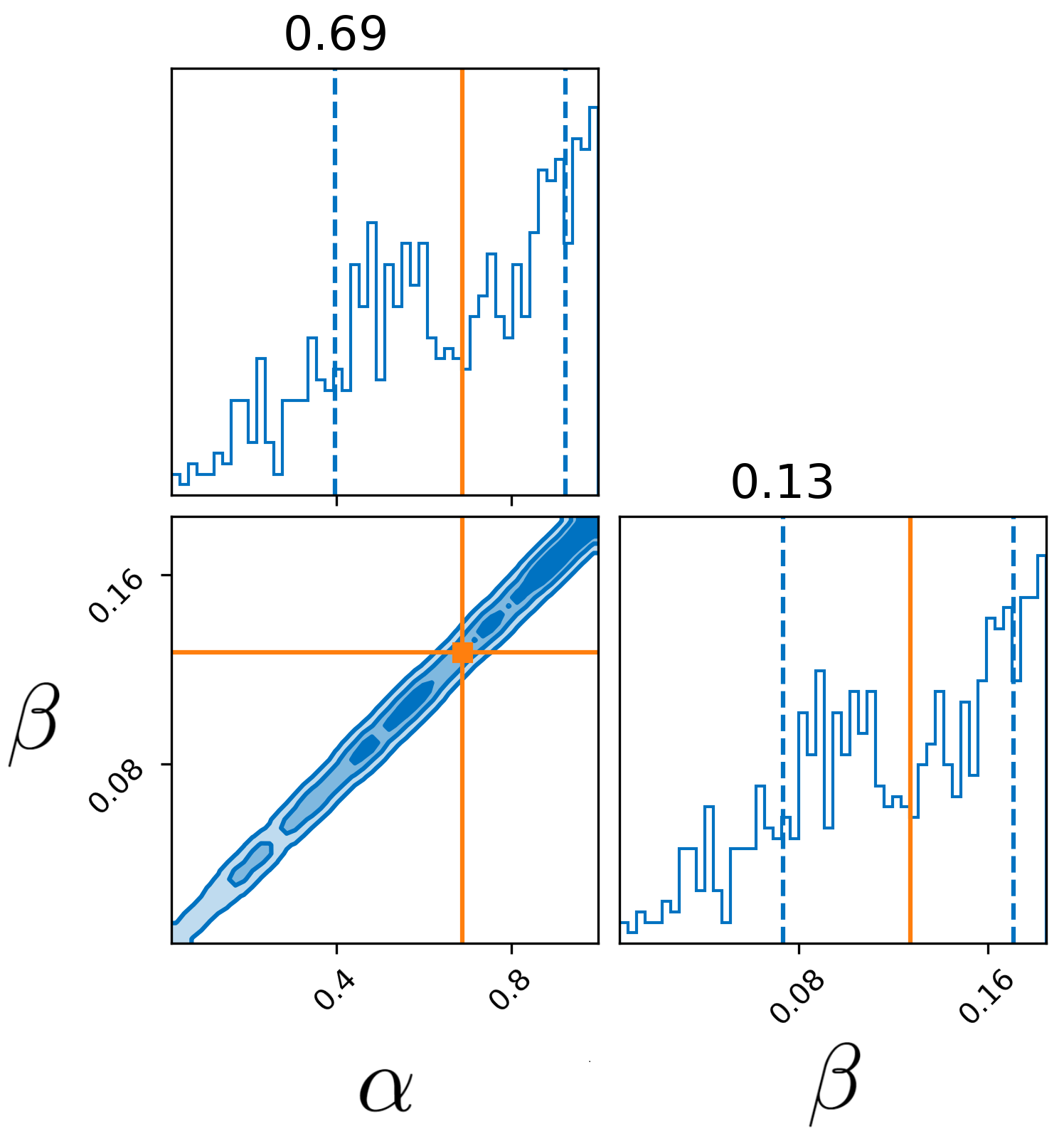}
        }  
    \subfloat[\label{fig:model3_snr8_corner}]
        {%
        \includegraphics[width=0.351\linewidth]{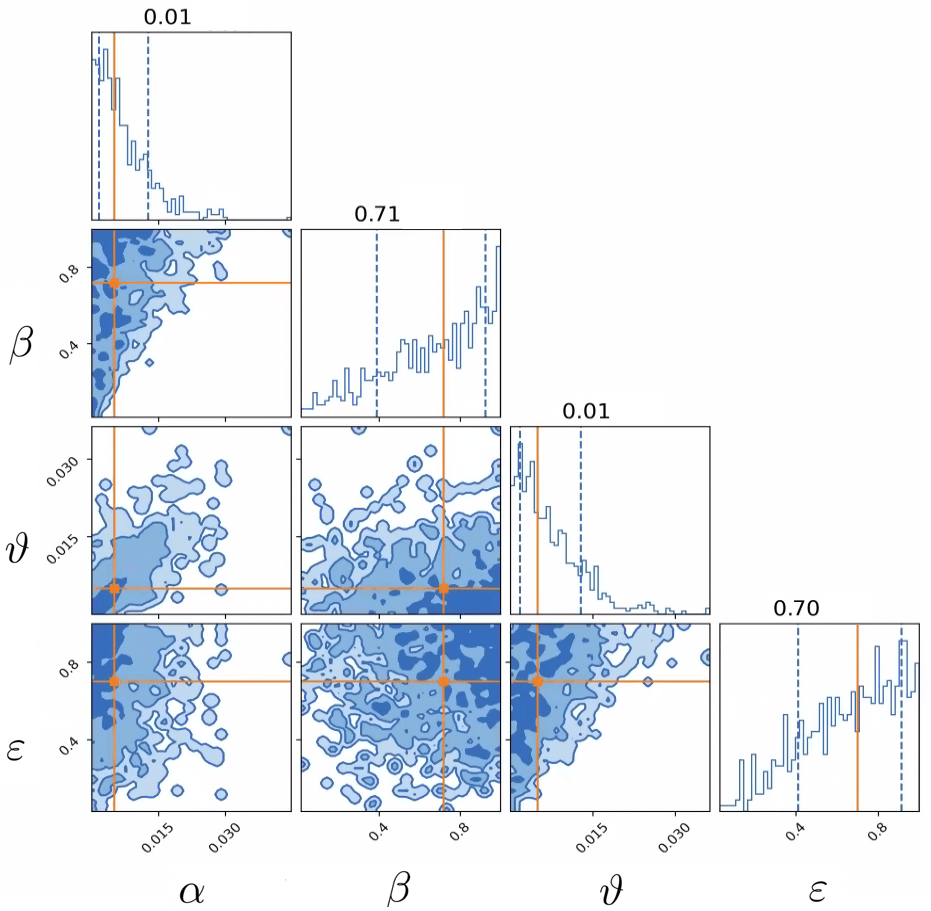}
        } 
    
    \caption{Corner plots of the inferred parameters ($\alpha$, $\beta$, $\vartheta$, and $\varepsilon$). The blue traces represent the obtained posteriors, while the orange lines signify the parameter values maximizing the posterior probability in the Bayesian time-lapse models. The upper row refers to the case with an SNR of 15 dB, the lower row to the case with an SNR of 8 dB. Panels (a)-(c) and panels (d)-(f) refer to the models $BW_1$, $BW_2$ and $BW_3$.}    
    \label{fig:Bayes_corner_Marmousi} 
\end{figure*}

To estimate posteriors and statistical evidence, we use a dynamic nested (dynesty) sampler \cite{Speagle_DYNESTY_2020}, which is integrated into the Bilby package \cite{BILBY_Ashton_et_al_2019, BILBY_2020}. This Bayesian analysis library provides a robust framework for performing efficient and accurate analysis. Figure \ref{fig:Bayes_corner_Marmousi} shows a corner diagram in which the inferred parameters ($\alpha$, $\beta$, $\vartheta$ and $\varepsilon$) are displayed.  The diagram shows the posterior probability calculated for all parameters together with contour plots illustrating the relationship between any two parameters (in solid blue contours), with the upper row corresponding to the time-lapse models derived from  the case with an SNR of 15 dB and the bottom row to the time-lapse models obtained from the case with an SNR of 8 dB. The orange lines represent the parameter values that maximize the posterior probability obtained with the Bayesian time-lapse models \textit{$\text{BW}_1$} (Figs. \ref{fig:model1_snr15_corner} and \ref{fig:model1_snr8_corner}), \textit{$\text{BW}_2$} (Figs. \ref{fig:model2_snr15_corner} and \ref{fig:model2_snr8_corner}), and \textit{$\text{BW}_3$} (Figs. \ref{fig:model3_snr15_corner} and \ref{fig:model3_snr8_corner}).

Analyzing Figs. \ref{fig:model1_snr15_corner} and \ref{fig:model1_snr8_corner}, in the case of the Bayesian-weighted model \textit{$\text{BW}_1$}, when the SNR value is relatively high, note that the parameter values determined through Bayesian analysis are similar ($\alpha = 0.68$ and $\beta = 0.70$), while in the case where the SNR is relatively low, a greater weight is assigned to the model obtained in the forward bootstrap ($\alpha = 0.70$ and $\beta = 0.40$). This indicates that when the background noise is less intense (higher SNR), forward and reverse bootstraps (Eqs. \eqref{eq:forward_bootstrap} and \eqref{eq:reverse_bootstrap}) models can generate similar time-lapse models, and therefore, combining them using an arithmetic average ($\alpha \approx \beta$) is a good approach. Figures \ref{fig:model1_snr15_corner} and \ref{fig:model1_snr8_corner} also suggests that when the background noise is more intense (lower SNR), the forward bootstrap modeling should carry a greater weight than the model generated by the reverse. In the case of the weighted Bayesian model \textit{$\text{BW}_2$}, where the time-lapse models are obtained by the linear combination between the parallel and sequential 4D FWI strategies \eqref{eq:delta_m__bw2}, we notice that the time-lapse model obtained by using the parallel 4D FWI strategy is assigned a greater weight regardless of the SNR scenario, as depicted in Figs. \ref{fig:model2_snr15_corner} and \ref{fig:model2_snr8_corner}. This result is consistent, since the sequential approach failed in all experiments performed and should therefore be weighted lower. Figures \ref{fig:model3_snr15_corner} and \ref{fig:model3_snr8_corner} depict the corner plots for the \textit{$\text{BW}_3$} case. In this strategy  \eqref{eq:delta_m__bw3}, the Bayesian analysis favored the second-stage time-lapse models ($m_m''$ and $m_b''$).

\begin{figure*}[!b]
    \centering
    \subfloat[\label{fig:high_srn_par_time_lapse_model}]{%
       \includegraphics[width=.48\linewidth]{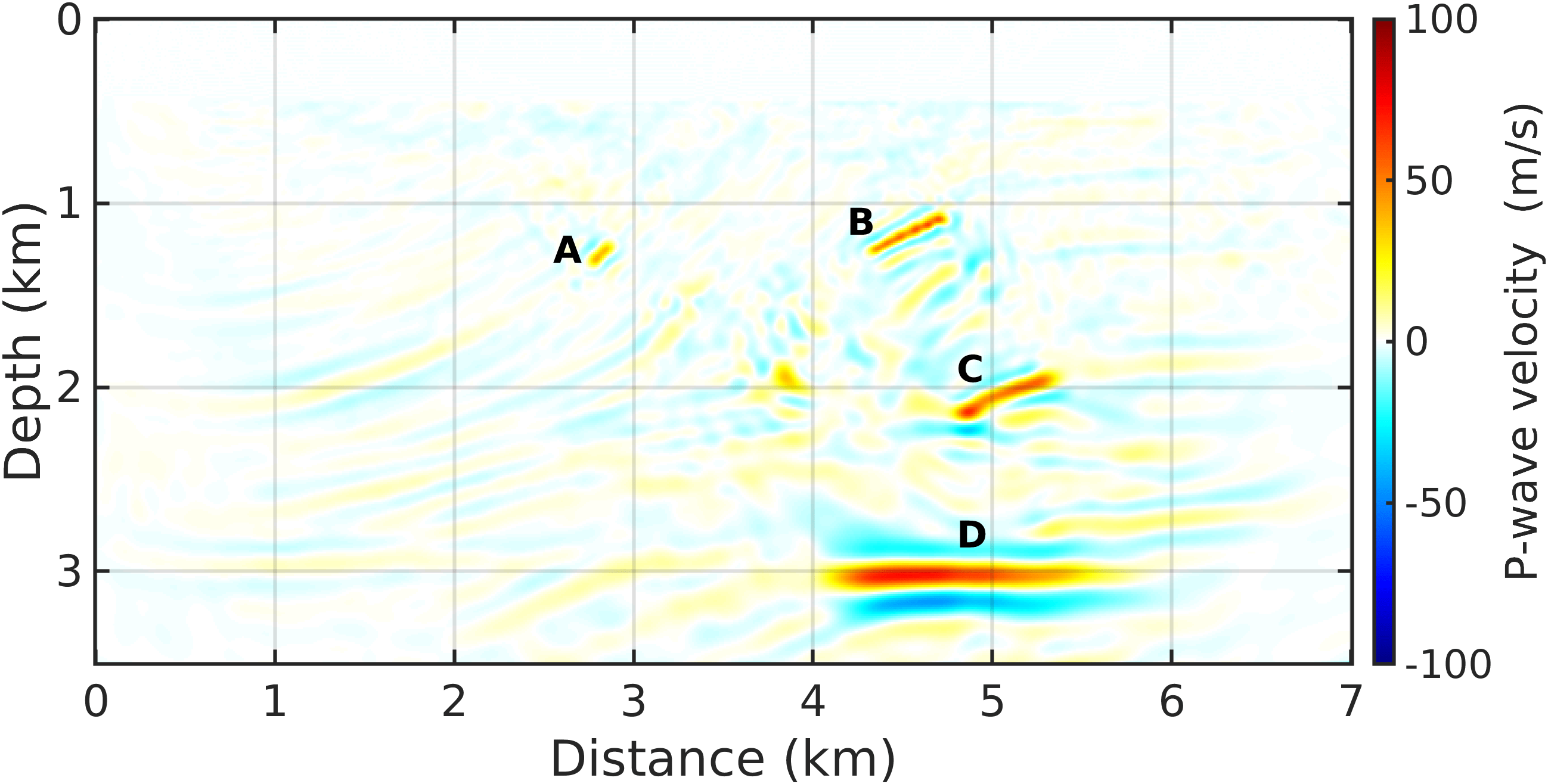}
       } 
       \hfill
    \subfloat[\label{fig:high_srn_bw1_time_lapse_model}]{%
       \includegraphics[width=.47\linewidth]{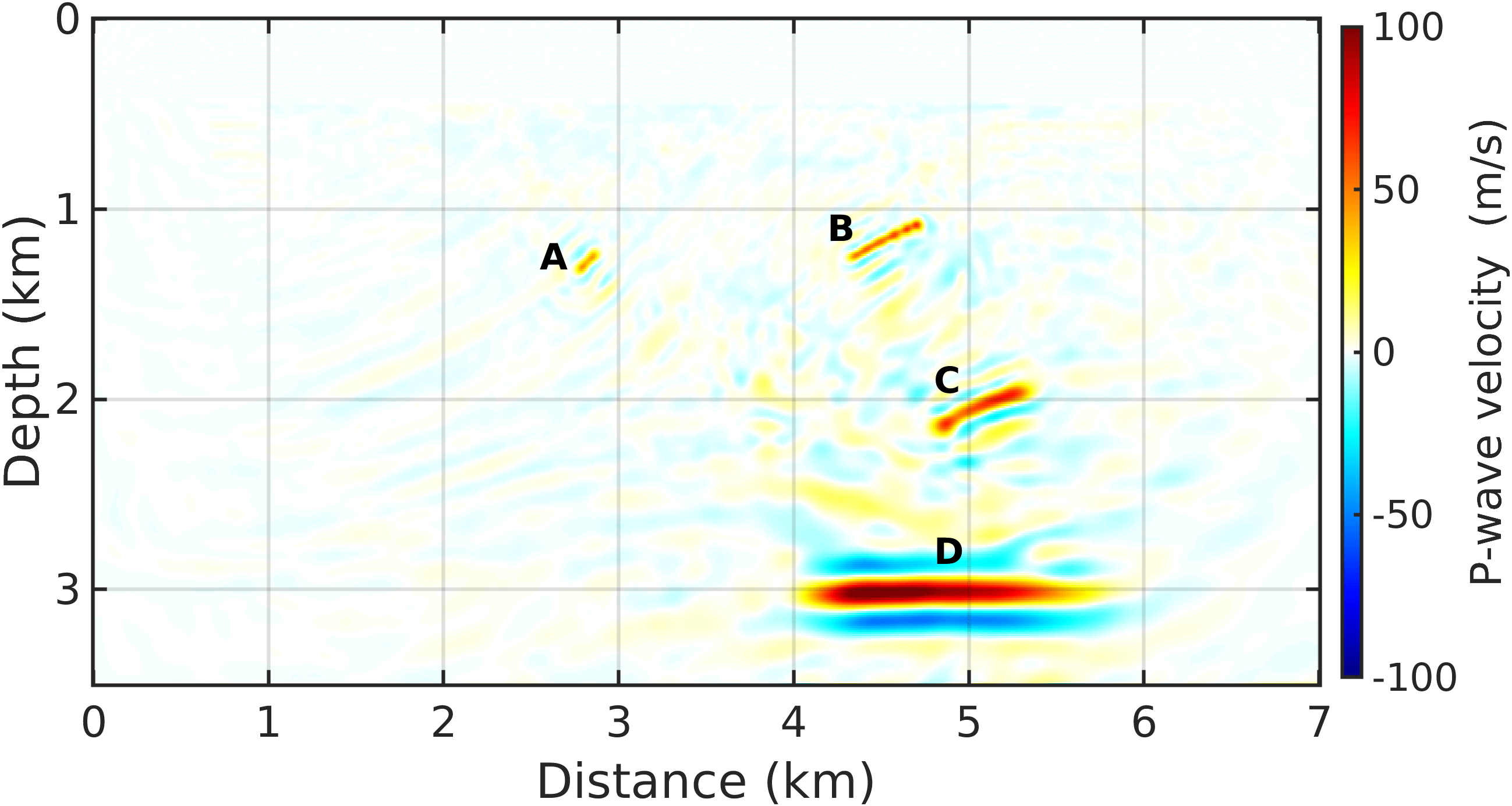}
       }
    \\
    \subfloat[\label{fig:high_srn_seq_time_lapse_model}]{%
       \includegraphics[width=.48\linewidth]{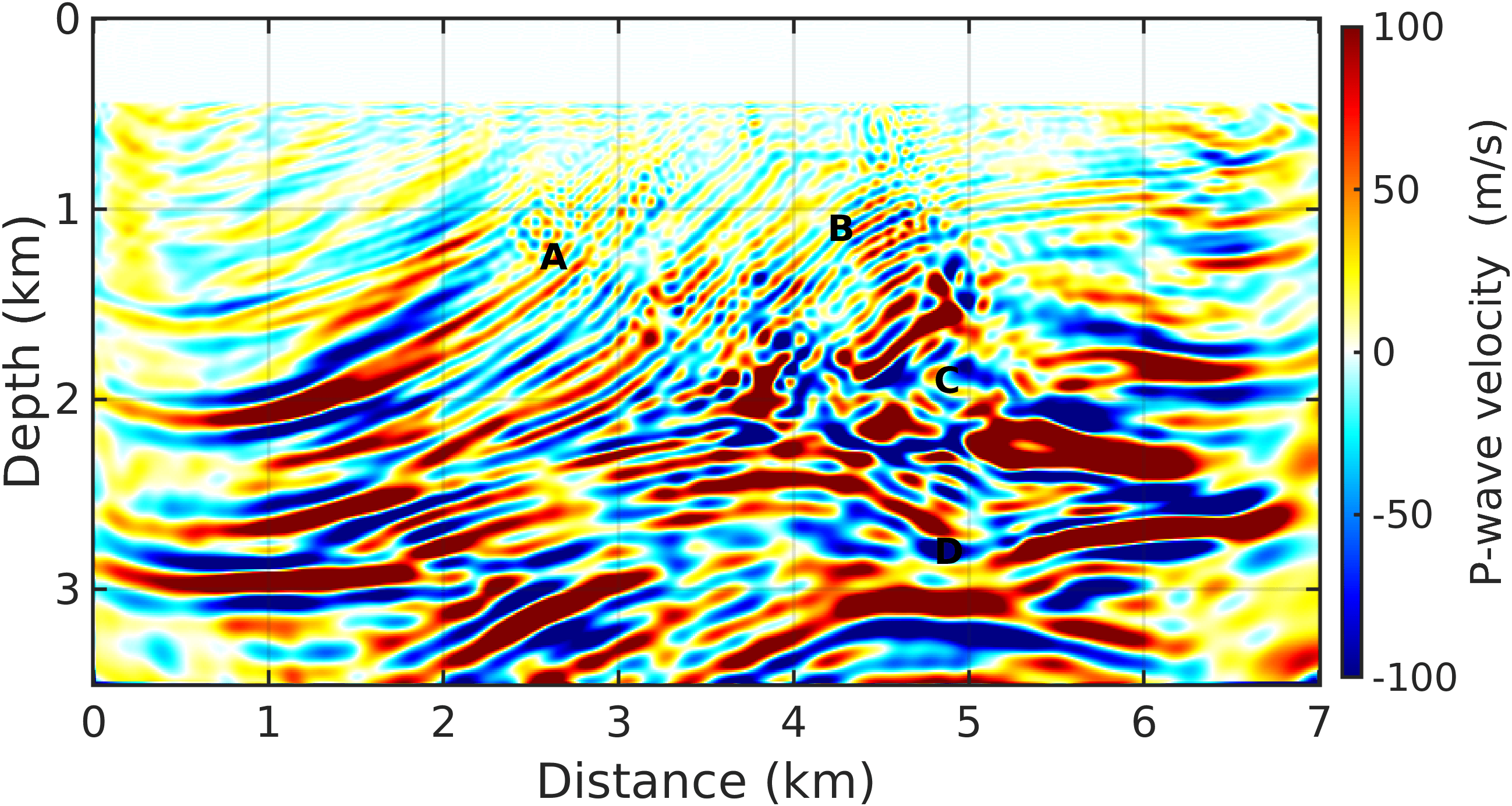}
       } 
       \hfill
    \subfloat[\label{fig:high_srn_bw2_time_lapse_model}]{%
       \includegraphics[width=.48\linewidth]{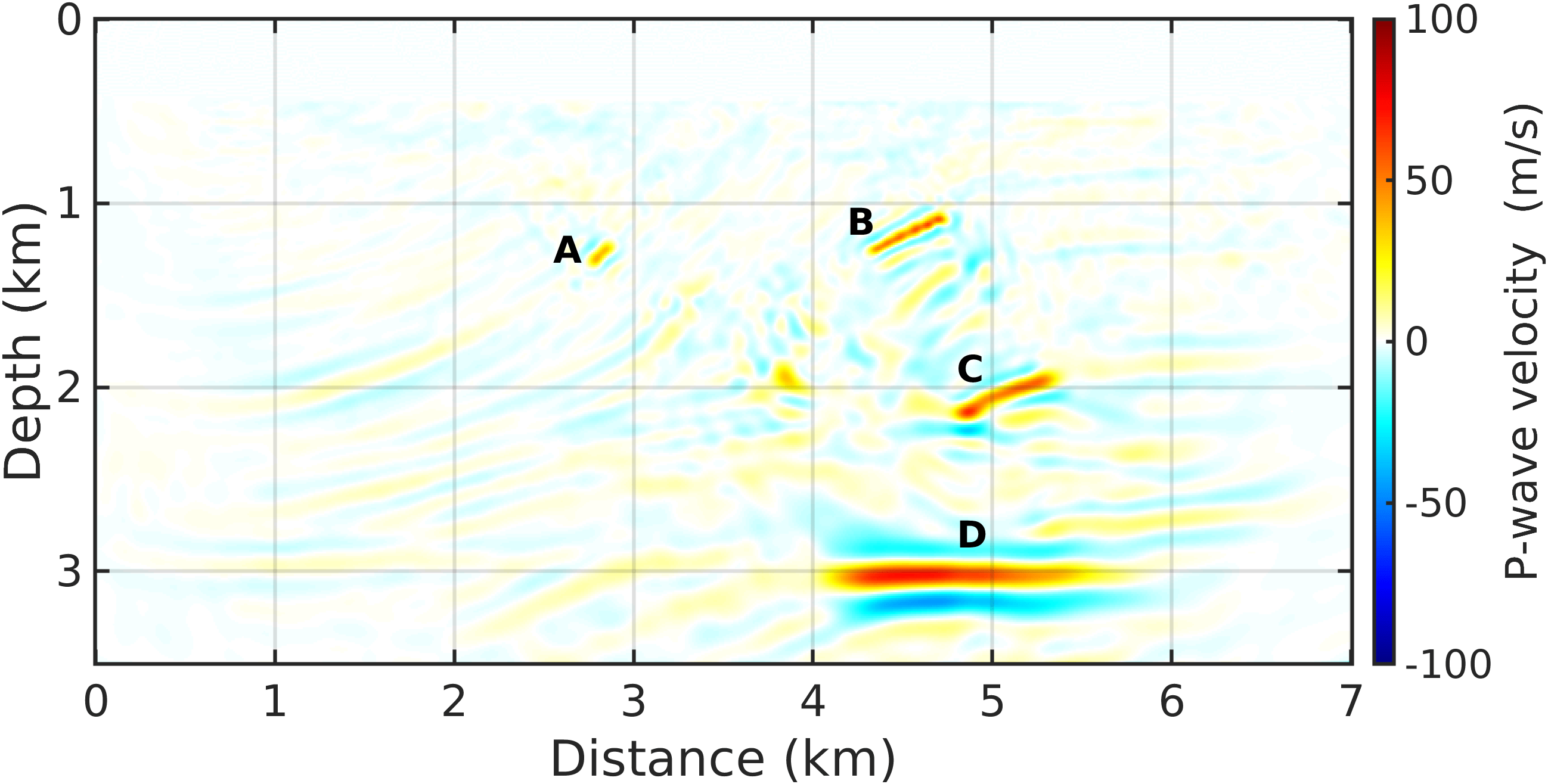}
       }
    \\
    \subfloat[\label{fig:high_srn_cd_time_lapse_model}]{%
       \includegraphics[width=.48\linewidth]{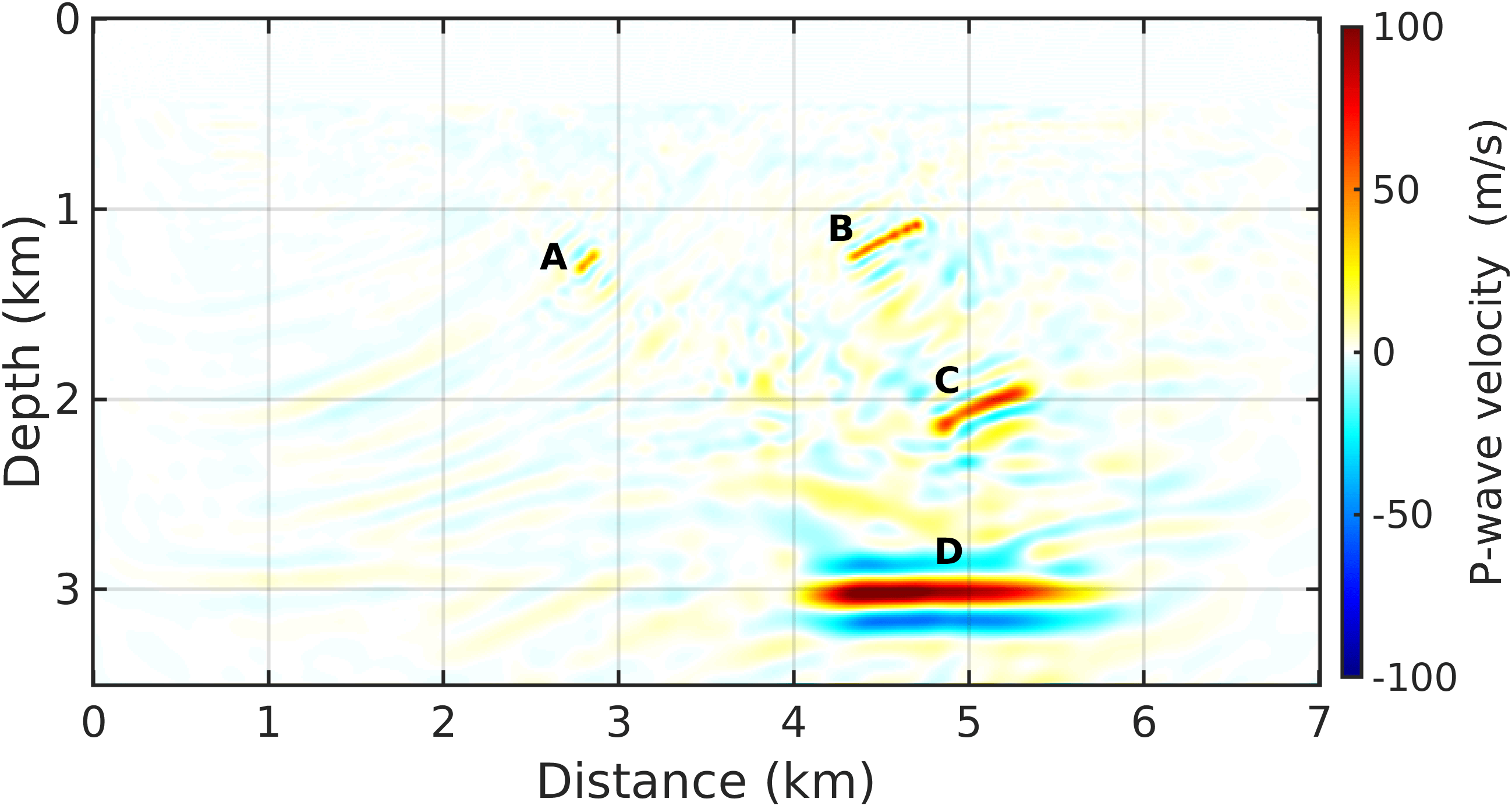}
       } 
       \hfill
    \subfloat[\label{fig:high_srn_bw3_time_lapse_model}]{%
       \includegraphics[width=.48\linewidth]{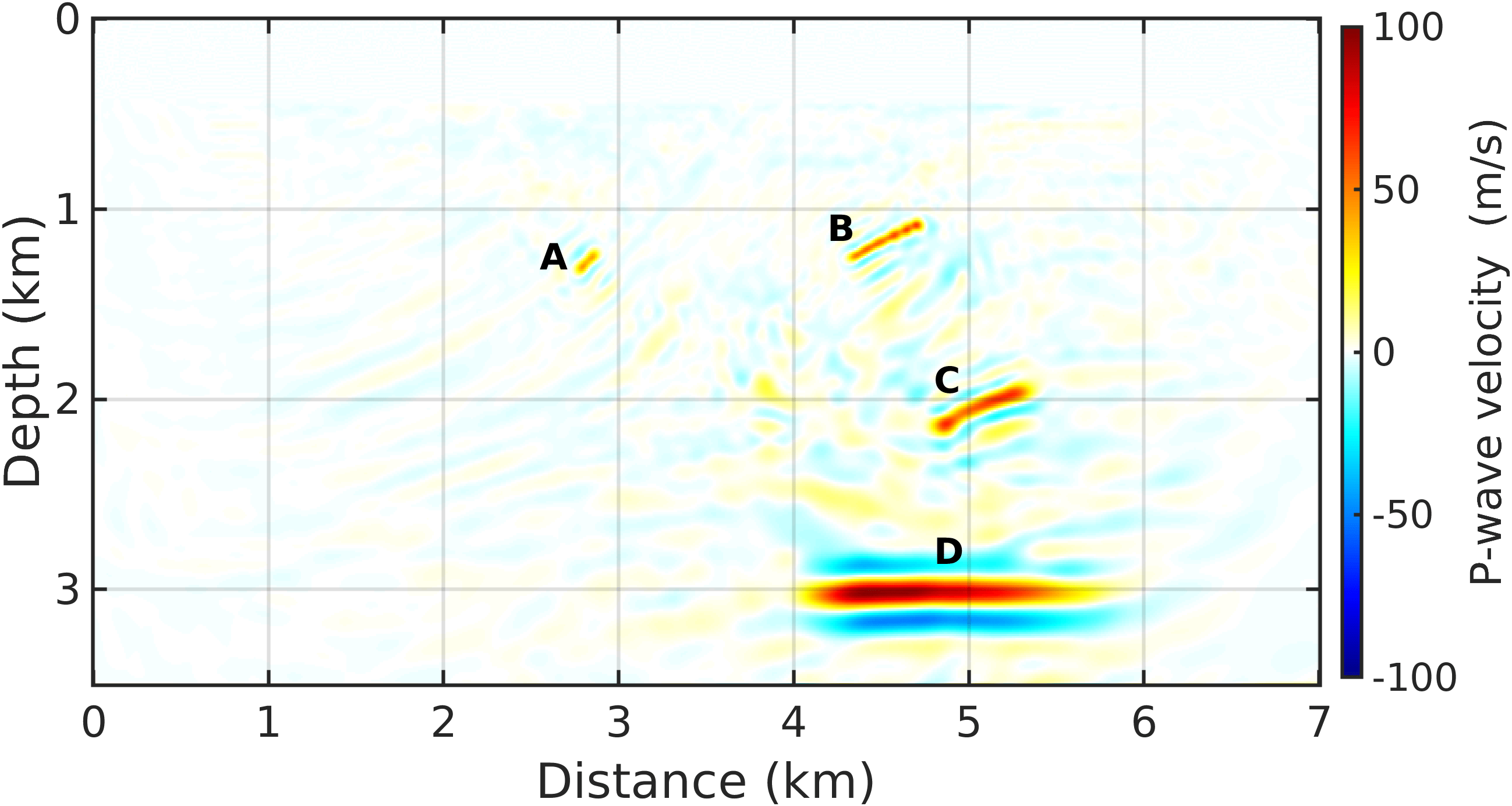}
       }
       \caption{Time-lapse estimates from the Marmousi case study by conventional time-lapse (left-hand column) and Bayesian-weighted (right column) strategies in the case with an SNR of 15 dB. Panels (a), (c) and (e) refer, respectively, to the parallel, sequential and central-difference 4D FWI schemes, while panels (b), (d) and (f) refer to Bayesian-weighted models \textit{$\text{BW}_1$}, \textit{$\text{BW}_2$} and \textit{$\text{BW}_3$}, respectively.}
    \label{fig:marmousi__ouput_models_highSNR}
\end{figure*}

\begin{figure*}[!htb]
    \centering
    \subfloat[\label{fig:low_srn_par_time_lapse_model}]{%
       \includegraphics[width=.48\linewidth]{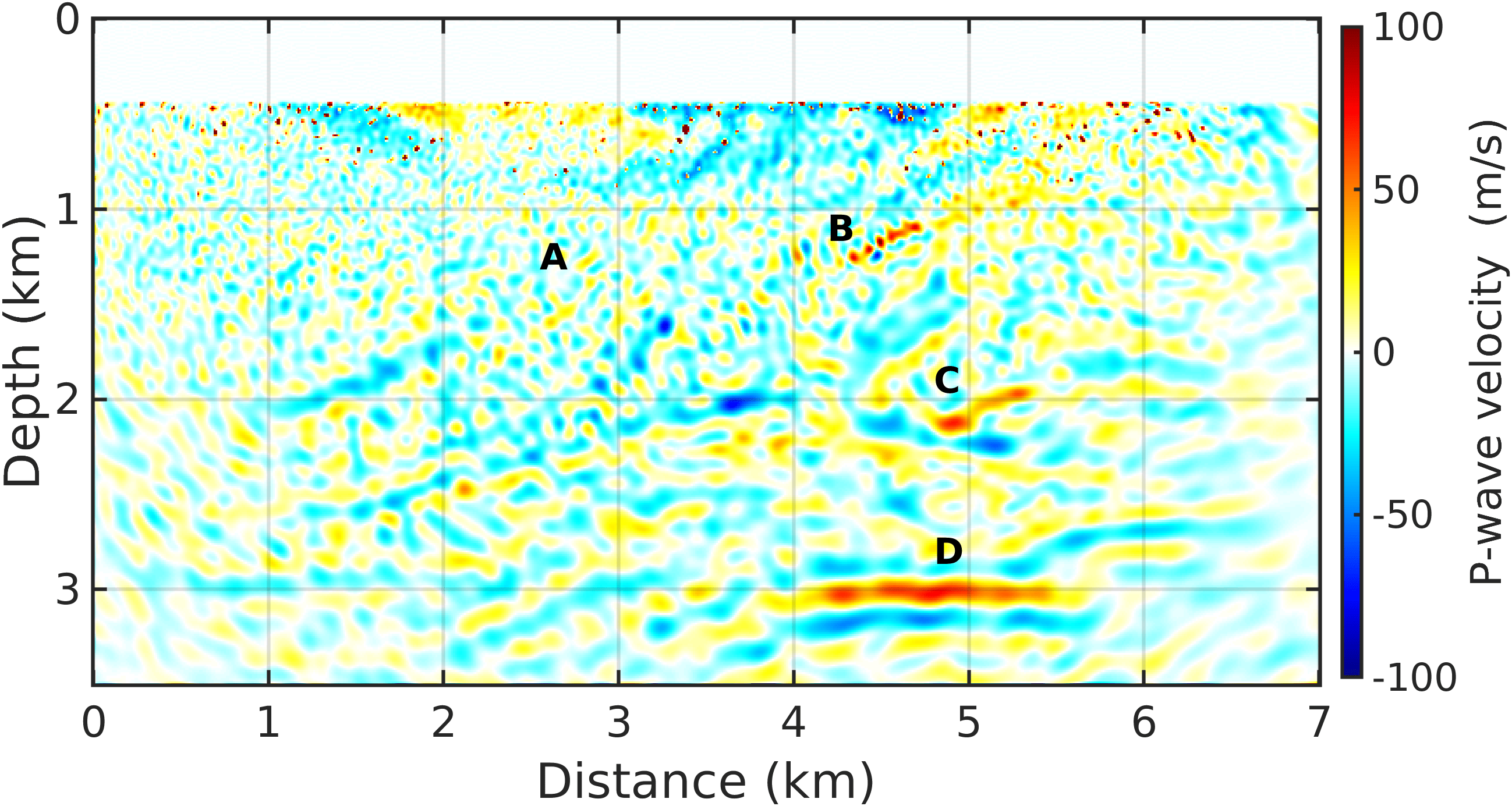}
       } 
       \hfill
    \subfloat[\label{fig:low_srn_bw1_time_lapse_model}]{%
       \includegraphics[width=.47\linewidth]{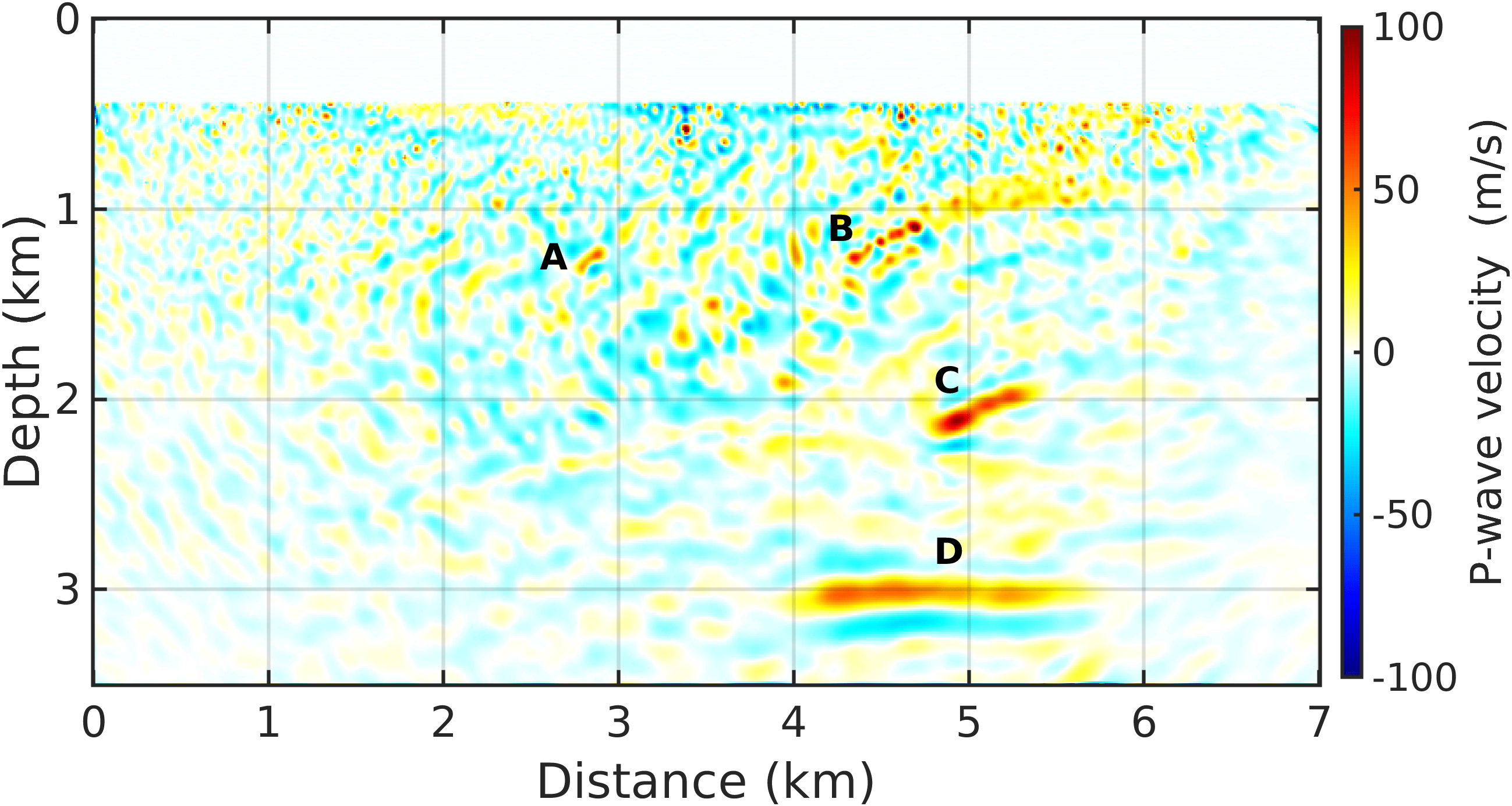}
       }
    \\
    \subfloat[\label{fig:low_srn_seq_time_lapse_model}]{%
       \includegraphics[width=.48\linewidth]{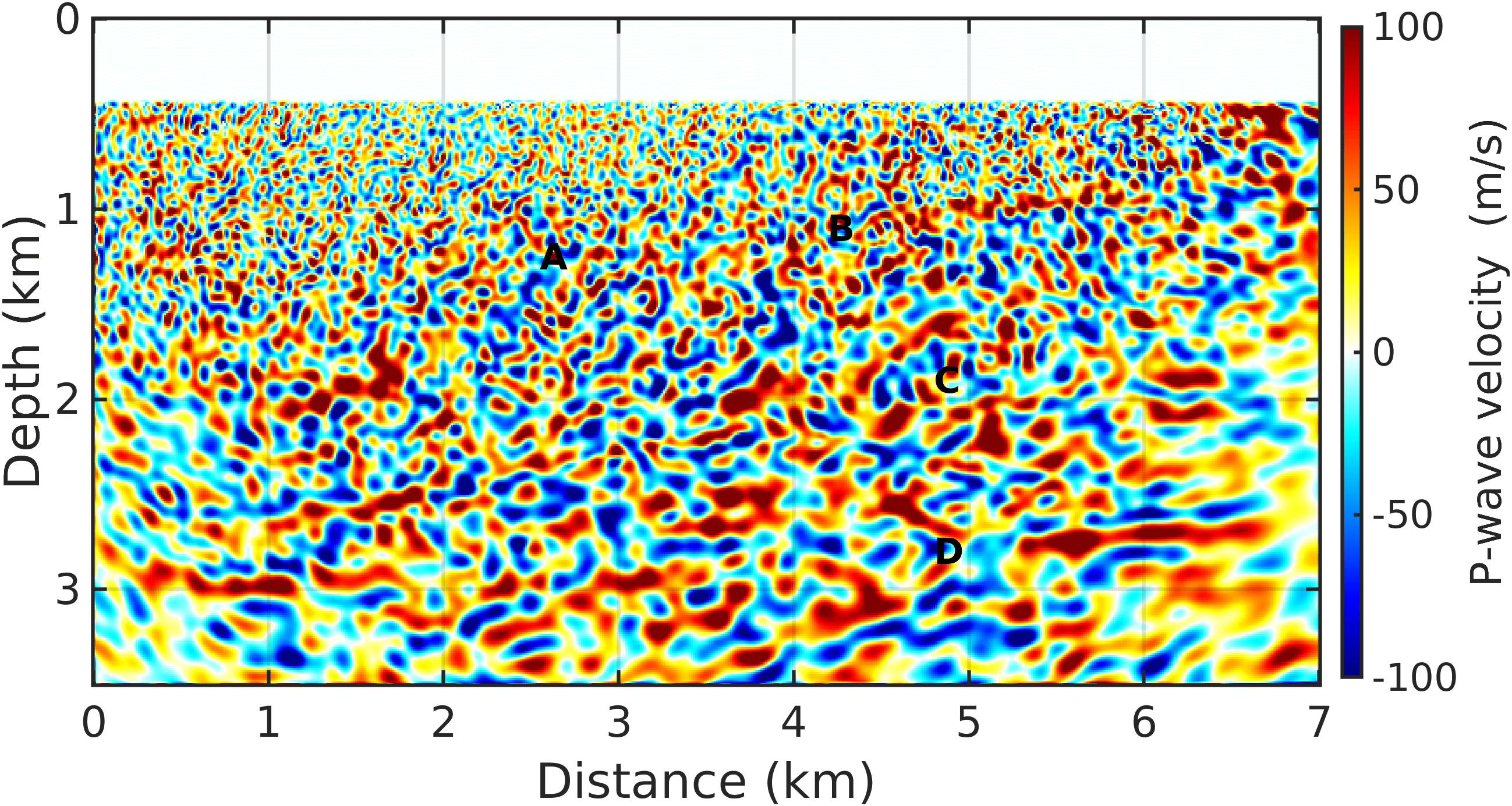}
       } 
       \hfill
    \subfloat[\label{fig:low_srn_bw2_time_lapse_model}]{%
       \includegraphics[width=.48\linewidth]{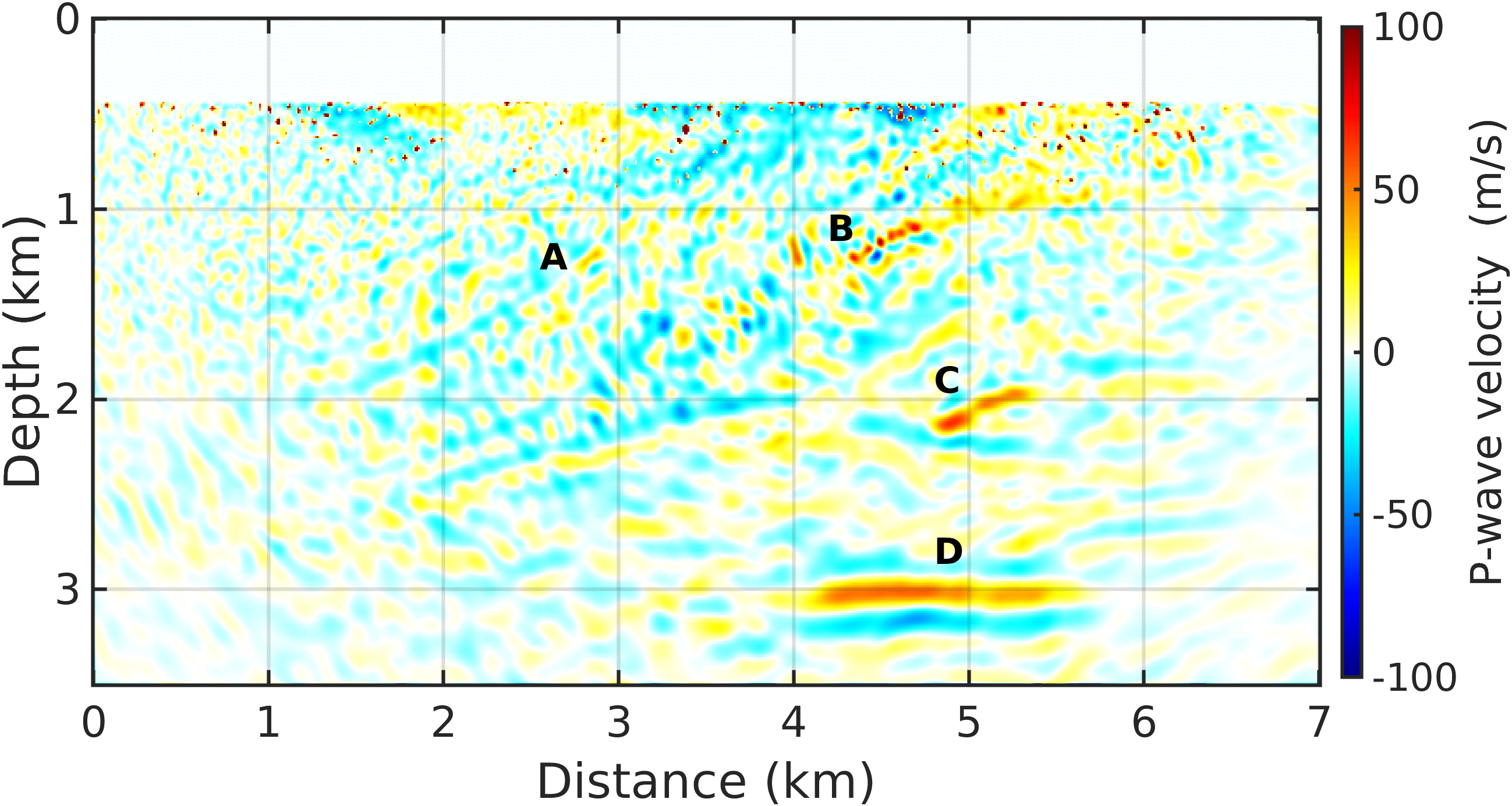}
       }
    \\
    \subfloat[\label{fig:low_srn_cd_time_lapse_model}]{%
       \includegraphics[width=.48\linewidth]{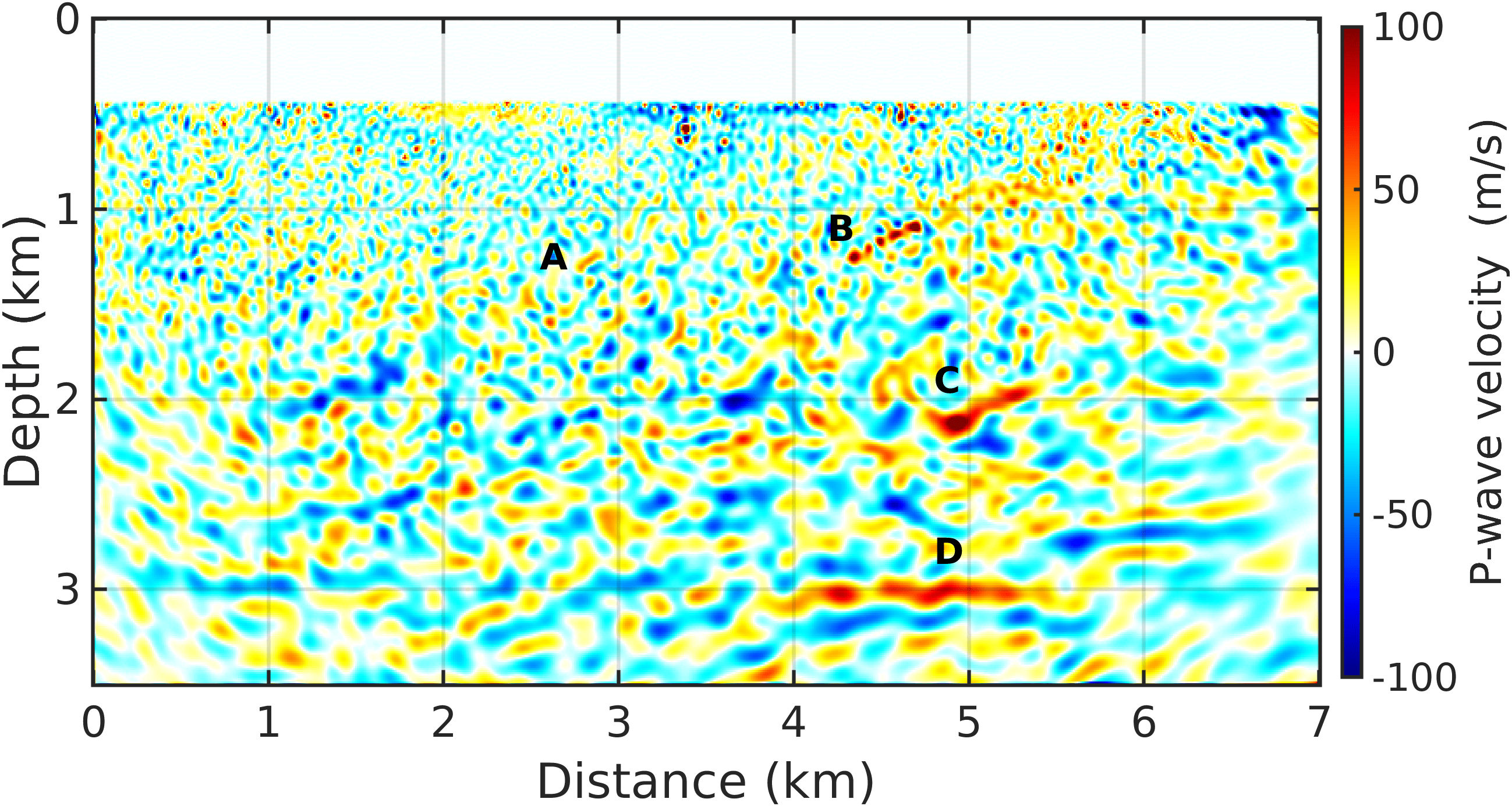}
       } 
       \hfill
    \subfloat[\label{fig:low_srn_bw3_time_lapse_model}]{%
       \includegraphics[width=.48\linewidth]{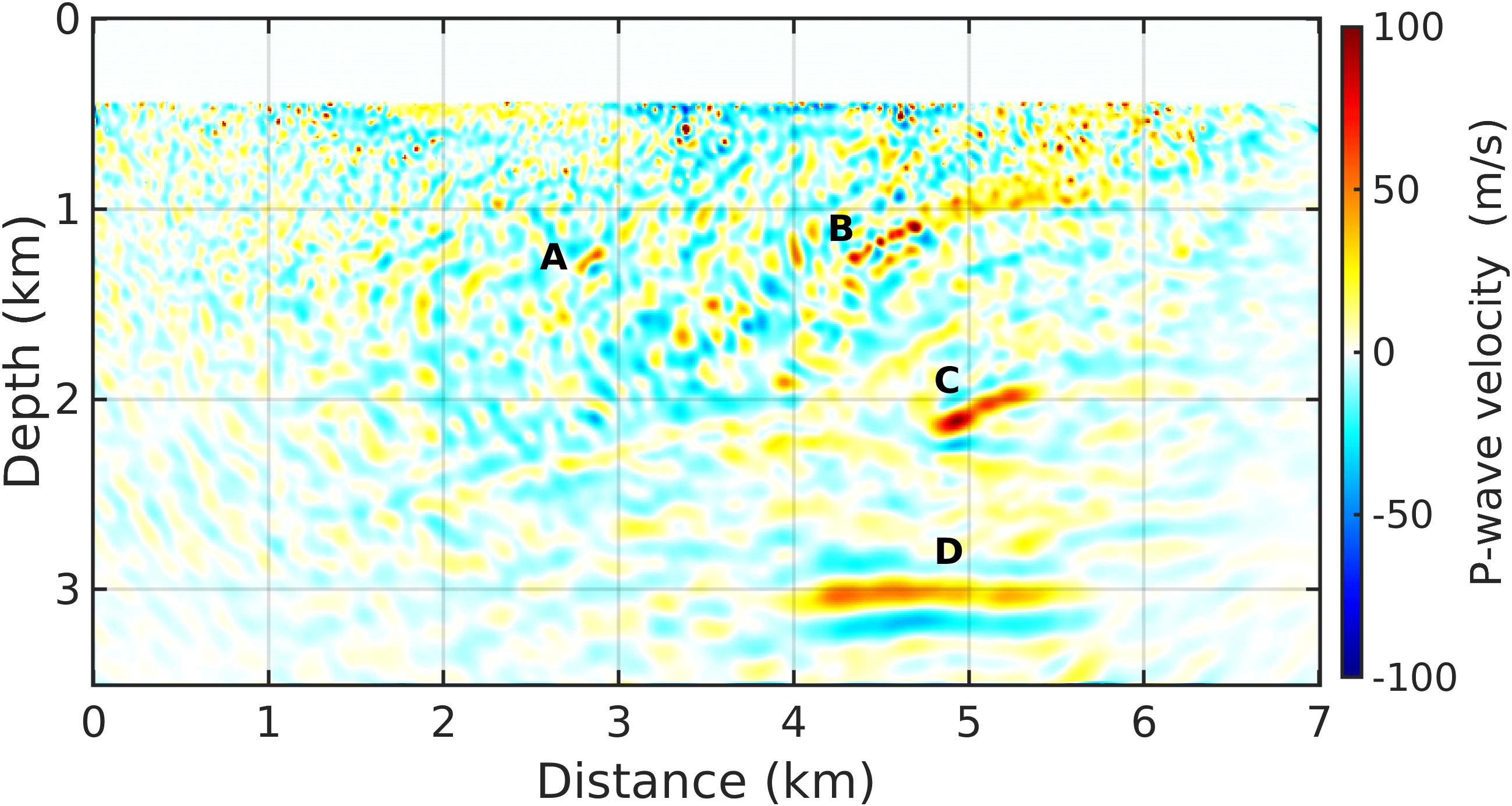}
       }
       \caption{Time-lapse estimates from the Marmousi case study by conventional time-lapse (left-hand column) and Bayesian-weighted (right column) strategies in the case with an SNR of 8 dB. Panels (a), (c) and (e) refer, respectively, to the parallel, sequential and central-difference 4D FWI schemes, while panels (b), (d) and (f) refer to Bayesian-weighted models \textit{$\text{BW}_1$}, \textit{$\text{BW}_2$} and \textit{$\text{BW}_3$}, respectively.}
    \label{fig:marmousi__ouput_models_lowSNR}
\end{figure*}

Figures \ref{fig:marmousi__ouput_models_highSNR} and \ref{fig:marmousi__ouput_models_lowSNR} show time-lapse estimates for the cases with an SNR of 15 dB and 8 dB, respectively. The left-hand column refers to the estimates derived by conventional time-lapse strategies (parallel 4D FWI, sequential 4D FWI and central central-difference 4D FWI), while the right-hand column refers to the estimates derived by the Bayesian-weighted models (\textit{$\text{BW}_1$}, \textit{$\text{BW}_2$} and \textit{$\text{BW}_3$}). It is noticeable that the sequential 4D FWI strategy completely fails to obtain a minimum interpretable time-lapse model, as depicted in Figs. \ref{fig:high_srn_seq_time_lapse_model} and \ref{fig:low_srn_seq_time_lapse_model}. When we consider the case with an SNR of 15 dB, all time-lapse strategies, except the sequential approach, have estimated satisfactory time-lapse models where it is possible to easily recognize the time-lapse anomalies, especially those located at shallower depths, denoted as A, B, and C, as depicted in Fig. \ref{fig:marmousi__ouput_models_highSNR}. Conversely, conventional time-lapse strategies produce numerous artifacts in the retrieved time-lapse models when we consider the case with an SNR of 8 dB, where the background noise is more intense. In contrast, Bayesian-weighted models tend to produce models with fewer artifacts, providing a more nuanced and reliable representation of the time-lapse estimates, as depicted in the right-hand of Fig. \ref{fig:marmousi__ouput_models_lowSNR}.

As we have access to the true time-lapse model, we perform a quantitative comparison with the retrieved time-lapse models to assess the quality of the results. To this end, we calculate two statistical measures: the normalized root-mean-square (NRMS) and Pearson's correlation coefficient (R). The Pearson's coefficient is based on the covariance between the two variables and their respective standard deviations, is defined as
\begin{equation}
    R = \frac{\sum_{i}(\Delta m_{i}^{\textrm{retr}} - \overbar{\Delta m}^{\textrm{retr}})(\Delta m_{i}^{\textrm{true}} - {\overbar{\Delta m}}^{\textrm{true}})}{\sqrt{\sum_{i}(\Delta m_{i}^{\textrm{retr}} - {\overbar{\Delta m}}^{\textrm{retr}})^2} \sqrt{\sum_{i}(\Delta m_{i}^{\textrm{true}} - {\overbar{\Delta m}}^{\textrm{true}})^2}},
\end{equation}
where $\Delta m^{\textrm{true}}$ and $\Delta m^{\textrm{retr}}$ are the true and retrieved time-lapse models, while $\overbar{\Delta m}^{\textrm{retr}}$ and $\overbar{\Delta m}^{\textrm{true}}$ represent the respective averages. The NRMS is based on the misfit between the true and retrieved time-lapse models and is defined as
\begin{equation}
    \textrm{NRMS} = \Bigg[\frac{\sum_i \big(\Delta m_{i}^{\textrm{true}}-\Delta m_i^{\textrm{retr}}\big)^2}{\sum_i \big(\Delta m_i^{\textrm{true}}\big)^2}\Bigg]^{1/2}.
    \label{eq:NRMS}
\end{equation}
The NRMS varies from 0 (indicating a perfect retrieved model) to $\infty$ (indicating a poor retrieved model). On the other hand, Pearson's correlation coefficient (R) measures the similarities between the true and retrieved time-lapse models and vary from $-1$ (indicating a poor retrieved model) to $1$  (indicating a perfect retrieved model). These statistical measures for comparing the time-lapse model are summarized in Table~\ref{tab:marmousi_table_errors}. The inefficiency of the sequential 4D FWI strategy is emphasized by these statistical error measures, as it produces time-lapse estimates with the highest NRMS error and the lowest R coefficient. On the other hand, from a quantitative perspective, the Bayesian-weighted models \textit{$\text{BW}_1$} and \textit{$\text{BW}_3$} have comparable error measures and produce similar NRMS error and R-coefficient values. However, the Bayesian-weighted model \textit{$\text{BW}_3$} is the most promising as it produces the best time-lapse estimates, has the lowest NRMS error and the highest R-coefficient, indicating its superiority over the other time-lapse strategies evaluated.

\begin{table}[!b]
\centering
\caption{Main statistics comparing the retrieved time-lapse models with the true model for the Marmousi case study. \label{tab:marmousi_table_errors}}
\begin{tabular}{*5c}
\toprule
 &  \multicolumn{2}{c}{15 dB SNR} & \multicolumn{2}{c}{8 dB SNR}\\
4D strategy   & NRMS  & R  & NRMS  & R  \\
\midrule
Parallel  & 0.7925  & 0.6121 & 1.7951  & 0.5437 \\
Sequential   &  7.2178 & 0.0592 & 5.5960  & 0.0023\\
Central-difference   &  0.7355  & 0.7056 &  2.4180    & 0.5021 \\
$BW_1$   &  0.6761  & 0.7416 &  1.1873     & 0.6531 \\
$BW_2$   &  0.7369  & 0.6744 &  1.2892     & 0.6217 \\
$BW_3$   &  \textbf{0.6728} & \textbf{0.7489} &  \textbf{1.1723}     & \textbf{0.6600} \\
\bottomrule
\end{tabular}\\
\vspace{.1cm} \centering {\footnotesize NRMS, normalized root-mean-square; R, Pearson's coefficient.}
\end{table}

\newpage

\subsection{Brazilian pre-salt case study\label{sec:numerical_experiments_Brazilian_pre-salt_case_study}}

\begin{figure}[!b]
    \centering
    \subfloat[\label{fig:true_velocity_model}]{%
       \includegraphics[width=.35\linewidth]{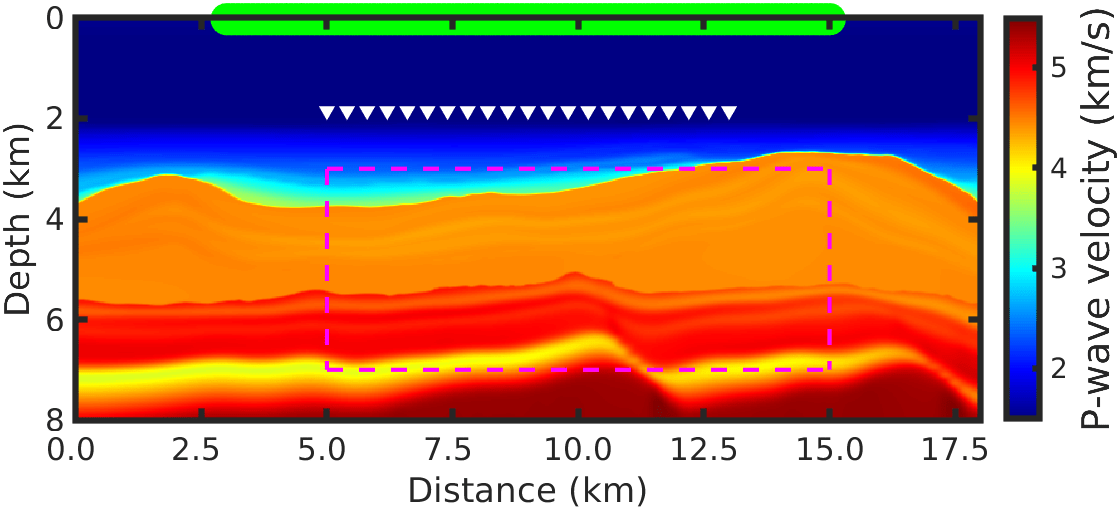}}    
    \\
  \subfloat[\label{fig:true_4D_model}]{%
       \includegraphics[width=.35\linewidth]{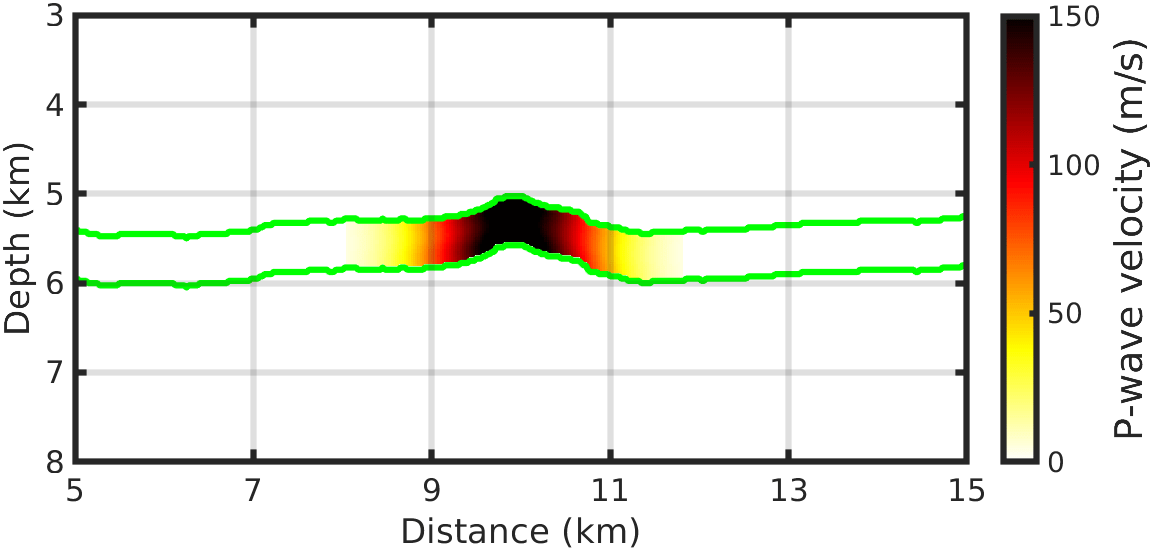}}
    \\
  \subfloat[\label{fig:initial_model}]{%
       \includegraphics[width=.35\linewidth]{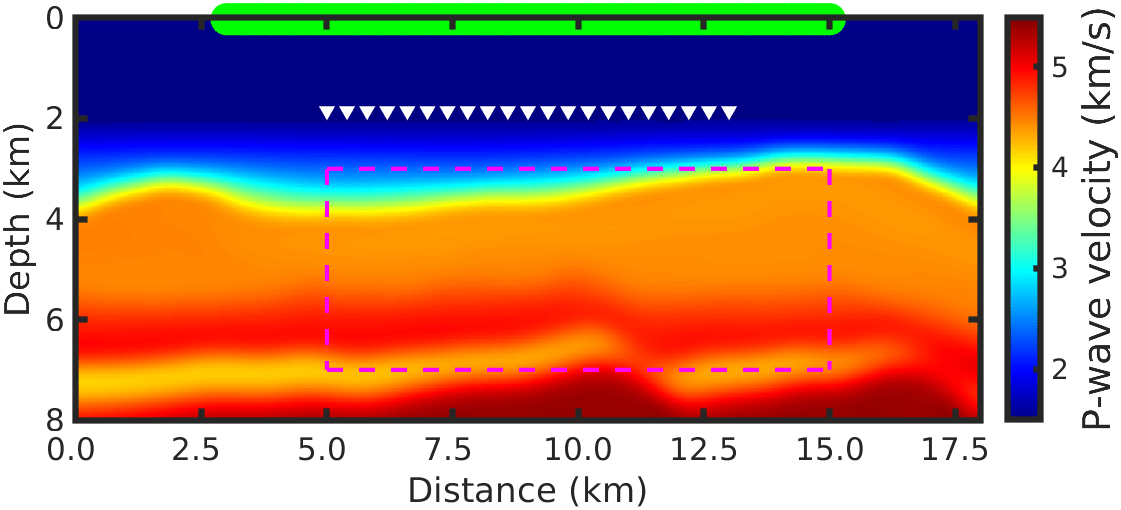}}
    \caption{Velocity models: \textbf{(a)} Baseline model considered in the Brazilian pre-salt case study; a realistic subsurface P-wave velocity model representing the deep-water Brazilian pre-salt region. The green line represents the seismic sources' positions, while the white triangles represent the locations of the nodes. The magenta dashed lines enclose the region where the velocity perturbation was applied to simulate the monitor acquisition. \textbf{(b)} The true time-lapse model (the difference between monitor and baseline models) related to the area enclosed by dashed lines in panel (a). \textbf{(c)} Initial model $m_0$ used in the forward and reverse bootstrap steps' first stages.}
    \label{fig:input_models}
\end{figure}

We investigate the potentialities of the receiver-extension strategy and Bayesian analysis by considering a realistic subsurface P-wave velocity model representing a deep-water Brazilian pre-salt region. The model is depicted in Fig. \ref{fig:true_velocity_model}; we consider it as the true baseline model in all numerical experiments. Such a model comprises a thick  water layer, followed by post-salt sediments, complex salt structures, the pre-salt area, and bedrock below, covering an area 19 km long and 8 km deep \cite{daSilva_et_al_2021_SEG_IMAGE_qFWI}. The Brazilian pre-salt oil region faces great challenges due to complex geological structures, high pressures, heterogeneous deposits and deep oil and gas reserves \cite{daSilva_et_al_2024_CircularShotOBN_GeophysProsp,Costa_et_al_2023_SEG_IMAGE_DataSelctionCircularShotOBN,Camargo_et_al_2023_PetroGeosci}.

Adopting ocean bottom node (OBN) acquisitions have significantly improved the characterization and monitoring process of deep oil target regions of Brazilian pre-salt fields \cite{Cruz_et_al_2021__TimeLapse_TupiNodesPilot}. For this reason, we consider a sparse OBN acquisition comprising 21 nodes regularly spaced every 400 m and 241 isotropic explosive sources at a depth of 10 m spaced every 50 m as a seismic survey. The locations of nodes and sources are shown in Figs. \ref{fig:true_velocity_model} and \ref{fig:initial_model} by the white triangles and green line, respectively. Every seismic source is a Ricker wavelet with a cut-off frequency of 15 Hz. The acquisition time is 7 s. 

We assemble the true monitor model by perturbing the true baseline model. In Fig. \ref{fig:true_velocity_model}, the magenta dashed lines enclosed the area where the velocity perturbation was operated to construct the true monitor model. We modify the pre-salt reservoir area, increasing P-wave velocities by no more than $3\%$, in order to simulate a realistic water injection into the reservoir to maintain reservoir pressure, i.e., simulating a voidage replacement. Figure \ref{fig:true_4D_model} shows the true time-lapse model. That is, the true monitor model is formed of the sum between the true baseline model (Fig. \ref{fig:true_velocity_model}) and the velocity anomaly shown in Fig. \ref{fig:true_4D_model}.

Since in OBN-type acquisitions the number of receivers is usually smaller than the number of seismic sources, we consider the principle of reciprocity. Thus, we interchange source and receiver locations by labeling receivers as sources and sources as receivers. In this scenario, the imprecision in the spatial coordinates are attributed to the source positions (referred to as receivers), whereas the NR issues related to the node positions are insignificant. This premise holds coherent as the identical group of nodes can be employed across various seismic surveys over time. Consequently, within the context of OBN operations, challenges related to NR are primarily linked to the positions of seismic sources.

To calculate the receiver correction parameter, we randomly sample 30 values of $\Delta x_{s,r}$ near each receiver position $x_{r}$ and evaluate the receiver-extension objective function. Figure \ref{fig:initial_model} illustrates the initial model $m_0$, which was generated through the application of a Gaussian smoothing operator featuring a standard deviation of $200m$ on the baseline model (Fig. \ref{fig:true_velocity_model}).

We divided this Brazilian pre-salt case study into two parts; the first involves assessing the receiver-extension strategy's effectiveness in mitigating time-lapse noises, considering the three existing 4D FWI strategies in the literature described in Section \ref{sec:time_lapse_FWI}. This examination encompasses several circumstances, including the (A) ideal case using time-lapse data with perfect repeatability and NR scenarios associated with (B) Global Positioning System (GPS) coordinate inaccuracies, (C) water velocity changes in the ocean layers, and (D) combined NR issues. In the second one, we will present the (E) Bayesian analysis, utilizing our proposed models described in Section \ref{sec:bayesian_analysis}, for the more intricate case related to the combined NR issues.

\subsubsection{Time-lapse data with perfect
repeatability}

By considering the true baseline and monitor models, we generate 'observed' data sets using the 2D time-domain acoustic wave equation given by Eq. \eqref{eq:wave_eq_time} and by employing the acquisition geometry as described before. In particular, we apply the finite difference method with $2nd$ and $8th$ order approximations for time and space, respectively, in a grid with regular discretization of 12.5 m. Furthermore, we introduce Gaussian noise with an SNR of 15.44 $dB$ to contaminate the data sets. It is worth noting that the seismic noise in the baseline data is distinct from the noise in the monitor data.

\begin{figure*}[!b]
    \centering
    \subfloat[\label{fig:seismogram_is_1_baseline}]{%
        \includegraphics[width=0.14\linewidth]{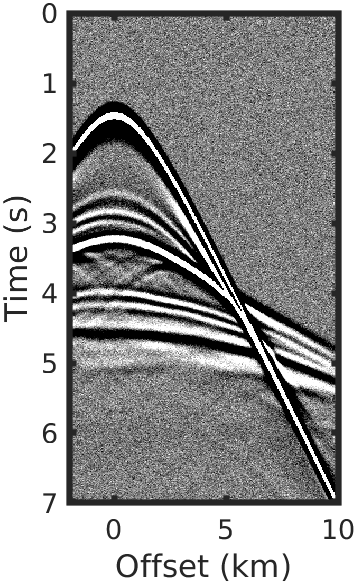}}
    \subfloat[\label{fig:seismogram_is_1_monitor}]{%
        \includegraphics[width=0.14\linewidth]{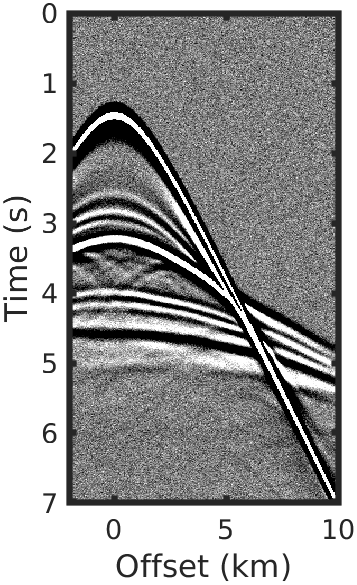}}
    \subfloat[\label{fig:seismogram_is_1_delta_m} {\tiny $\text{NRMS} = 0.0145$}
    ]{%
        \includegraphics[width=0.14\linewidth]{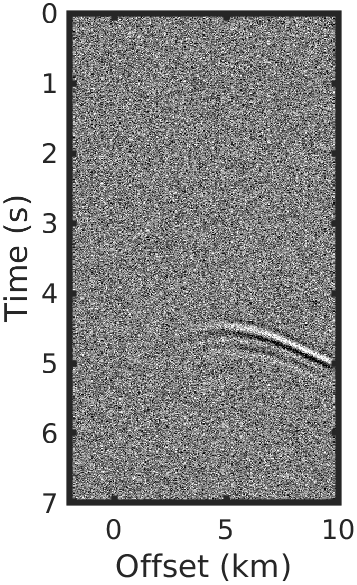}
    }
    \subfloat[{\tiny $\text{NRMS} = 0.7461$}\label{fig:seismogram_is_1_delta_m_GPS}]{%
        \includegraphics[width=0.14\linewidth]{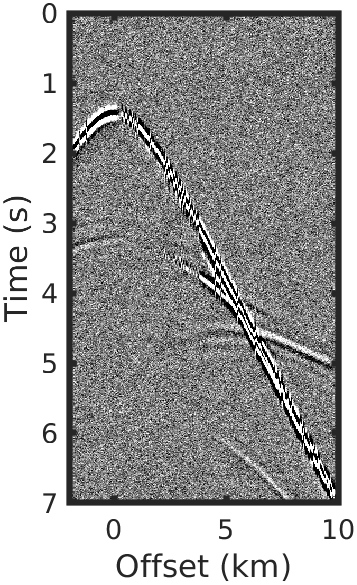}}
    \subfloat[{\tiny $\text{NRMS} = 1.4074$} \label{fig:seismogram_is_1_delta_m_we}]{%
        \includegraphics[width=0.14\linewidth]{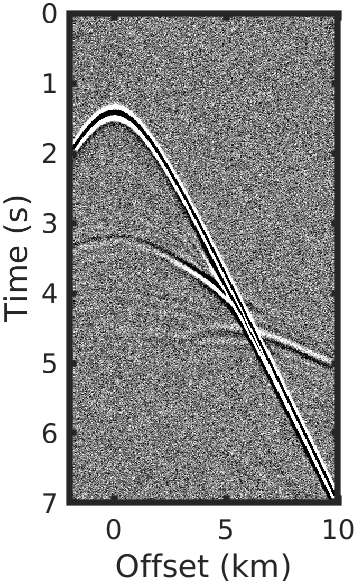}}
    \subfloat[{\tiny $\text{NRMS} = 0.8456$} \label{fig:seismogram_is_1_delta_m_we_lateral}]{%
        \includegraphics[width=0.14\linewidth]{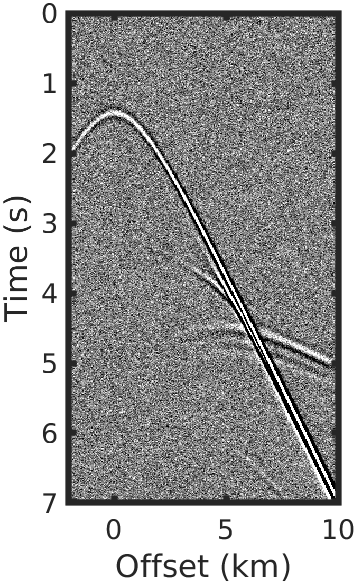}}
    \subfloat[{\tiny $\text{NRMS} = 0.7831$}\label{fig:seismogram_is_1_delta_m_fullNR}]{%
        \includegraphics[width=0.14\linewidth]{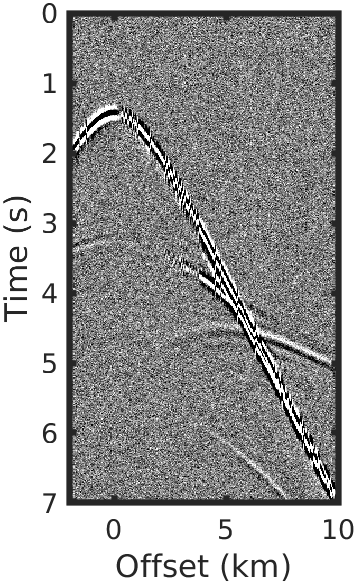}}
    \\  
    \centering
    \subfloat[\label{fig:seismogram_is_11_baseline}]{%
        \includegraphics[width=0.14\linewidth]{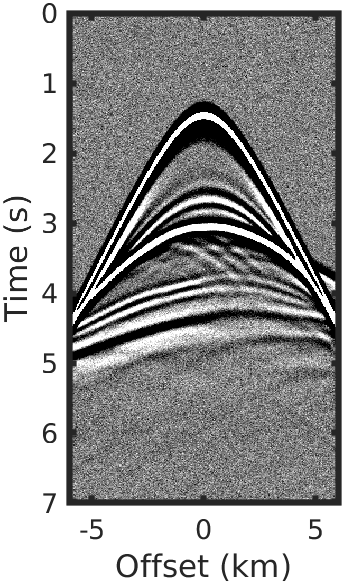}}
    \subfloat[\label{fig:seismogram_is_11_monitor}]{%
        \includegraphics[width=0.14\linewidth]{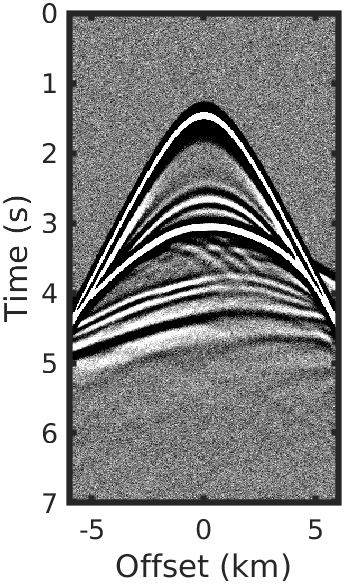}}
    \subfloat[{\tiny $\text{NRMS} = 0.0133$}\label{fig:seismogram_is_11_delta_m}]{%
        \includegraphics[width=0.14\linewidth]{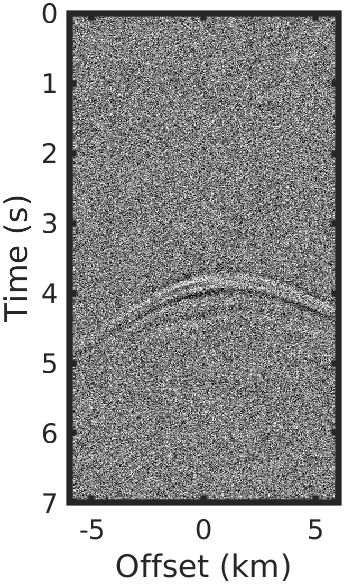}}
    \subfloat[{\tiny $\text{NRMS} = 0.8098$}\label{fig:seismogram_is_11_delta_m_GPS}]{%
        \includegraphics[width=0.14\linewidth]{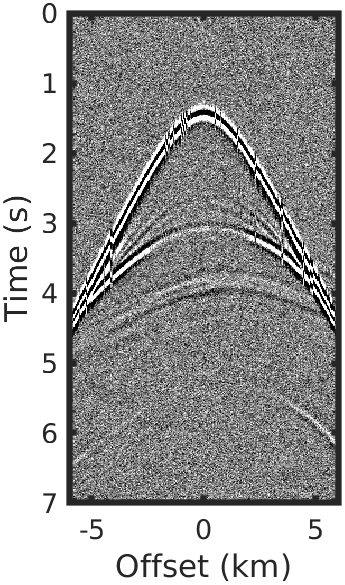}}
    \subfloat[{\tiny $\text{NRMS} = 1.4105$}\label{fig:seismogram_is_11_delta_m_we}]{%
        \includegraphics[width=0.14\linewidth]{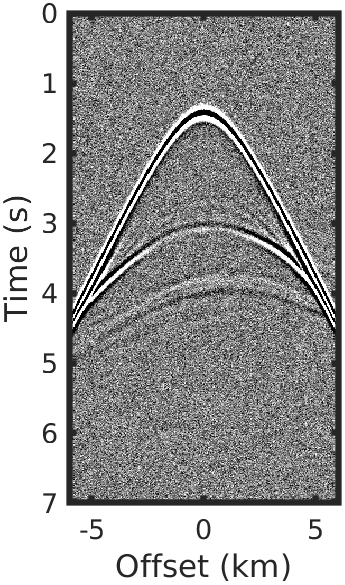}}
    \subfloat[{\tiny $\text{NRMS} = 0.8893$}\label{fig:seismogram_is_11_delta_m_we_lateral}]{%
        \includegraphics[width=0.14\linewidth]{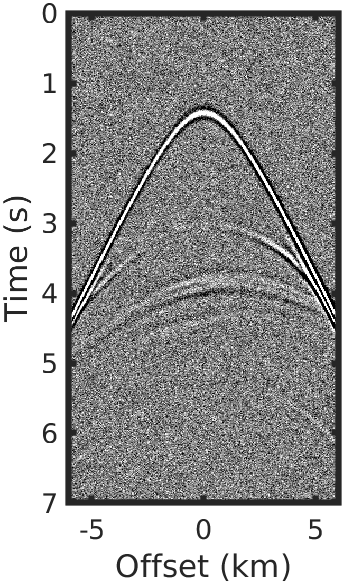}}
    \subfloat[{\tiny $\text{NRMS} = 0.8602$}\label{fig:seismogram_is_11_delta_m_fullNR}]{%
        \includegraphics[width=0.14\linewidth]{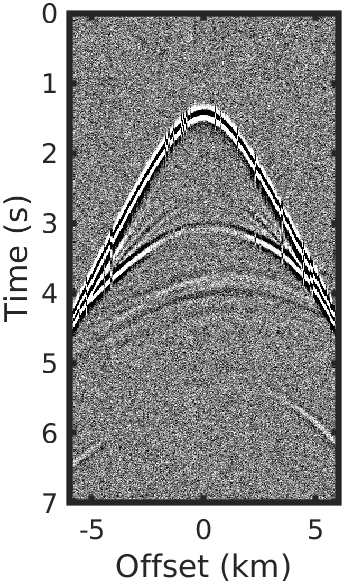}}
  \caption{Receiver-gathers and time-lapse data for different OBNs. The top line corresponds to the first left-side node, while the bottom line represents the central node. Panels (a) and (h) display receiver-gathers associated with the baseline acquisition, and panels (b) and (i) showcase the monitor data. Panels (c) and (j) illustrate time-lapse data with perfect repeatability, while panels (d) and (k) show time-lapse data accounting for GPS coordinate inaccuracies. Panels (e) and (l) depict time-lapse data considering water velocity changes in ocean layers due to seasonal variations, and panels (f) and (m) display time-lapse data considering shallow lateral water velocity changes caused by \textit{warm} temperature variations. Panels (g) and (n) illustrate time-lapse data addressing combined non-repeatability (NR) issues. The NRMS values were determined by comparing the receiver gathers of the baseline acquisition with the receiver gathers of the monitor acquisition.}
  \label{fig:seismograms} 
\end{figure*}

Figure \ref{fig:seismograms} shows receiver-gathers and time-lapse data associated with different nodes. The top row corresponds to the first left-side node, while the bottom row displays data from the central node. Figures \ref{fig:seismogram_is_1_baseline} and \ref{fig:seismogram_is_11_baseline} depict receiver-gathers from the baseline acquisition, while Figs. \ref{fig:seismogram_is_1_monitor} and \ref{fig:seismogram_is_11_monitor} show the monitor data. The discrepancies between the baseline and monitor data are very subtle and may not be readily noticeable. Consequently, the remaining panels in Fig. \ref{fig:seismograms} highlight the time-lapse data for each scenario investigated in this work. Using Eq. \eqref{eq:NRMS}, but assuming that $\Delta m^{\textrm{true}}$ denotes the receiver-gathers from the baseline acquisition and $\Delta m^{\textrm{retr}}$ denotes the receiver gathers from the monitor acquisition, we calculate the NRMS between the receiver gathers to quantify the data discrepancies for different NRs issues. The NRMS values are shown in the corresponding panels. Figures \ref{fig:seismogram_is_1_delta_m} and \ref{fig:seismogram_is_11_delta_m} illustrate the time-lapse data for the perfect repeatability case. These figures reveal that the time-lapse data are intermixed with different waveforms of the receiver gathers, highlighting the great challenge of time-lapse analysis. While we label this scenario as "perfect repeatability," it is essential to clarify that, in this context, this term exclusively pertains to the repeatability of acquisition parameters. Thus, it is important to note that the background noise is not annulled. This is because the background noise varies from one acquisition to another, and as a result, it is not effectively cancelled out in the time-lapse data. A subsequent discussion will pertain to the seismograms depicted in Figs. \ref{fig:seismogram_is_1_delta_m_GPS}-\ref{fig:seismogram_is_1_delta_m_fullNR} and \ref{fig:seismogram_is_11_delta_m_GPS}-\ref{fig:seismogram_is_11_delta_m_fullNR}, detailing their origin. 

\begin{figure*}[!htb]
    \centering
    \subfloat[\label{resulting_noNR_par_classical}]
        {%
        \includegraphics[width=0.48\linewidth]{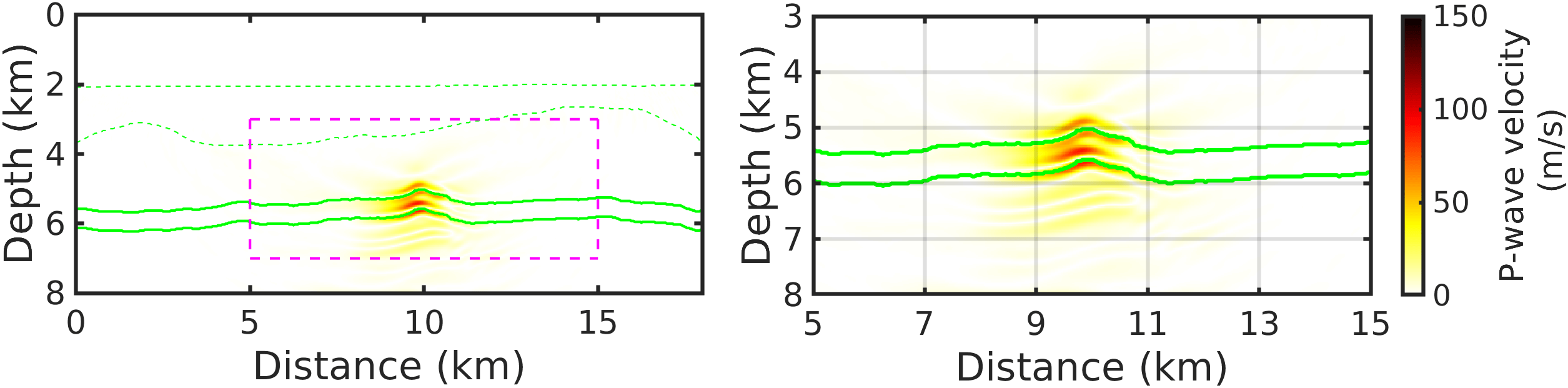}
        }  
    \hfill
    \subfloat[\label{resulting_noNR_par_recExt}]
        {%
        \includegraphics[width=0.48\linewidth]{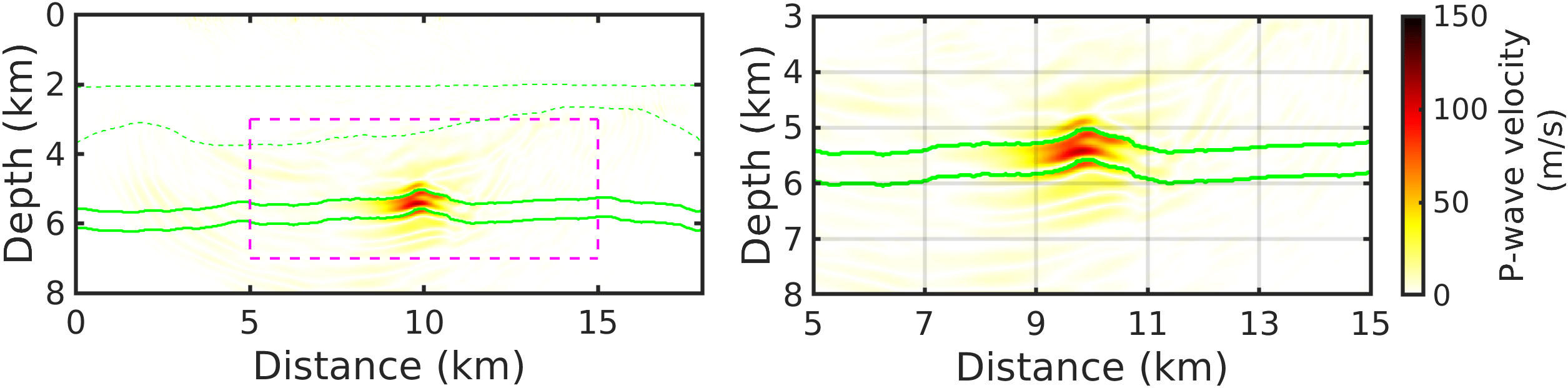}
    
        }
    \\
    \subfloat[\label{resulting_noNR_seq_classical}]
        {%
        \includegraphics[width=0.48\linewidth]{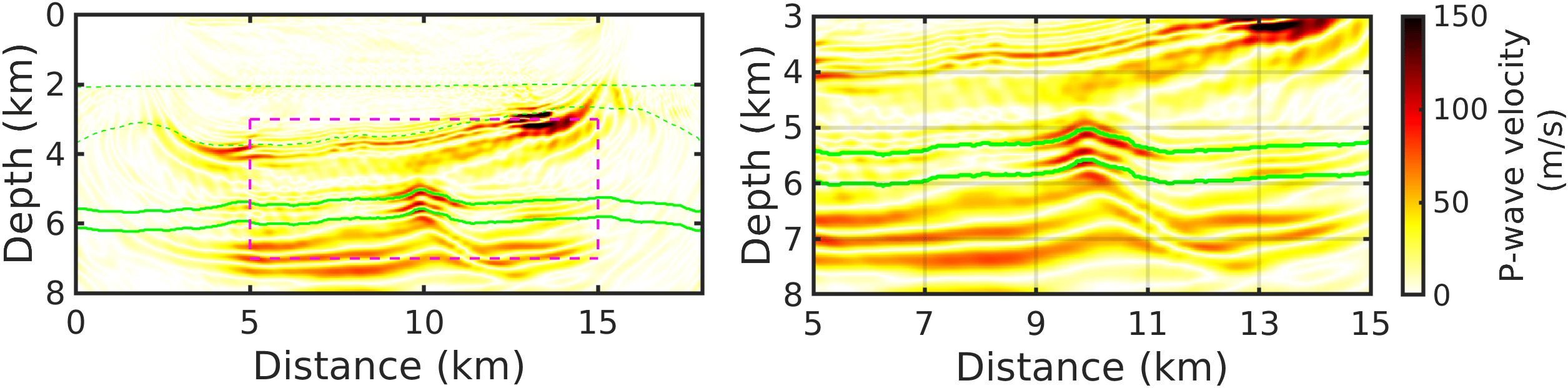}
        }  
    \hfill
    \subfloat[\label{resulting_noNR_seq_recExt}]
        {%
        \includegraphics[width=0.48\linewidth]{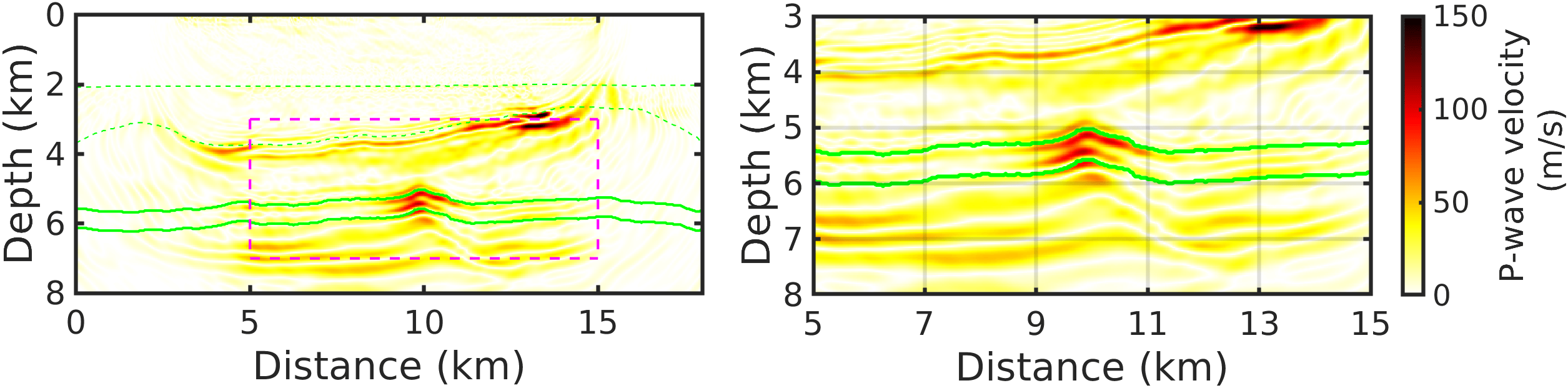}
        }
    \\
    \subfloat[\label{resulting_noNR_cd_classical}]
        {%
        \includegraphics[width=0.48\linewidth]{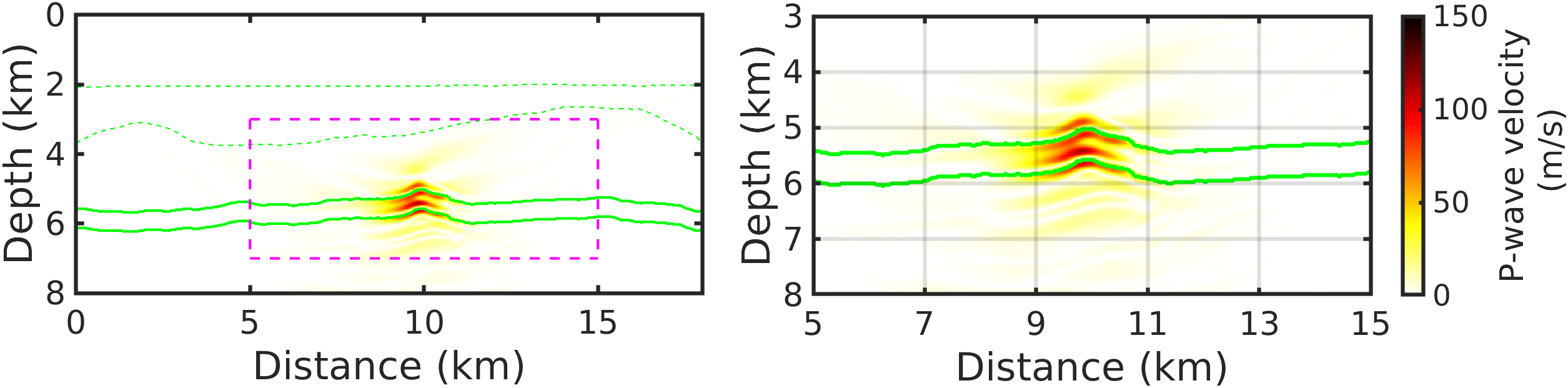}
        }  
    \hfill
    \subfloat[\label{resulting_noNR_cd_recExt}]
        {%
        \includegraphics[width=0.48\linewidth]{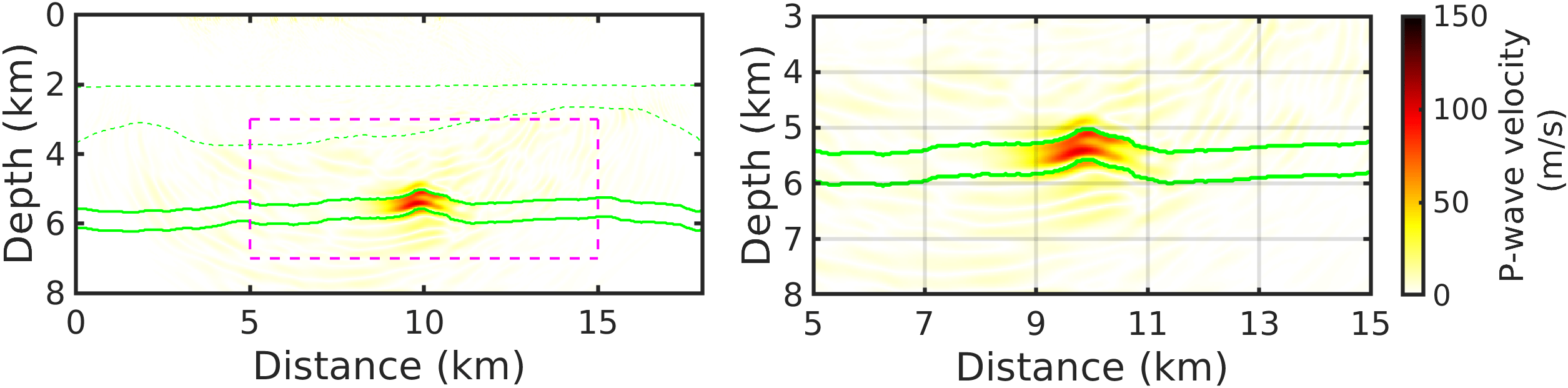}
        }
  \caption{Time-lapse estimates by the conventional approach (left-hand column) and the receiver-extension strategy (right-hand column) in the \textbf{perfect repeatability case}. Panels (a) and (b) refer to the parallel, (c) and (d) to the sequential, and (e) and (f) to the central-difference 4D FWI schemes.}
  \label{fig:resulting_noNR} 
\end{figure*}

Figure \ref{fig:resulting_noNR} shows time-lapse estimates, and a view of the area enclosed by the magenta rectangle, obtained through the conventional FWI approach (left-hand column) and the receiver-extension strategy (right-hand column) in the perfect repeatability scenario. Figures \ref{resulting_noNR_par_classical} and \ref{resulting_noNR_par_recExt} depict the time-lapse estimates using the parallel 4D FWI strategy, Figs. \ref{resulting_noNR_seq_classical} and \ref{resulting_noNR_seq_recExt} illustrate the estimates from the sequential 4D FWI strategy, and panels \ref{resulting_noNR_cd_classical} and \ref{resulting_noNR_cd_recExt} show the central-difference 4D FWI resulting models. Remarkably, the sequential 4D FWI approach introduces many artifacts unrelated to the true time-lapse differences (Fig. \ref{fig:true_4D_model}), regardless of whether we apply the conventional FWI (Fig. \ref{resulting_noNR_seq_classical}) or the receiver-extension strategy (Fig. \ref{resulting_noNR_seq_recExt}). These artifacts are notably pronounced, particularly in geological layers between regions characterized by significant contrasts in P-wave velocities, such as the upper boundary of the salt body. In contrast, the parallel and central-difference 4D FWI approaches produce time-lapse estimates that closely approximate the true time-lapse model and exhibit fewer artifacts. However, it is important to highlight that while the parallel and central-difference 4D FWI approaches offer valuable insights into their effectiveness in the perfect repeatability case, the receiver-extension FWI outperforms the conventional FWI by accurately reconstructing the true time-lapse anomaly. It is worth emphasizing that this perfect repeatability case highlights FWI's potential to achieve satisfactory time-lapse estimates, and underlines that the designated time-lapse strategy significantly impacts these models' accuracy.

The statistical measures comparing the time-lapse model for the perfect repeatability case are summarized in Table~\ref{tab:perfect_repeatability_case}, where C-FWI represents the conventional FWI approach and RE-FWI refers to receiver-extension FWI. It is important to note that the central-difference 4D FWI, when combined with the receiver-extension FWI, produced the best time-lapse model by yielding estimates with the lowest NRMS error and the highest R coefficient. 

\begin{table}[!htbp]
\centering
\caption{Main statistics comparing the retrieved time-lapse models with the true model for the perfect repeatability case. \label{tab:perfect_repeatability_case}}
\begin{tabular}{*5c}
\toprule
 &  \multicolumn{2}{c}{NRMS} & \multicolumn{2}{c}{R}\\
4D strategy   & C-FWI   & RE-FWI   & C-FWI   & RE-FWI \\
\midrule
Parallel  &  0.8367 & 0.7084 & 0.5519  & 0.7226 \\
Sequential   &  2.0646 & 1.5204 & 0.1685  & 0.2845\\
Central-difference   &  0.7781  & \textbf{0.6681} &  0.6286     & \textbf{0.7803}\\
\bottomrule
\end{tabular} \\
\vspace{.1cm} {\footnotesize NRMS, normalized root-mean-square; R, Pearson's coefficient; \\ C-FWI, conventional FWI; RE-FWI, receiver-extension FWI.}
\end{table}

\newpage
\subsubsection{GPS coordinate inaccuracies}

\begin{figure}[!b]   
\centering
\subfloat[\label{fig:gps_error_distribution_histogram}]
        {%
        \includegraphics[width=.5\columnwidth]{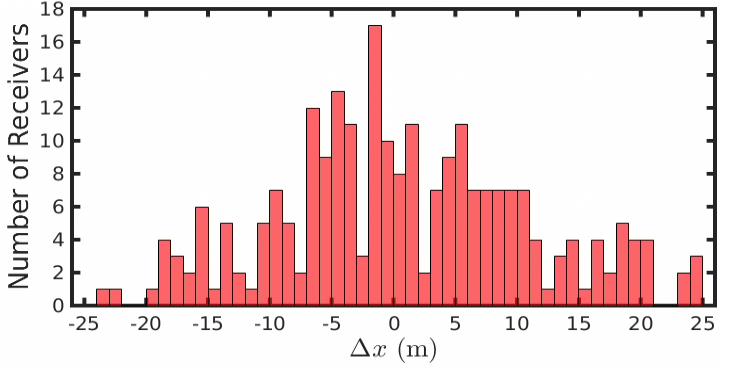}
        }
    \\
    \subfloat[\label{fig:gps_error_distribution}]
        {%
        \includegraphics[width=.5\columnwidth]{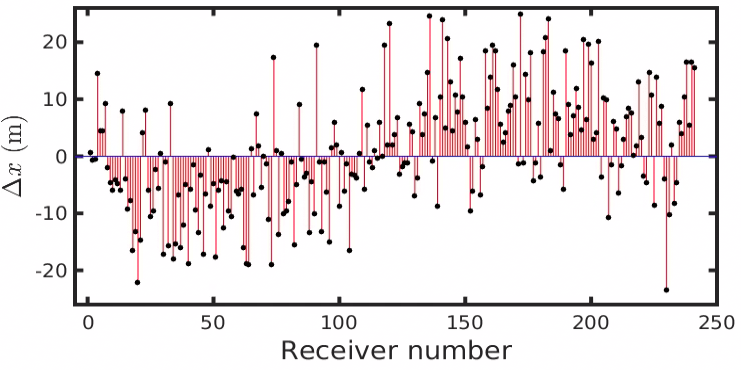}
        }
  \caption{Receiver positioning error statistics.(a) Histogram of coordinate accuracy error distributions of the receivers. (b)  Coordinate errors for each receiver; these errors are correlated and follow a sinusoidal behavior.}
  \label{fig:gps_errors} 
\end{figure}

Although the Global Positioning System (GPS) has revolutionized locating objects or people, its measurements are subject to inaccuracies caused by several factors, such as atmospheric interference and inaccuracies in satellite orbit data and their geometry concerning the receiver. Thus, the locations of receivers and sources are subject to errors. In this work, we consider GPS-related positioning uncertainties similar to Ref. \cite{Reasnor_et_al_2010_SEG_OBN} to simulate realistic error patterns. In this context, we assume that coordinate inaccuracies follow a Gaussian distribution with a mean of $2.8$ m and a standard deviation of $2.5$ m. Figure \ref{fig:gps_error_distribution_histogram} depicts the histogram of these errors, which are exclusively pertain to lateral discrepancies. Furthermore, we account for spatially correlated errors characterized by a sinusoidal mean, as depicted in Fig. \ref{fig:gps_error_distribution}. Receiver coordinate errors are independently computed for both baseline and monitor acquisitions. Figures \ref{fig:seismogram_is_1_delta_m_GPS} and \ref{fig:seismogram_is_11_delta_m_GPS} illustrate the time-lapse data in scenarios involving GPS coordinate inaccuracies. We notice that numerous waveforms not related to time-lapse changes (as seen in Figs. \ref{fig:seismogram_is_1_delta_m} and \ref{fig:seismogram_is_11_delta_m}) are also present in the time-lapse data, highlighting the negative impact of GPS inaccuracies. 

\begin{figure*}[!htb]
    \centering
    \subfloat[\label{fig:resulting_GPS_par_classical}]
        {%
        \includegraphics[width=0.48\linewidth]{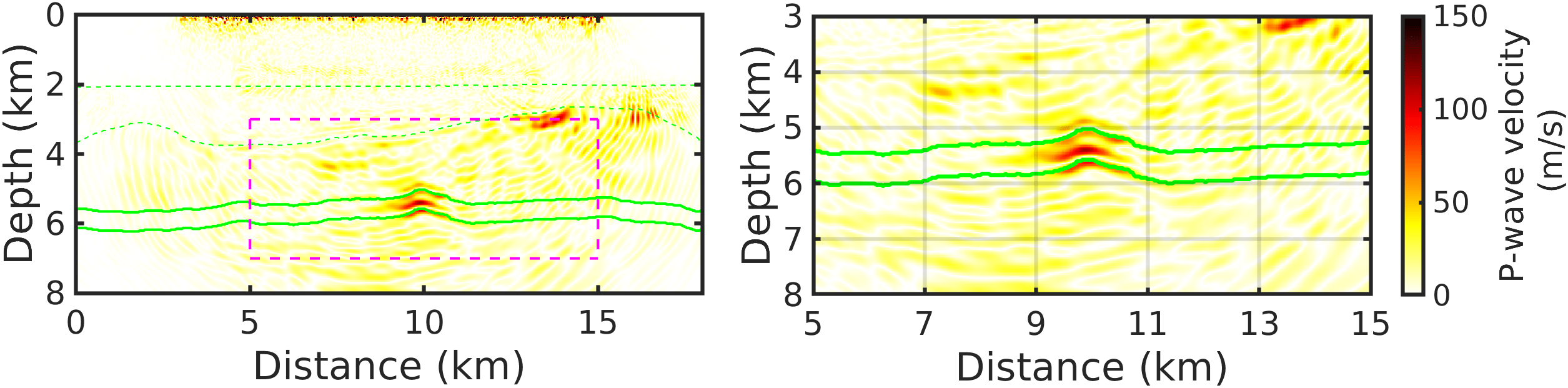}
        }  
    \hfill
    \subfloat[\label{fig:resulting_GPS_par_recExt}]
        {%
        \includegraphics[width=0.48\linewidth]{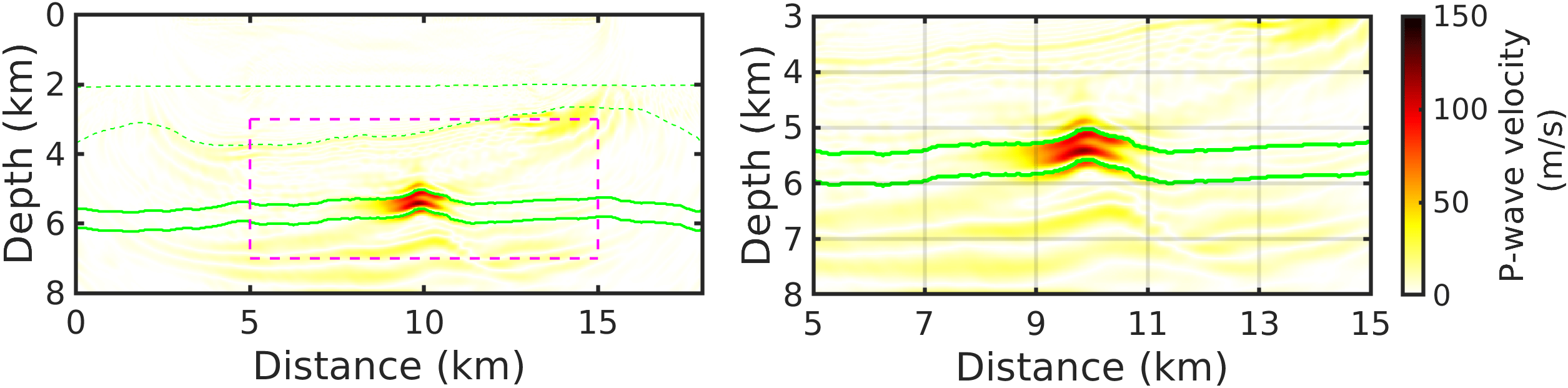}
        }
    \\
    \subfloat[\label{fig:resulting_GPS_seq_classical}]
        {%
        \includegraphics[width=0.48\linewidth]{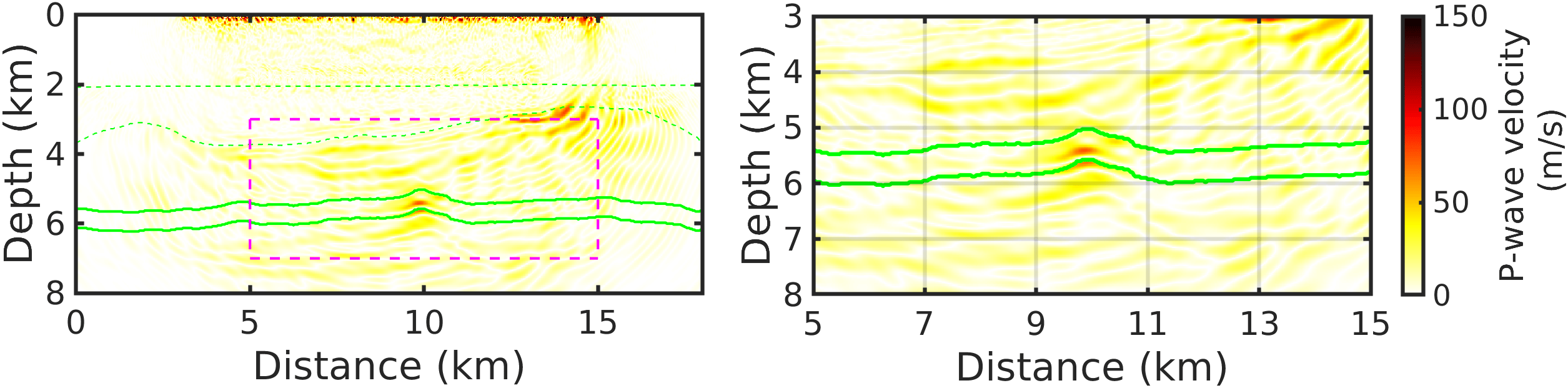}
        }  
    \hfill
    \subfloat[\label{fig:resulting_GPS_seq_recExt}]
        {%
        \includegraphics[width=0.48\linewidth]{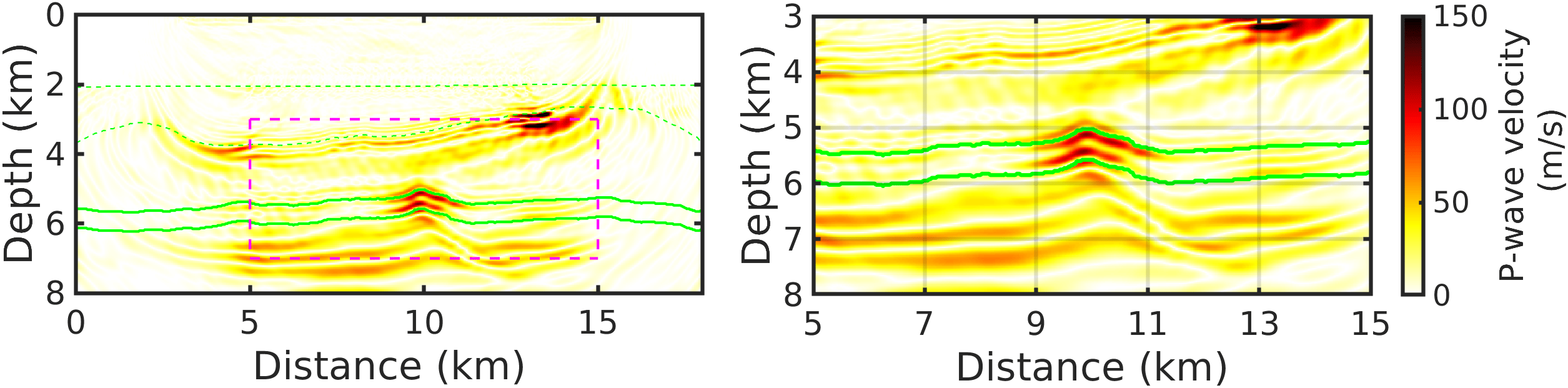}
        }
    \\
    \subfloat[\label{fig:resulting_GPS_cd_classical}]
        {%
        \includegraphics[width=0.48\linewidth]{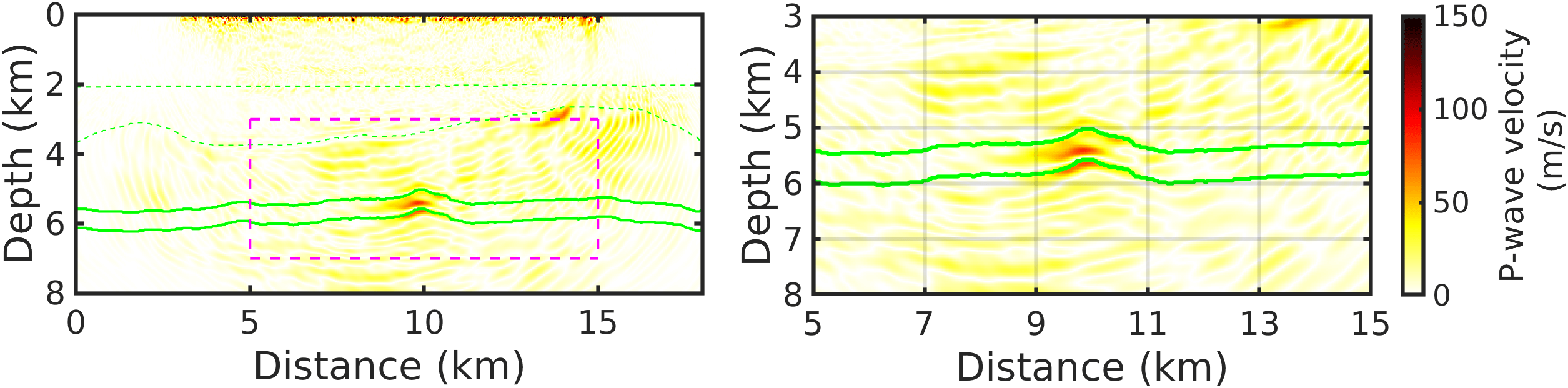}
        }  
    \hfill
    \subfloat[\label{fig:resulting_GPS_cd_recExt}]
        {%
        \includegraphics[width=0.48\linewidth]{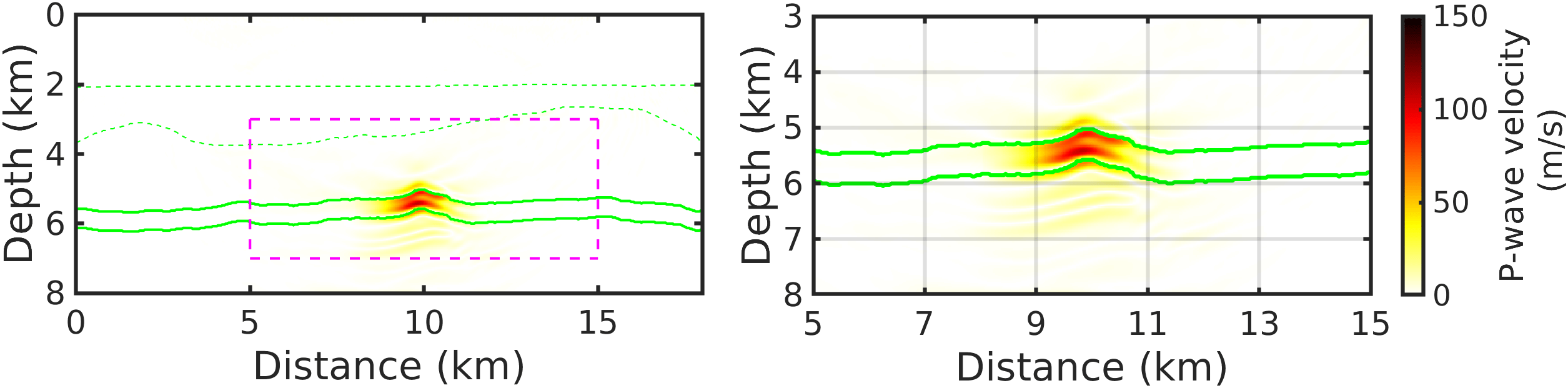}
        }
  \caption{Time-lapse estimates by the conventional approach (left-hand column) and the receiver-extension strategy (right-hand column) in the \textbf{GPS coordinate inaccuracy case}. Panels (a) and (b) refer to the parallel, (c) and (d) to the sequential, and (e) and (f) to the central-difference 4D FWI schemes.}
  \label{fig:resulting_GPS} 
\end{figure*}

Figure \ref{fig:resulting_GPS} shows time-lapse estimates and presents a zoomed view of the target region enclosed by the magenta rectangle, in which the left-hand column refers to the conventional FWI approach application and the right-hand column refers to the receiver-extension strategy. In particular, Figs. \ref{fig:resulting_GPS_par_classical} and \ref{fig:resulting_GPS_par_recExt} present time-lapse estimates using the parallel 4D FWI strategy, while Figs. \ref{fig:resulting_GPS_seq_classical} and \ref{fig:resulting_GPS_seq_recExt} illustrate estimates from the sequential 4D FWI strategy, and Figs. \ref{fig:resulting_GPS_cd_classical} and \ref{fig:resulting_GPS_cd_recExt} show the central-difference 4D FWI resulting time-lapse models. Notably, time-lapse estimates resulting from the combination of the sequential 4D FWI approach and the receiver-extension strategy exhibit several artifacts, as depicted in Fig. \ref{fig:resulting_GPS_seq_recExt}, particularly prominent in the top of salt layer and immediately following the pre-salt reservoirs, where wavepaths imprints become prominent. Interestingly, the time-lapse models obtained through the combination of the sequential 4D FWI approach and conventional FWI (Fig. \ref{fig:resulting_GPS_seq_classical}) display fewer artifacts than in the perfect repeatability case (Fig. \ref{resulting_noNR_seq_classical}). It is also worth noting that the time-lapse model resulting from this combination (the sequential 4D FWI strategy with the conventional FWI approach) exhibits fewer artifacts in the area enclosed by the magenta rectangle (\ref{fig:resulting_GPS_seq_classical}) than the time-lapse model resulting from the combination between the sequential 4D FWI strategy and the receiver-extension FWI approach (Fig. \ref{fig:resulting_GPS_seq_recExt}). However, the obtained time-lapse anomaly is not as good retrieved as in the perfect repeatability case and may be misinterpreted as noise rather than an indicative of P-wave velocity changes within the reservoirs.

The results also reveal that time-lapse models obtained through conventional FWI (left-hand column of Fig. \ref{fig:resulting_GPS}) do not clearly delineate the time-lapse anomaly, potentially complicating seismic interpretations. Furthermore, several artifacts are observed surrounding the seismic source (labelled as receiver) positions at the top of the models. In contrast, when the receiver-extension strategy is combined with the parallel 4D FWI (Fig. \ref{fig:resulting_GPS_par_recExt}) and central-difference 4D FWI (Fig. \ref{fig:resulting_GPS_cd_recExt}) approaches, the retrieved time-lapse anomaly is satisfactorily outlined while mitigating time-lapse noises throughout the time-lapse estimates. These combinations result in more accurate estimates, closely reaching the true model. In this NR scenario, we notice that the combination of central difference 4D FWI and receiver-extension-based FWI has the ability to generate more accurate time-lapse models (Fig. \ref{fig:resulting_GPS_cd_recExt}) according to the measurements presented in Table \ref{tab:GPS_coordinate_inaccuracies}.

\begin{table}[!b]
\centering
\caption{Main statistics comparing the retrieved time-lapse models with the true model for the GPS coordinate inaccuracies case. \label{tab:GPS_coordinate_inaccuracies}}
\begin{tabular}{*5c}
\toprule
 &  \multicolumn{2}{c}{NRMS} & \multicolumn{2}{c}{R}\\
4D strategy   & C-FWI   & RE-FWI   & C-FWI   & RE-FWI \\
\midrule
Parallel  &  1.9109 & 0.7718 & 0.1651  & 0.6481 \\
Sequential   &  1.8103 & 1.7171 & 0.0965  & 0.2638\\
Central-difference   &  1.6647  & \textbf{0.6291} &  0.1470     & \textbf{0.8322}\\
\bottomrule
\end{tabular}\\
\vspace{.1cm} {\footnotesize NRMS, normalized root-mean-square; R, Pearson's coefficient; \\ C-FWI, conventional FWI; RE-FWI, receiver-extension FWI.}
\end{table}

\subsubsection{Water velocity changes in the ocean layers}

\begin{figure}[!htb]    
\centering
\subfloat[\label{fig:water_variation_model_season}]
        {%
        \includegraphics[width=.5\columnwidth]{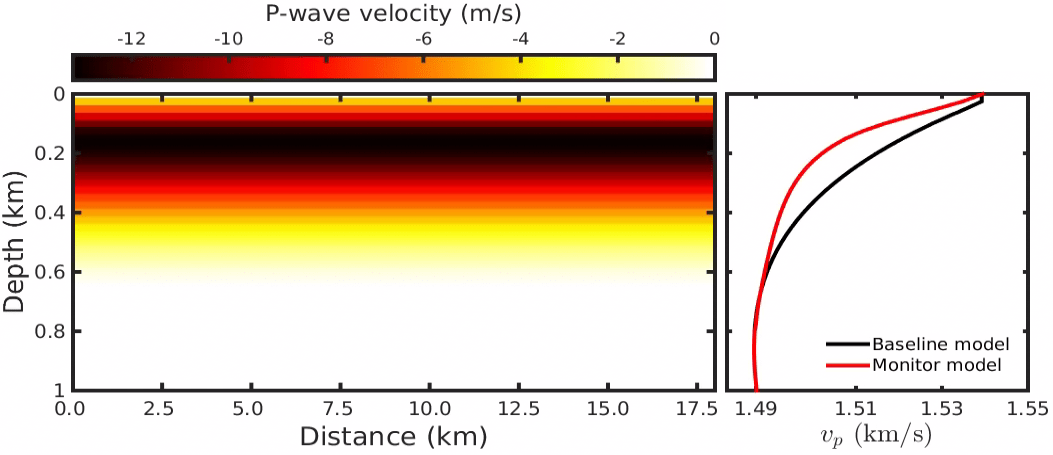}
        }
    \\
    \subfloat[\label{fig:water_variation_model_lateral}]
        {%
        \includegraphics[width=.5\columnwidth]{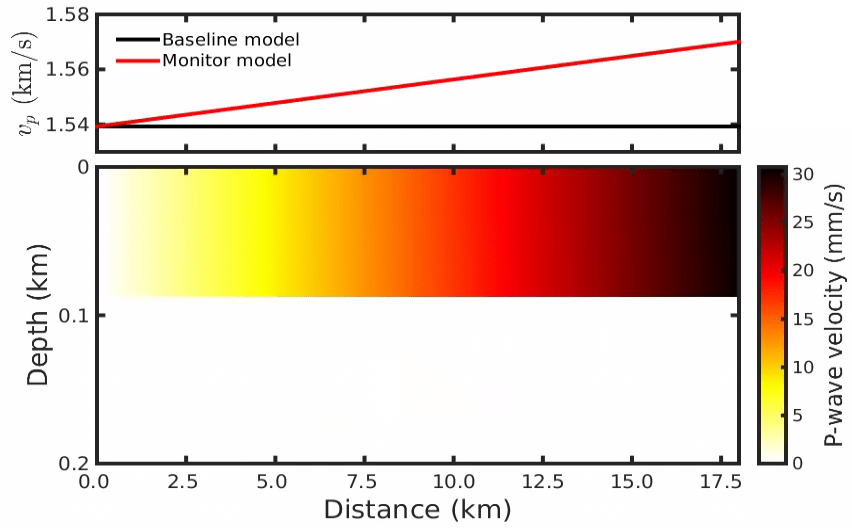}
        }
  \caption{The ocean water layer's (a) seasonal and (b) \textit{warm} temperature perturbations considered in this work. }
  \label{fig:water_errors} 
\end{figure}

In this section we address NR challenges related to velocity variations within the ocean water layer. We investigate two specific scenarios. Firstly, we explore the impact of seasonal variations in ocean water properties driven by natural phenomena influenced by changes in temperature, exposure to sunlight, and wind patterns throughout the year. In this context, the monitor model includes, in addition to the time-lapse anomaly, the perturbation depicted in Fig. \ref{fig:water_variation_model_season}. In the second one, we investigate a scenario involving a shallow oceanic sublayer characterized by lateral velocity variations, featuring a gradual increase in horizontal P-wave velocity gradients to simulate the effect of a \textit{warm} temperature. This variation is confined within an upper limit of $2\%$. The perturbation included in the monitor model for this scenario is illustrated in Fig. \ref{fig:water_variation_model_lateral}. 

Figures \ref{fig:seismogram_is_1_delta_m_we} and \ref{fig:seismogram_is_11_delta_m_we} depict the time-lapse data related to the seasonal variations case in the ocean layer, while Figs. \ref{fig:seismogram_is_1_delta_m_we_lateral} and \ref{fig:seismogram_is_11_delta_m_we_lateral} illustrate the time-lapse data related to the lateral variations case. It is important to observe that several waveform phases are unrelated to reservoir responses (as evident by comparing it with the data in Figs. \ref{fig:seismogram_is_1_delta_m} and \ref{fig:seismogram_is_11_delta_m}). This feature highlights the adverse impacts of water layer velocity variations on the time-lapse data interpretation. 

\begin{figure*}[!htb]
    \centering
    \subfloat[\label{fig:resulting_seasonalWater_par_classical}]
        {%
        \includegraphics[width=0.48\linewidth]{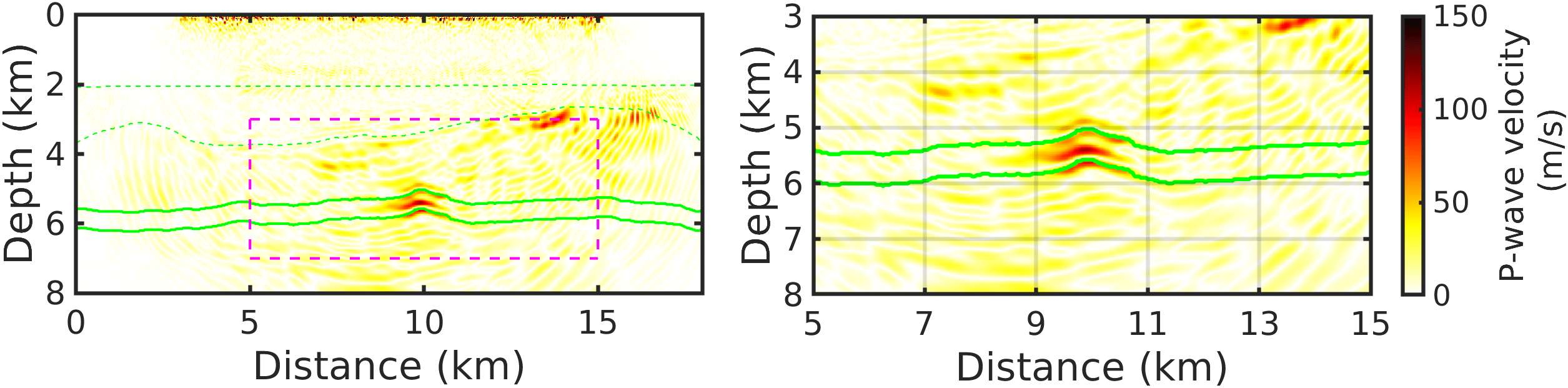}

        }  
    \hfill
    \subfloat[\label{fig:resulting_seasonalWater_par_recExt}]
        {%
        \includegraphics[width=0.48\linewidth]{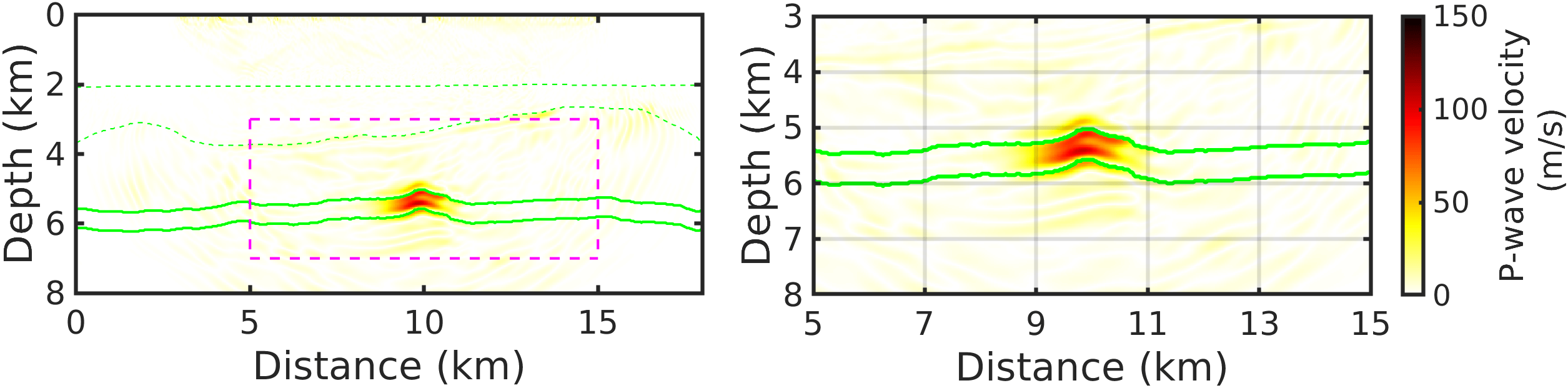}
        }
    \\
    \subfloat[\label{fig:resulting_seasonalWater_seq_classical}]
        {%
        \includegraphics[width=0.48\linewidth]{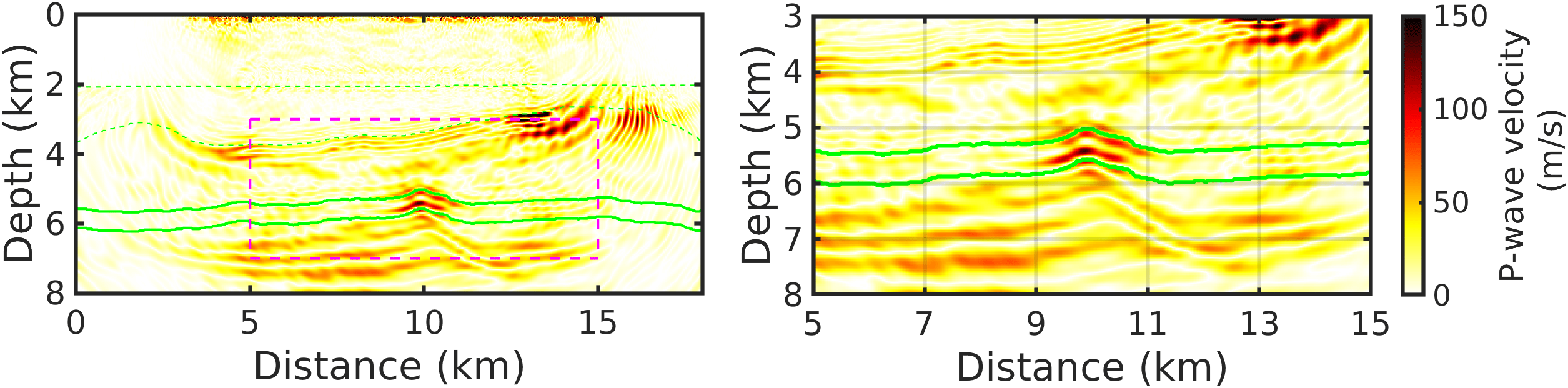}
        }  
    \hfill
    \subfloat[\label{fig:resulting_seasonalWater_seq_recExt}]
        {%
        \includegraphics[width=0.49\linewidth]{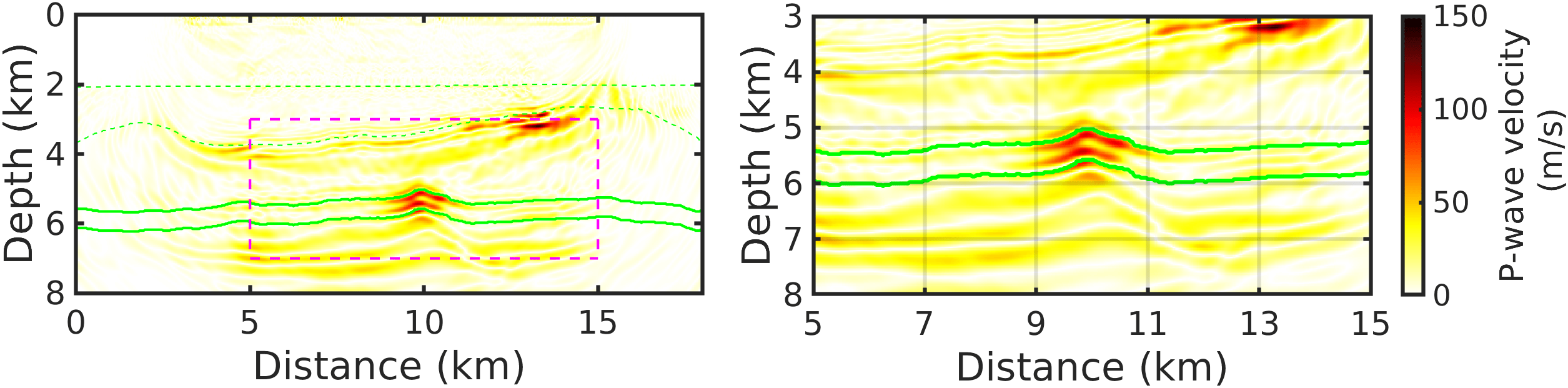}
        }
    \\
    \subfloat[\label{fig:resulting_seasonalWater_cd_classical}]
        {%
        \includegraphics[width=0.48\linewidth]{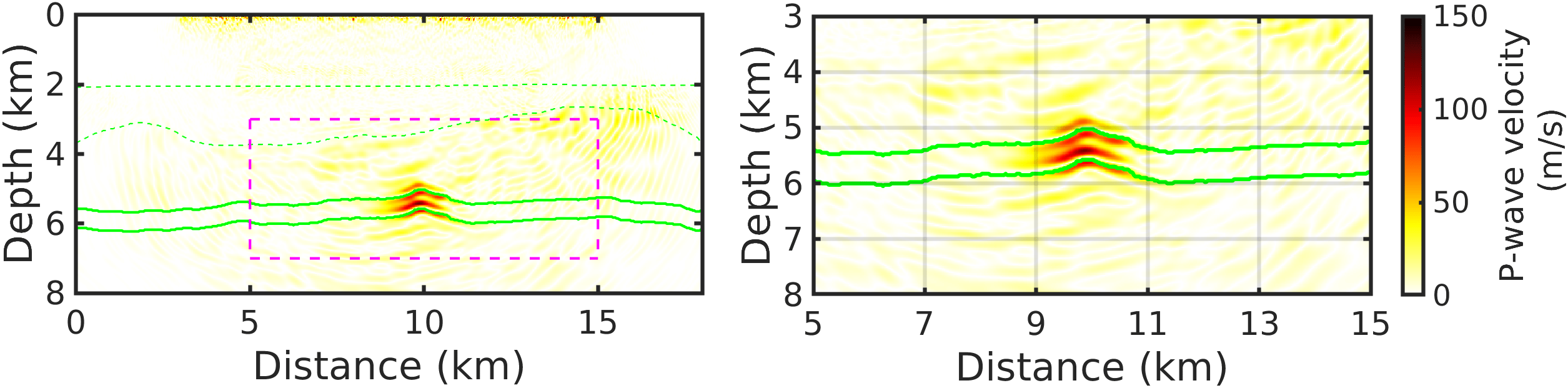}
        
        }  
    \hfill
    \subfloat[\label{fig:resulting_seasonalWater_cd_recExt}]
        {%
        \includegraphics[width=0.48\linewidth]{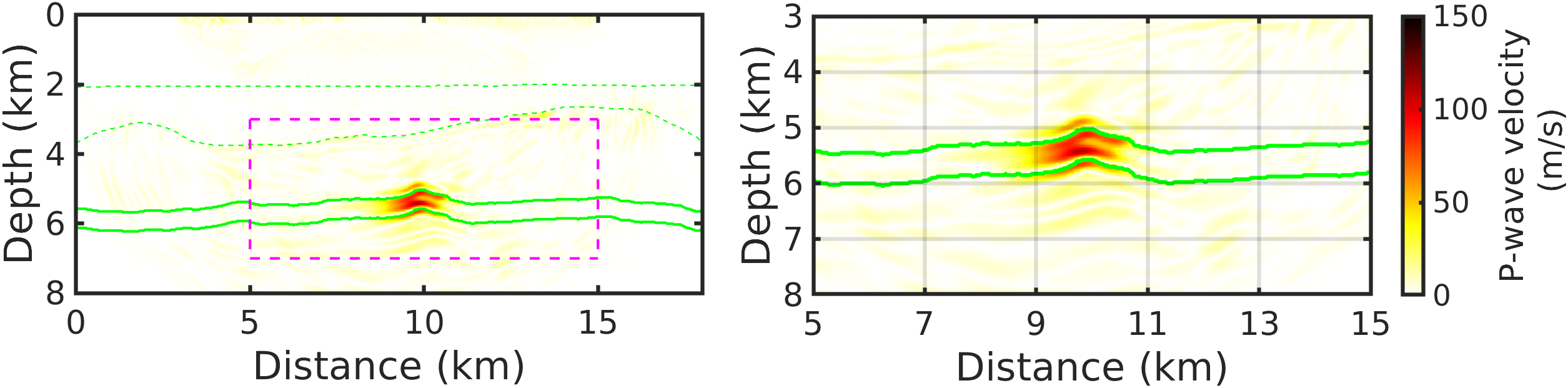}
        }
    \caption{Time-lapse estimates by the conventional approach (left-hand column) and the receiver-extension strategy (right-hand column) in the \textbf{water velocity changes in the sea layer case related to seasonal variability}. Panels (a) and (b) refer to the parallel, (c) and (d) to the sequential, and (e) and (f) to the central-difference 4D FWI schemes.}    
    \label{fig:resulting_seasonalWater} 
\end{figure*}

\begin{figure*}[!htb]
    \centering
    \subfloat[\label{fig:resulting_lateralWater_par_classical}]
        {%
        \includegraphics[width=0.48\linewidth]{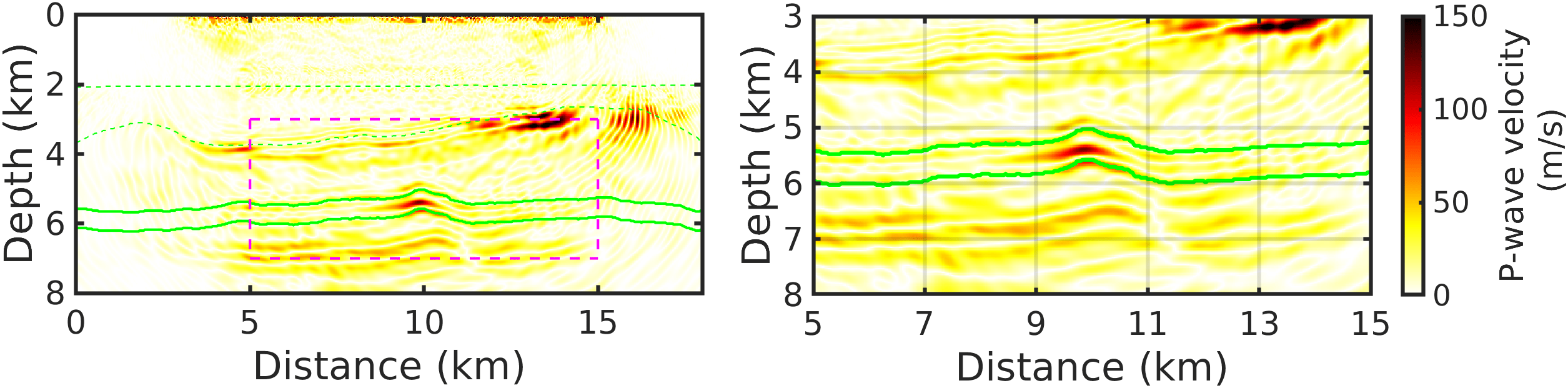}
        }  
    \hfill
    \subfloat[\label{fig:resulting_lateralWater_par_recExt}]
        {%
        \includegraphics[width=0.48\linewidth]{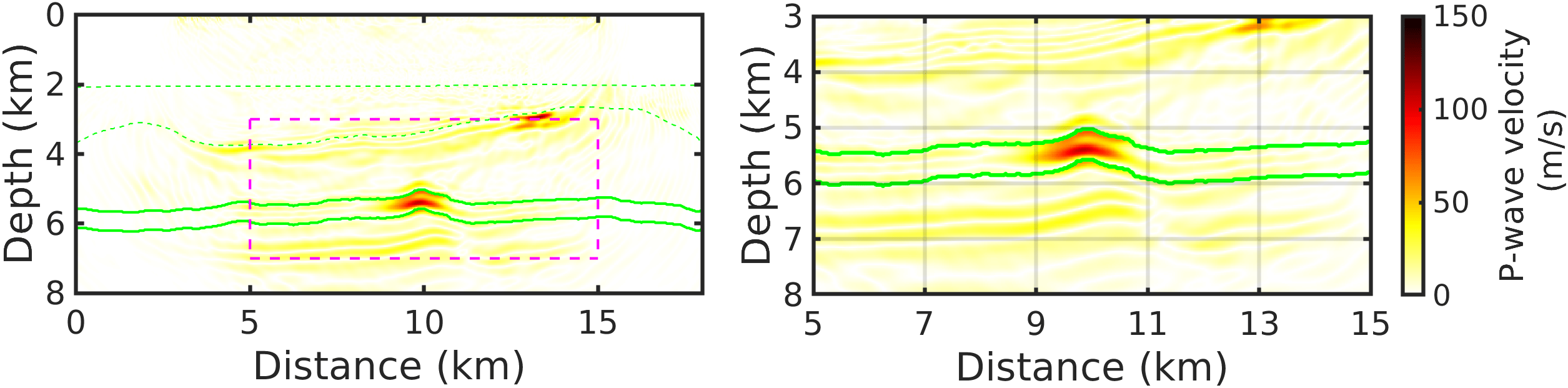}
        }
    \\
    \subfloat[\label{fig:resulting_lateralWater_seq_classical}]
        {%
        \includegraphics[width=0.48\linewidth]{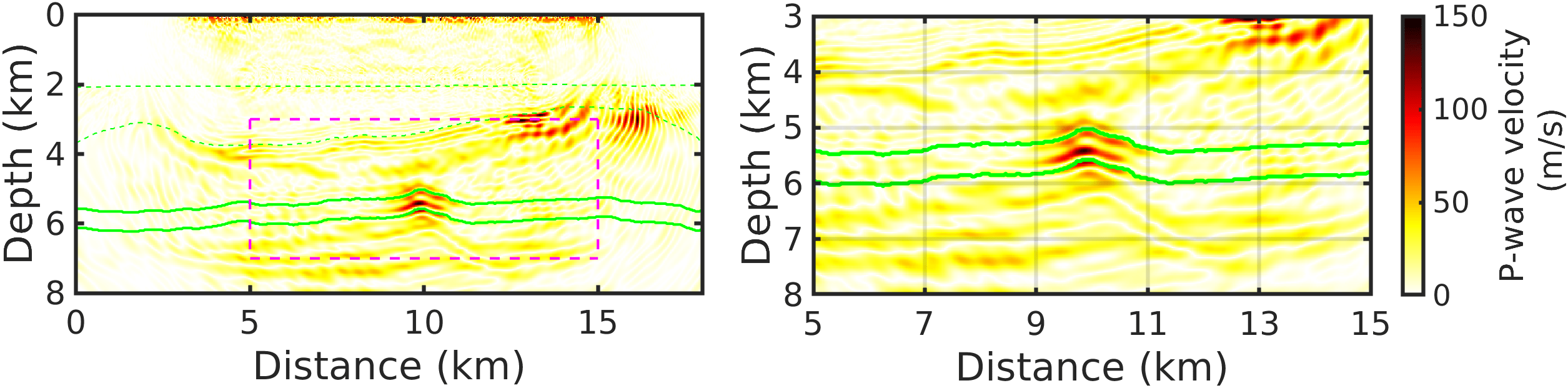}
        }  
    \hfill
    \subfloat[\label{fig:resulting_lateralWater_seq_recExt}]
        {%
        \includegraphics[width=0.48\linewidth]{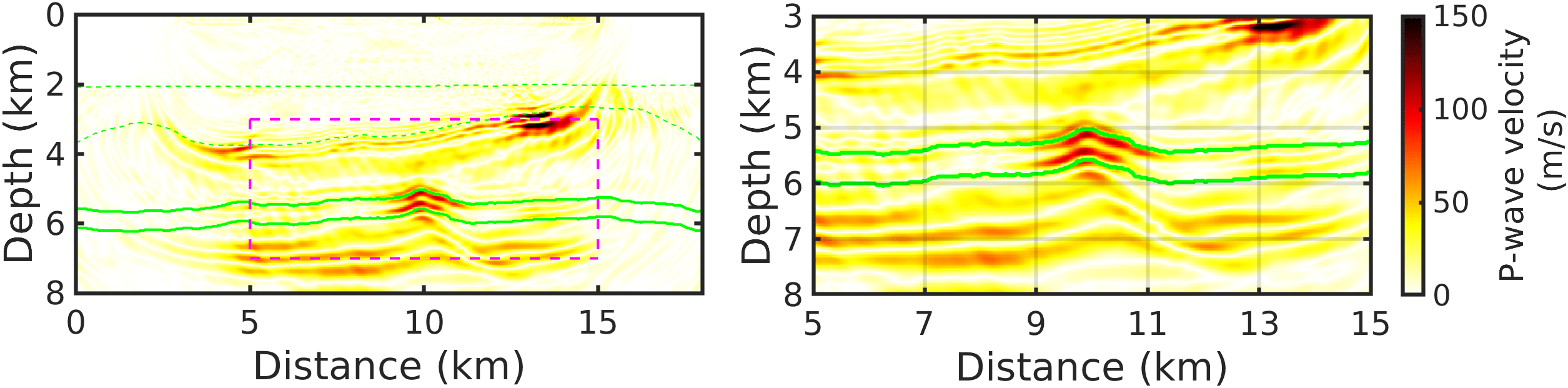}
        }
    \\
    \subfloat[\label{fig:resulting_lateralWater_cd_classical}]
        {%
        \includegraphics[width=0.48\linewidth]{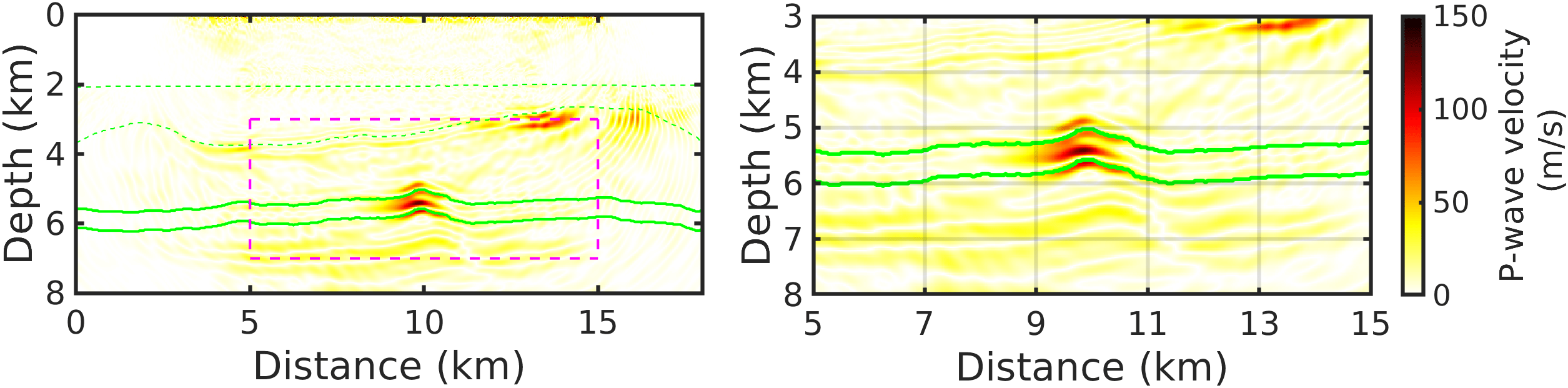}
        }  
    \hfill
    \subfloat[\label{fig:resulting_lateralWater_cd_recExt}]
        {%
        \includegraphics[width=0.48\linewidth]{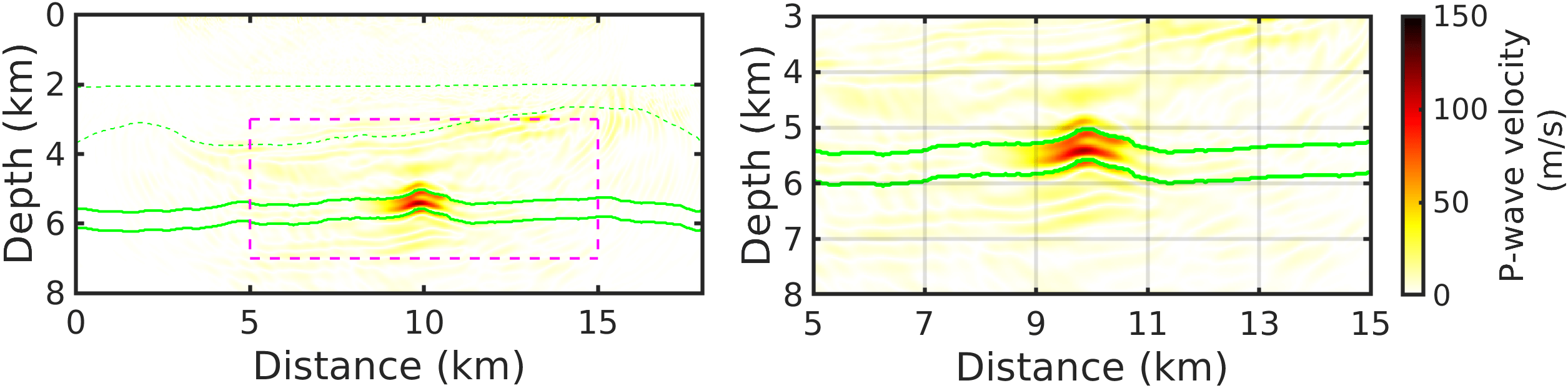}
        }
    \caption{Time-lapse estimates by the conventional approach (left-hand column) and the receiver-extension strategy (right-hand column) in the \textbf{water velocity lateral changes case}. Panels (a) and (b) refer to the parallel, (c) and (d) to the sequential, and (e) and (f) to the central-difference 4D FWI schemes.}    \label{fig:resulting_lateralWater} 
\end{figure*}

Figures \ref{fig:resulting_seasonalWater} and \ref{fig:resulting_lateralWater} show time-lapse estimates and present a zoomed view of the target region enclosed by the magenta rectangle for the NR issues related to water velocity changes in the sea layer due to seasonal variability and the lateral changes case, respectively; the left-hand column refers to the conventional FWI approach application, while the right-hand column refers to the receiver-extension strategy. Once more, the sequential 4D FWI strategy produces time-lapse models with the most significant time-lapse noises, manifesting as a substantial presence of artifacts in the entire models, as depicted in panels (c) and (d) of Figs. \ref{fig:resulting_seasonalWater} and \ref{fig:resulting_lateralWater}. When the sequential strategy is employed, the resulting time-lapse model via the conventional FWI approach (Fig. \ref{fig:resulting_lateralWater_seq_classical}) exhibits fewer artifacts in the area enclosed by the magenta rectangle than in the case of the receiver-extension FWI approach (Fig. \ref{fig:resulting_lateralWater_seq_recExt}). In the cases of the parallel 4D FWI and central-difference 4D FWI strategies, it is worth noting that the receiver-extension approach excels in generating cleaner time-lapse models compared to the conventional FWI approach. However, combining the receiver-extension FWI and the central-difference 4D FWI approach generated the best time-lapse model with the lowest NRMS error and highest similarity to the true model (highest R-value), as summarized in Table \ref{tab:Water_velocity_changes}, in both water variations cases.

\begin{table}[!b]
\centering
\caption{Main statistics comparing the retrieved time-lapse models with the true model for the water velocity changes in the ocean layers case. \label{tab:Water_velocity_changes}}
\flushleft{\hspace{.55cm} \textit{Seasonal
variations}} \\ \centering
\begin{tabular}{*5c}
\toprule
 &  \multicolumn{2}{c}{NRMS} & \multicolumn{2}{c}{R}\\
4D strategy   & C-FWI   & RE-FWI   & C-FWI   & RE-FWI \\
\midrule
Parallel  & 1.9137  & 0.7215 & 0.1679  & 0.6995 \\
Sequential   &  2.3383 & 1.3445 & 0.1539  & 0.3222\\
Central-difference   &  1.1403  & \textbf{0.6728} &  0.3710     & \textbf{0.7634} \\
\bottomrule
\end{tabular}
\flushleft{\hspace{.55cm} \textit{Lateral
variations}} \\ \centering
\begin{tabular}{*5c}
\toprule
 &  \multicolumn{2}{c}{NRMS} & \multicolumn{2}{c}{R}\\
4D strategy   & C-FWI   & RE-FWI   & C-FWI   & RE-FWI \\
\midrule
Parallel  & 2.1168  & 0.9382 & 0.1223  & 0.4814 \\
Sequential   &  1.9822 & 1.7290 & 0.1768  & 0.2551\\
Central-difference   &  1.2656  & \textbf{0.7472} &  0.3052     & \textbf{0.6636} \\
\bottomrule
\end{tabular}\\
\vspace{.1cm} \centering {\footnotesize NRMS, normalized root-mean-square; R, Pearson's coefficient; \\ C-FWI, conventional FWI; RE-FWI, receiver-extension FWI.}
\end{table}

\subsubsection{Combined NR issues}

Now, we analyze the impact of combined NR factors on time-lapse FWI to simulate a more realistic and challenging situation. In this regard, instead of investigating the NR effects of GPS inaccuracies and time-lapse changes in water velocities in the ocean layers separately, we consider them collectively. Consequently, the monitor model is now a composite of the perturbations depicted in Fig. \ref{fig:water_errors}, superimposed onto the baseline model, all contributing to the complexity of the time-lapse seismic analysis. It is worth highlighting that, to simplify the model, we are assuming a linear superposition of these NR effects, especially in the presence of temperature gradients. Figures \ref{fig:seismogram_is_1_delta_m_fullNR} and \ref{fig:seismogram_is_11_delta_m_fullNR} illustrate the time-lapse data associated with the scenario involving combined NR issues.

\begin{figure*}[!htb]
    \centering
    \subfloat[\label{fig:resulting_fullNR_par_classical}]
        {%
        \includegraphics[width=0.48\linewidth]{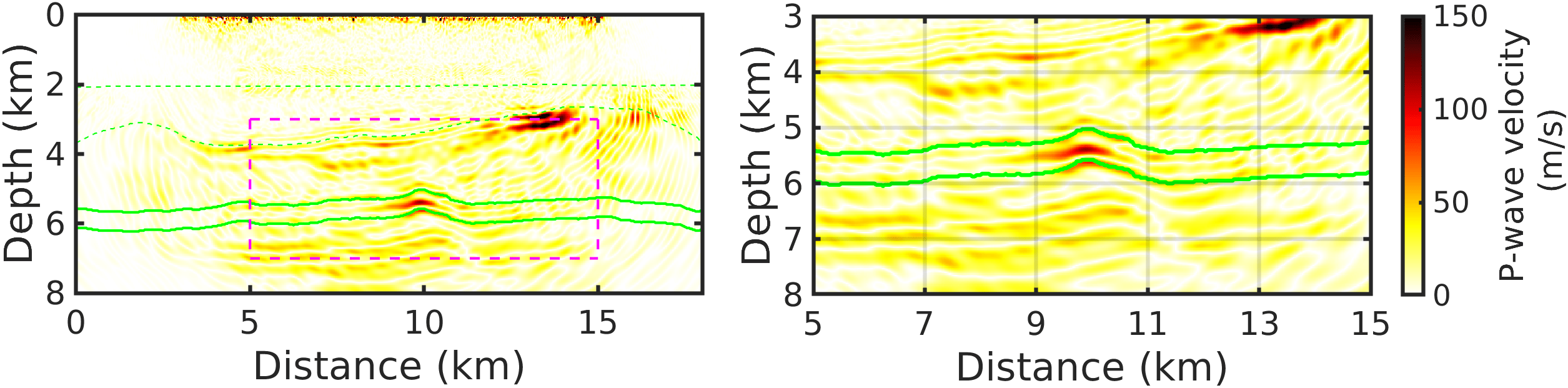}
        }  
    \hfill
    \subfloat[\label{fig:resulting_fullNR_par_recExt}]
        {%
        \includegraphics[width=0.48\linewidth]{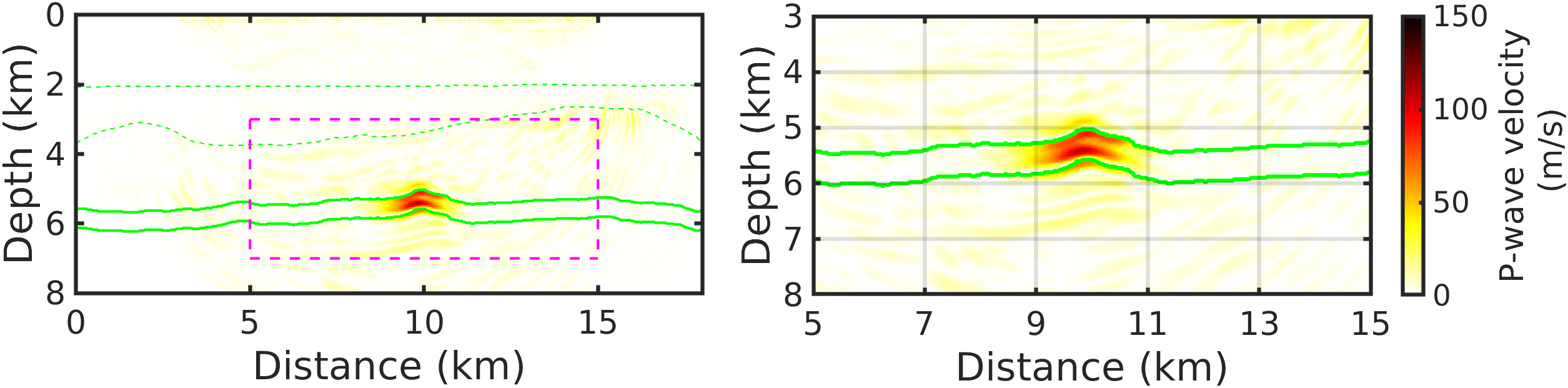}
        }
    \\
    \subfloat[\label{fig:resulting_fullNR_seq_classical}]
        {%
        \includegraphics[width=0.48\linewidth]{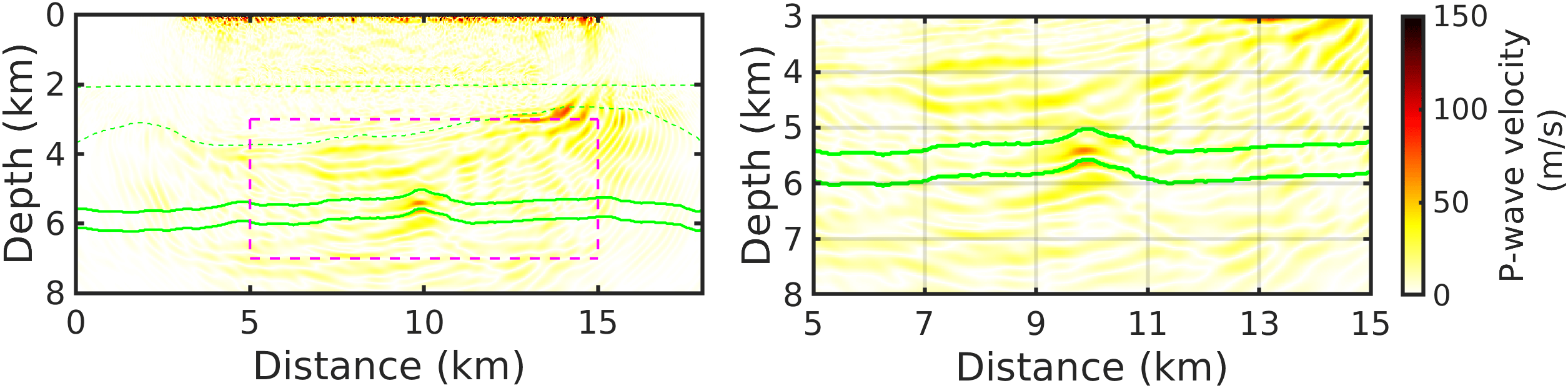}
        }  
    \hfill
    \subfloat[\label{fig:resulting_fullNR_seq_recExt}]
        {%
        \includegraphics[width=0.48\linewidth]{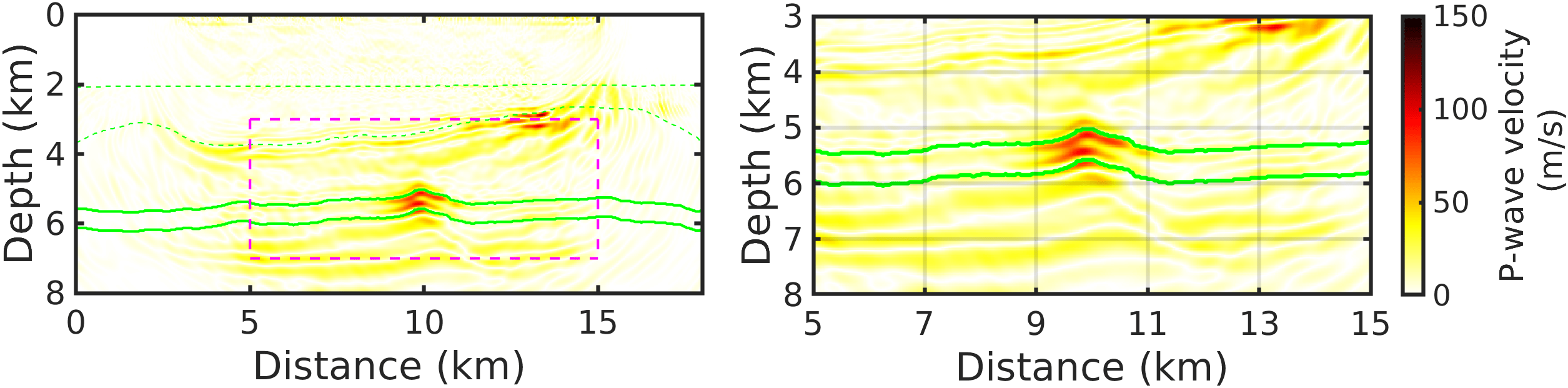}
        }
    \\
    \subfloat[\label{fig:resulting_fullNR_cd_classical}]
        {%
        \includegraphics[width=0.48\linewidth]{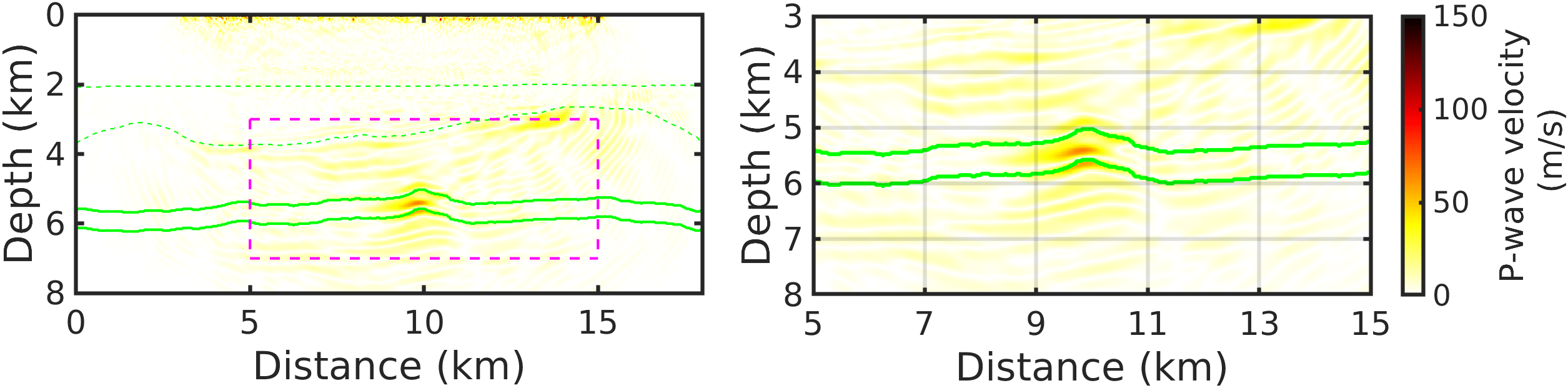}
        }  
    \hfill
    \subfloat[\label{fig:resulting_fullNR_cd_recExt}]
        {%
        \includegraphics[width=0.48\linewidth]{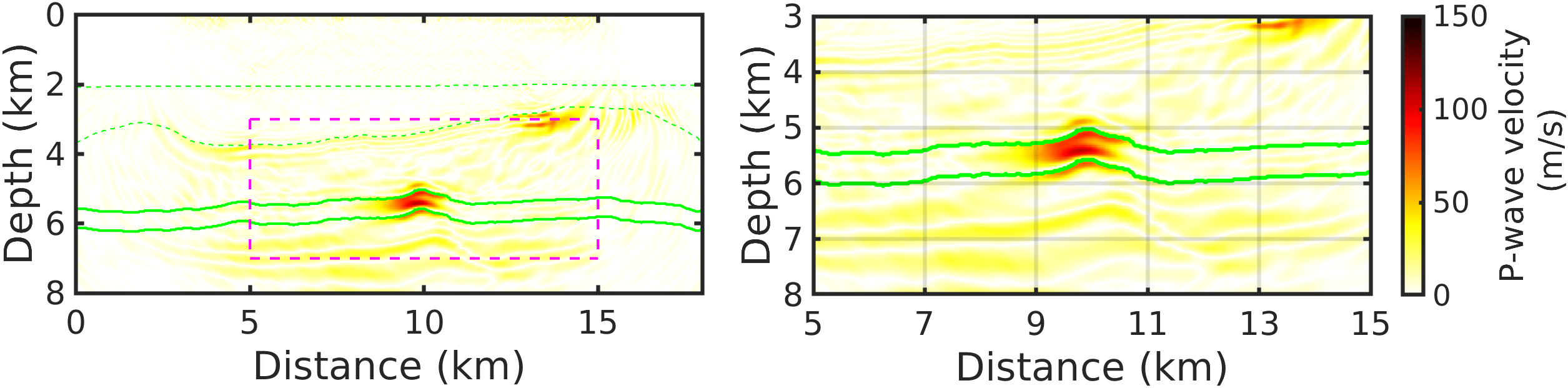}
        }
    \caption{Time-lapse estimates by the conventional approach (left-hand column) and the receiver-extension strategy (right-hand column) in the \textbf{combined NR case}. Panels (a) and (b) refer to the parallel, (c) and (d) to the sequential, and (e) and (f) to the central-difference 4D FWI schemes.}    
    \label{fig:resulting_full_NR} 
\end{figure*}

Figure \ref{fig:resulting_full_NR} shows the estimated models. When employing the parallel 4D FWI strategy, we notice that the time-lapse model obtained via the conventional FWI procedure exhibits considerable time-lapse noises (Fig. \ref{fig:resulting_fullNR_par_classical}), whereas the receiver-extension FWI effectively mitigates the time-lapse noises (Fig. \ref{fig:resulting_fullNR_par_recExt}). Besides, when employing the sequential 4D FWI and central-difference 4D FWI strategies, the retrieved time-lapse models exhibit fewer artifacts, but the time-lapse anomaly is not satisfactorily recovered, as depicted in Figs. \ref{fig:resulting_fullNR_seq_classical} and \ref{fig:resulting_fullNR_cd_classical}. In contrast, the time-lapse models retrieved using the receiver-extension FWI are notably cleaner when compared to the conventional FWI results, featuring a well-defined time-lapse anomaly that closely matches the true time-lapse model. Although, from a visual inspection, the combination of the receiver-extension FWI approach with the parallel 4D FWI strategy appears to deliver the most accurate time-lapse model, again, from a quantitative point of view, combining the receiver-extension FWI approach with the central-difference 4D FWI strategy generated time-lapse models with lower error and more significant similarity to the true model, as summarized in Table \ref{tab:combined_NR_issues}. 

\begin{table}[!htbp]
\centering
\caption{Main statistics comparing the retrieved time-lapse models with the true model for the combined NR issues case. \label{tab:combined_NR_issues}}
\begin{tabular}{*5c}
\toprule
 &  \multicolumn{2}{c}{NRMS} & \multicolumn{2}{c}{R}\\
4D strategy   & C-FWI   & RE-FWI   & C-FWI   & RE-FWI \\
\midrule
Parallel  & 2.1626  & 0.9584 & 0.1230  & 0.4926 \\
Sequential   &  1.7871 & 1.1730 & 0.0883  & 0.3477\\
Central-difference   &  1.1870  & \textbf{0.7040} & 0.1837 & \textbf{0.7170} \\
\bottomrule
\end{tabular}\\
\vspace{.1cm} {\footnotesize NRMS, normalized root-mean-square; R, Pearson's coefficient; \\ C-FWI, conventional FWI; RE-FWI, receiver-extension FWI.}
\end{table}

\subsubsection{Bayesian Analysis}

To demonstrate the robustness and time-lapse noise-reduction capabilities of Bayesian analysis, utilizing our proposed models described in Section \ref{sec:bayesian_analysis}, we consider the combined NR issues case, which is the most challenging and realistic scenario. As mentioned in the preceding section, these effects include GPS inaccuracies and changes in water velocities in the ocean layer. 

\begin{figure*}[!b]
    \centering
    \subfloat[\label{fig:model1_tst0_0_corner}]
        {%
        \includegraphics[width=0.351\linewidth]{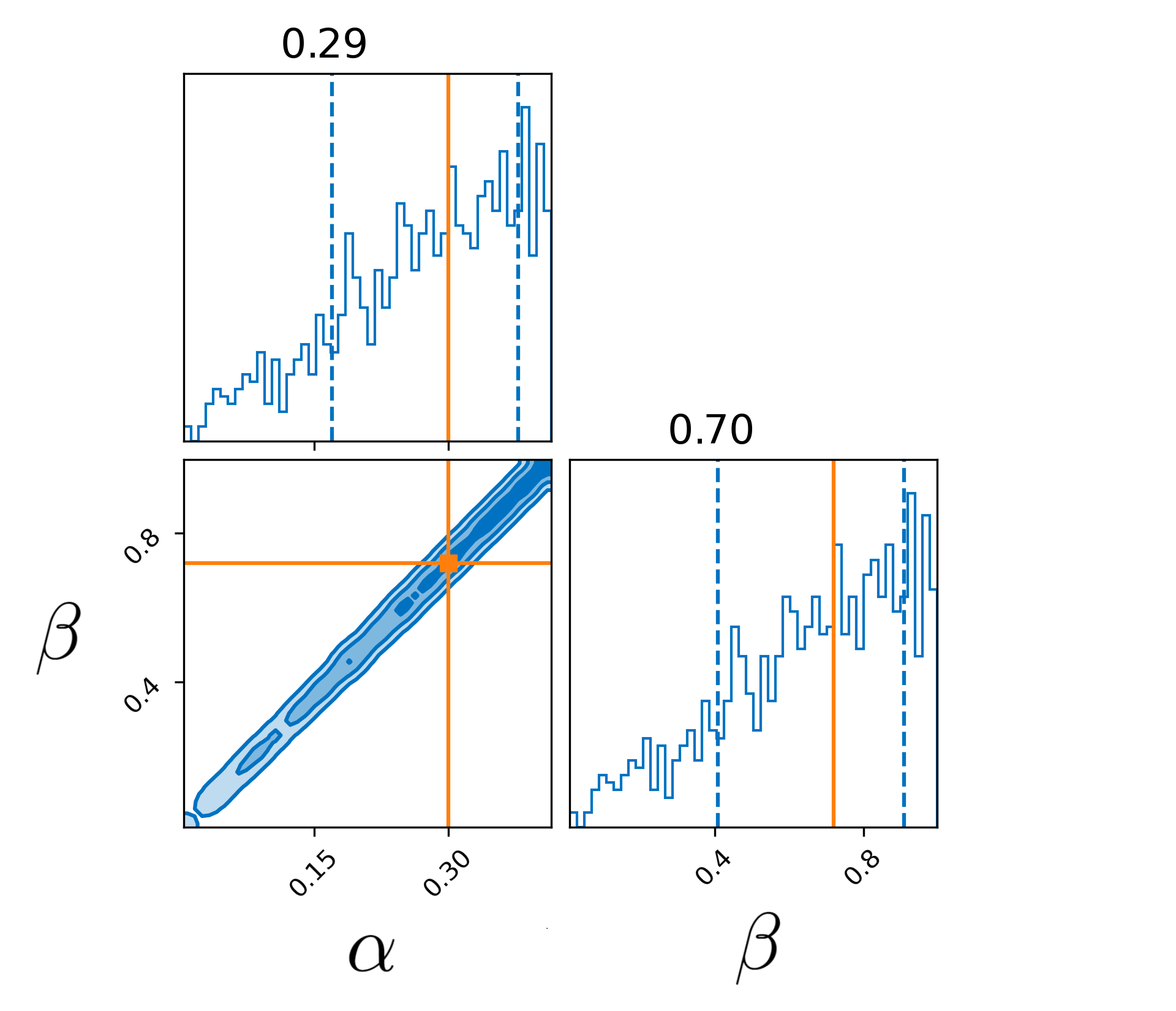}
        }
    \subfloat[\label{fig:model2_tst0_0_corner}]
        {%
        \includegraphics[width=0.301\linewidth]{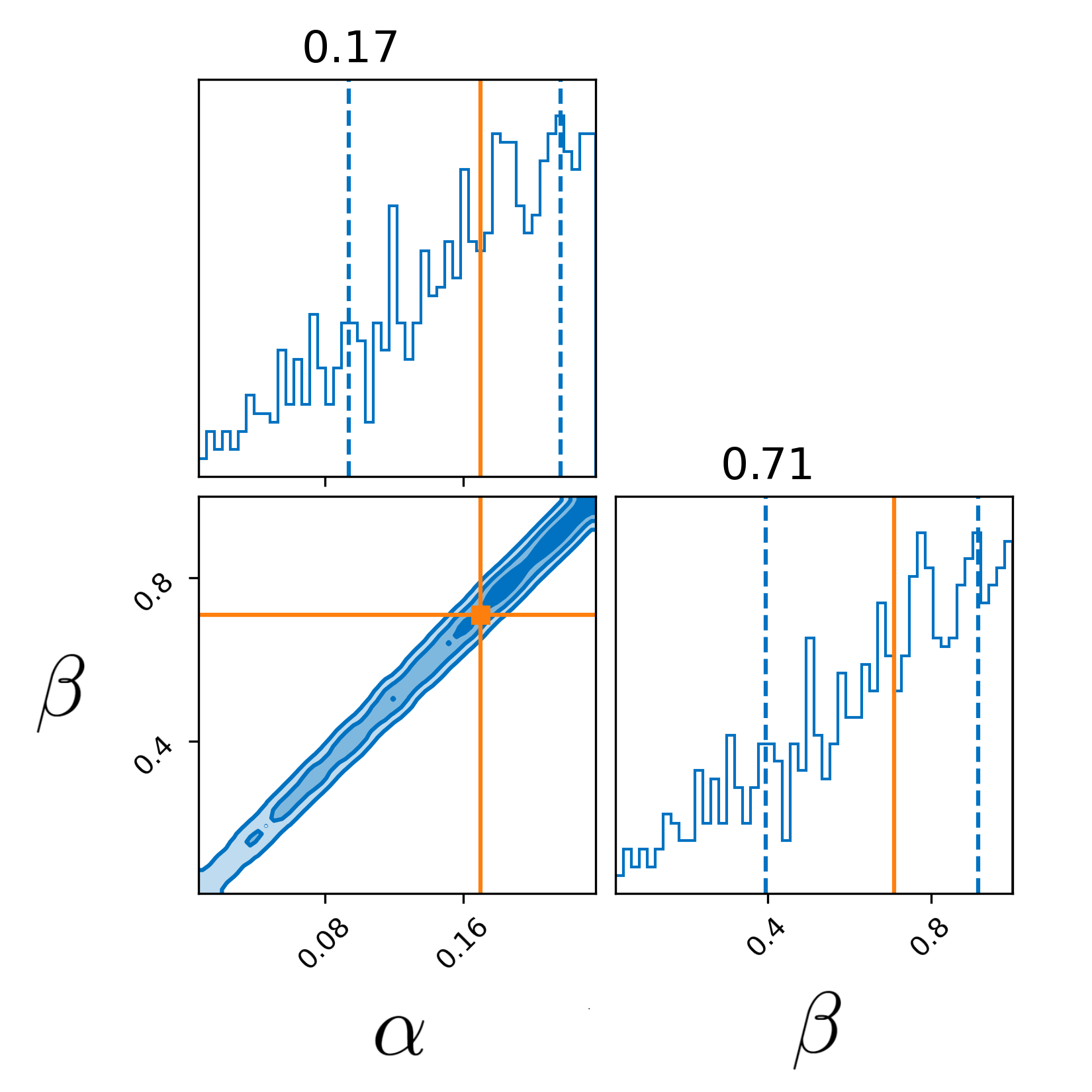}
        }  
    \subfloat[\label{fig:model3_tst0_0_corner}]
        {%
        \includegraphics[width=0.301\linewidth]{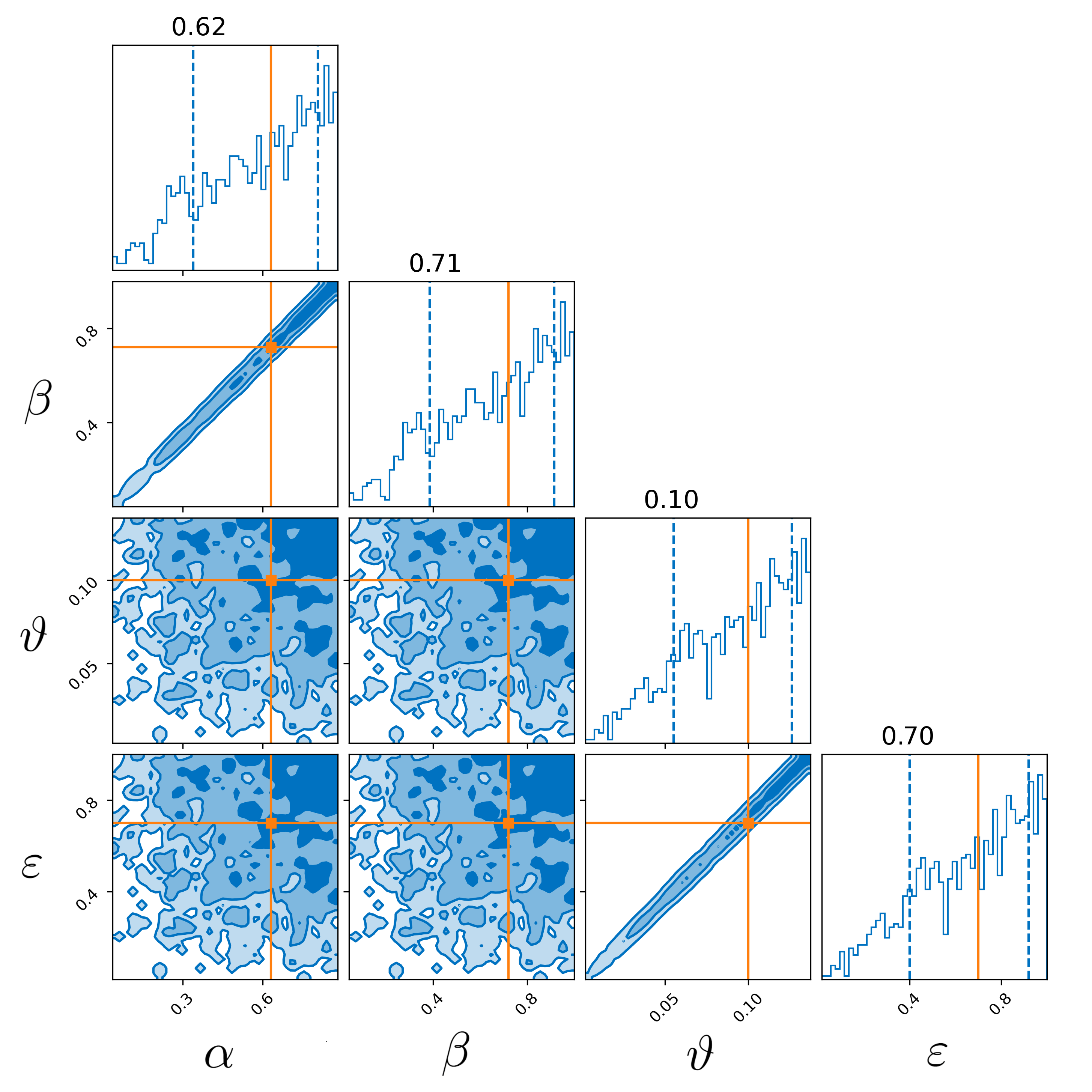}
        }  
        \\
    \subfloat[\label{fig:model1_tst0_corner}]
        {%
        \includegraphics[width=0.301\linewidth]{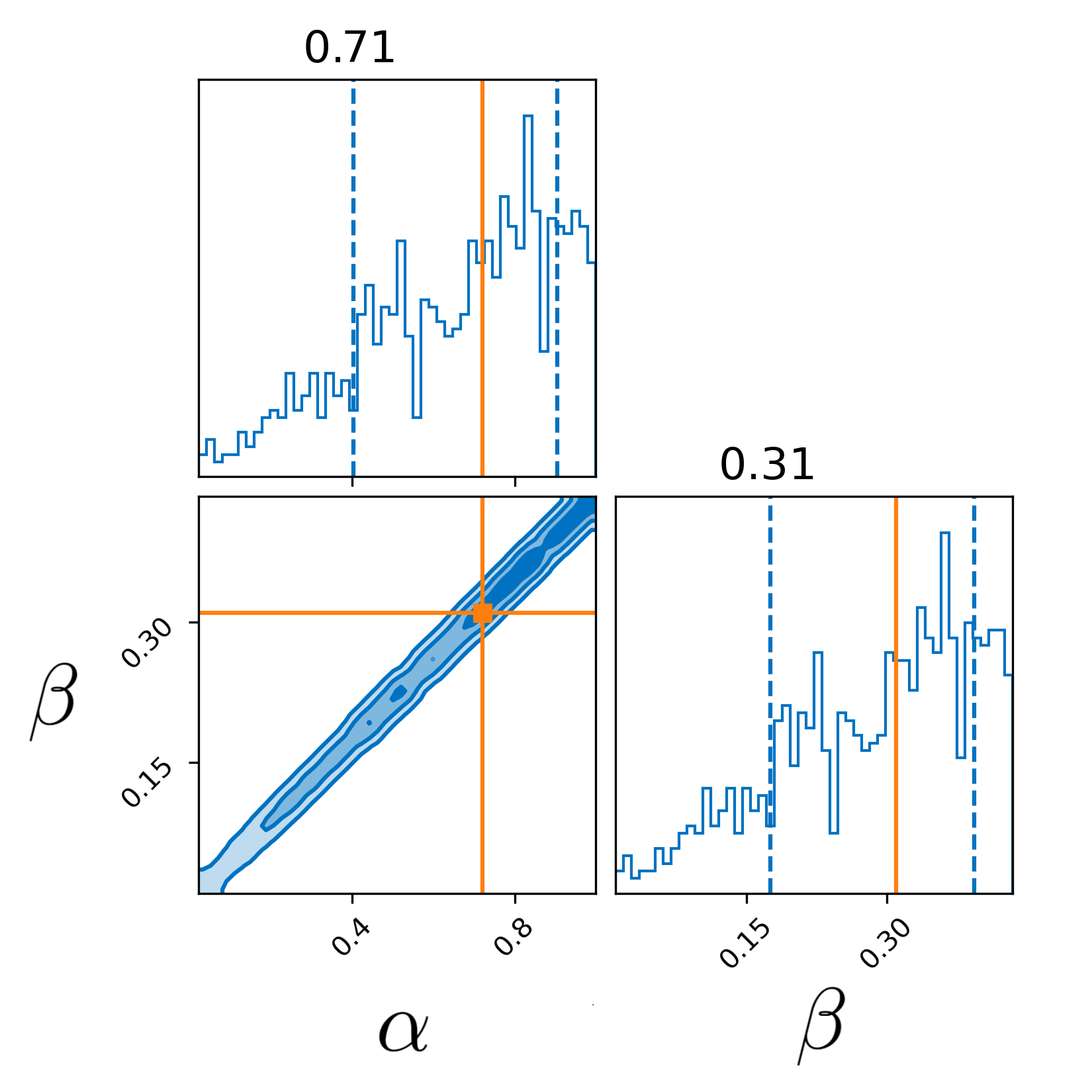}
        }
    \subfloat[\label{fig:model2_tst0_corner}]
        {%
        \includegraphics[width=0.301\linewidth]{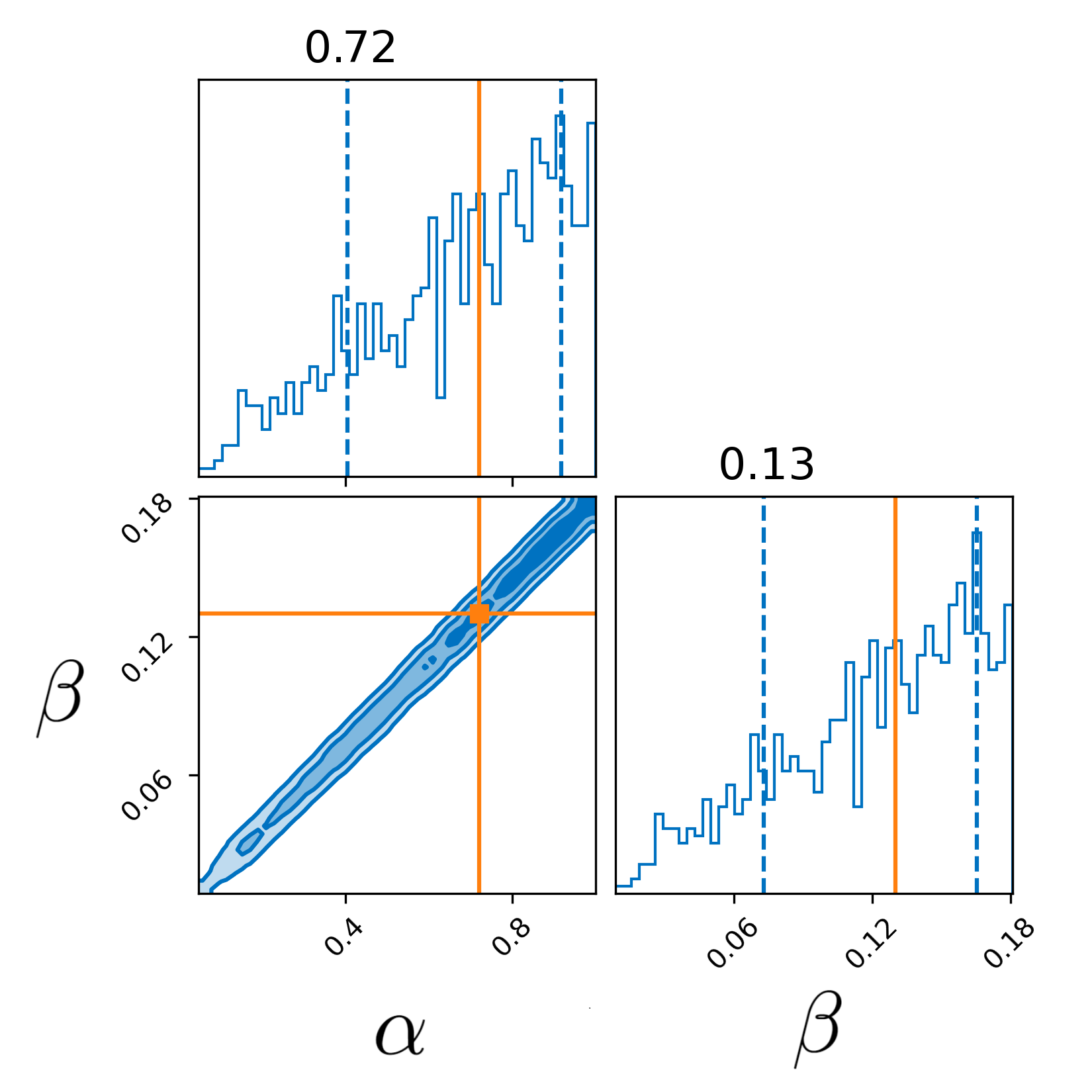}
        }  
    \subfloat[\label{fig:model3_tst0_corner}]
        {%
        \includegraphics[width=0.301\linewidth]{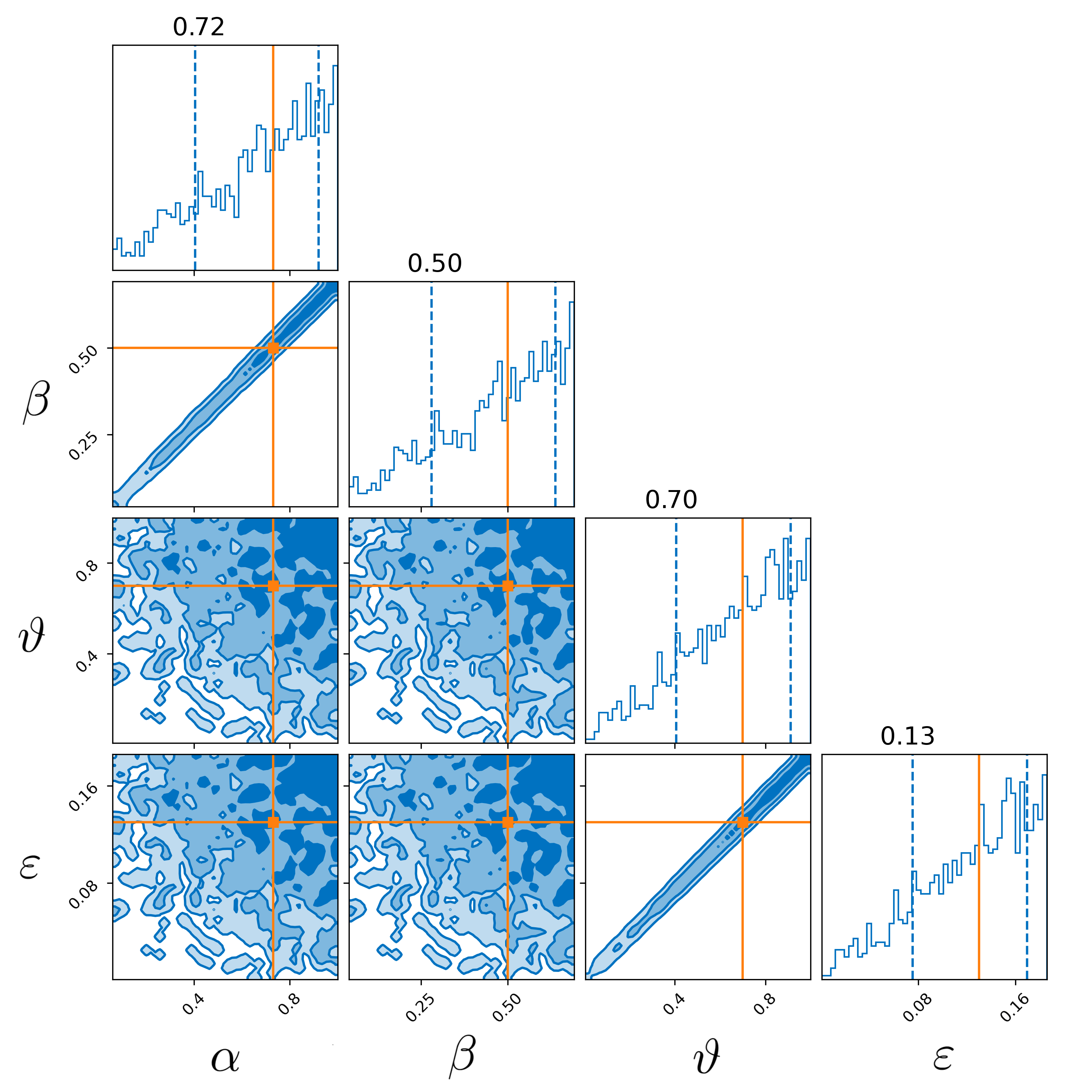}
        }  
    
    \caption{Corner plots of the inferred parameters ($\alpha$, $\beta$, $\vartheta$, and $\varepsilon$). The blue traces represent the obtained posteriors, while the orange lines signify the parameter values maximizing the posterior probability in the Bayesian time-lapse models. Panels (a)-(c) refer to the $BW_1$, $BW_2$ and $BW_3$ models applied to the resulting models from the conventional FWI approach, while panels (d)-(f) refer to the $BW_1$, $BW_2$ and $BW_3$ models applied to the resulting models from the receiver-extension FWI procedures. }    
    \label{fig:Bayes_corner} 
\end{figure*}

To estimate posteriors and statistical evidences, we employ a dynesty sampler \cite{Speagle_DYNESTY_2020}, incorporated within the Bilby package \cite{BILBY_Ashton_et_al_2019, BILBY_2020}. Figure \ref{fig:Bayes_corner} presents a corner plot displaying the inferred parameters ($\alpha$, $\beta$, $\vartheta$, and $\varepsilon$). The plot depicts the obtained posterior probability for all parameters, along with contour plots illustrating the relationship between any two parameters in solid (blue) contours, with the upper row corresponding to time-lapse models derived from conventional FWI applications and the lower row representing the time-lapse models recovered through receiver-extension FWIs. The orange lines represent the parameter values that maximize the posterior probability associated with the Bayesian time-lapse models \textit{$\text{BW}_1$} (Figs. \ref{fig:model1_tst0_0_corner} and \ref{fig:model1_tst0_corner}), \textit{$\text{BW}_2$} (Figs. \ref{fig:model2_tst0_0_corner} and \ref{fig:model2_tst0_corner}), and \textit{$\text{BW}_3$} (Figs. \ref{fig:model3_tst0_0_corner} and \ref{fig:model3_tst0_corner}).

\begin{figure*}[!b]
    \centering
    \subfloat[\label{fig:resulting_fullNR_Bayes_mod1_classical}]
        {%
        \includegraphics[width=0.48\linewidth]{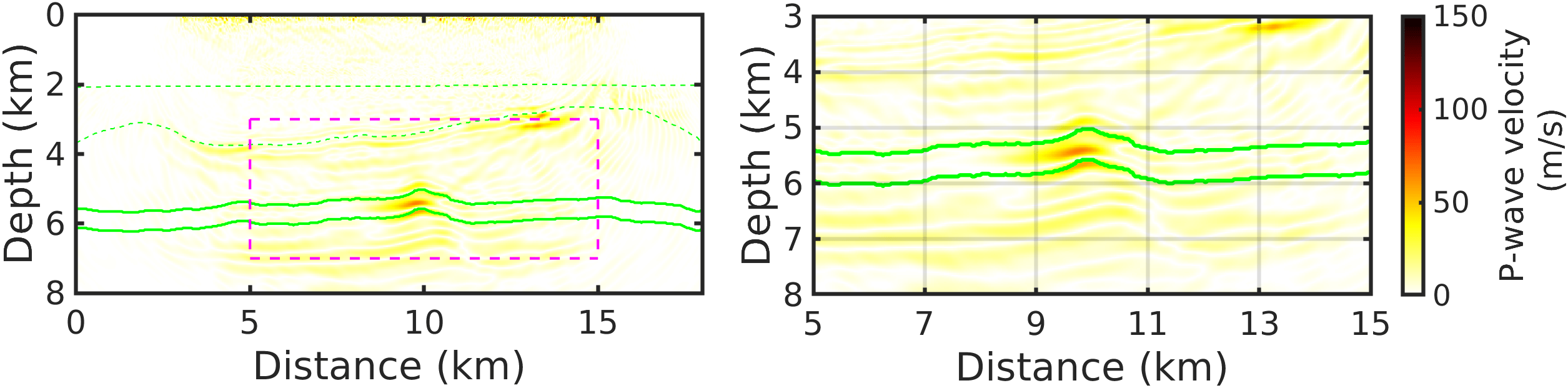}
        }  
    \hfill
    \subfloat[\label{fig:resulting_fullNR_Bayes_mod1_recExt}]
        {%
        \includegraphics[width=0.48\linewidth]{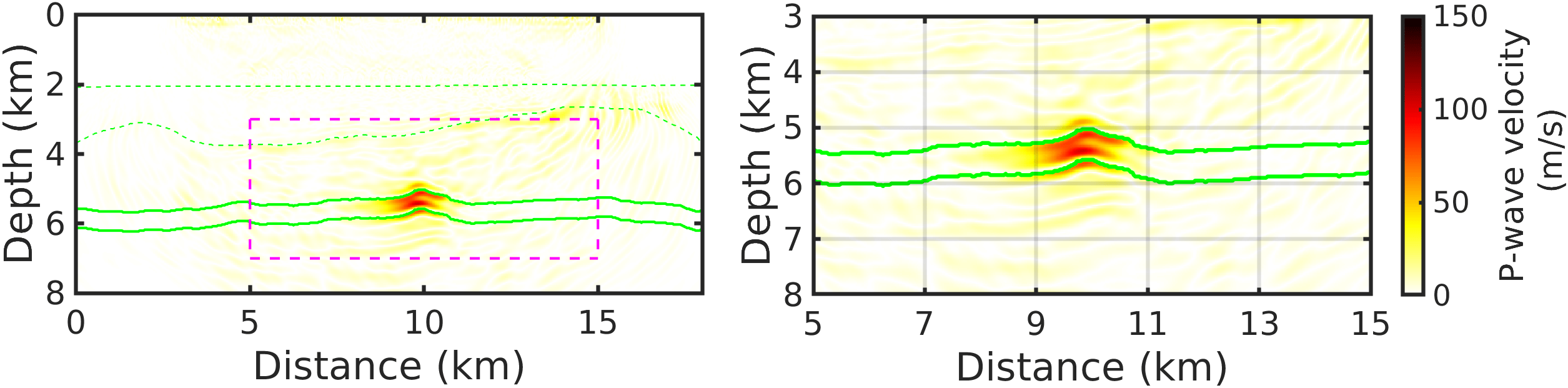}
        }
    \\
    \subfloat[\label{fig:resulting_fullNR_Bayes_mod2_classical}]
        {%
        \includegraphics[width=0.48\linewidth]{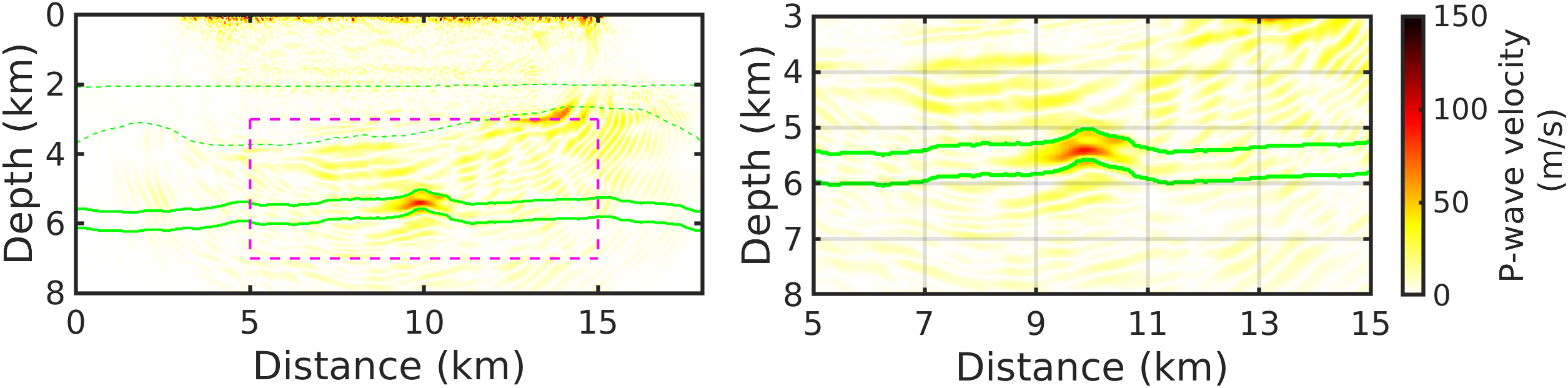}
        }  
    \hfill
    \subfloat[\label{fig:resulting_fullNR_Bayes_mod2_recExt}]
        {%
        \includegraphics[width=0.48\linewidth]{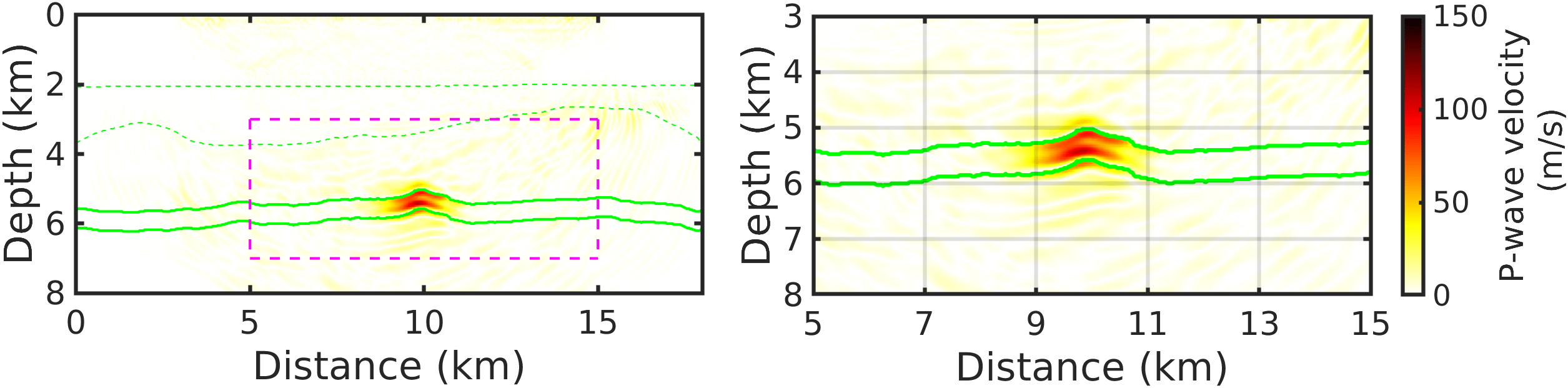}
        }
    \\
    \subfloat[\label{fig:resulting_fullNR_Bayes_mod3_classical}]
        {%
        \includegraphics[width=0.48\linewidth]{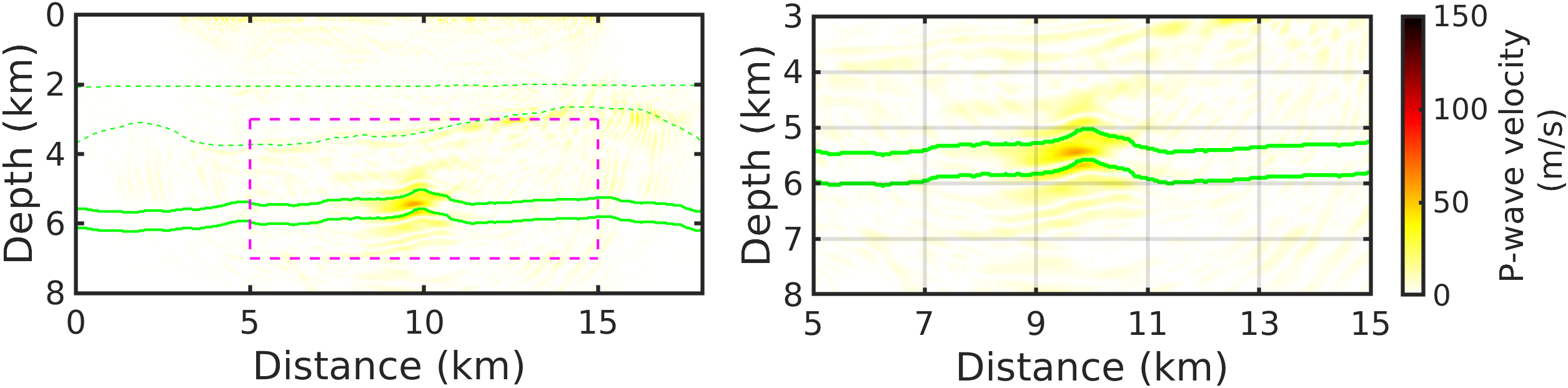}
        }  
    \hfill
    \subfloat[\label{fig:resulting_fullNR_Bayes_mod3_recExt}]
        {%
         \includegraphics[width=0.48\linewidth]{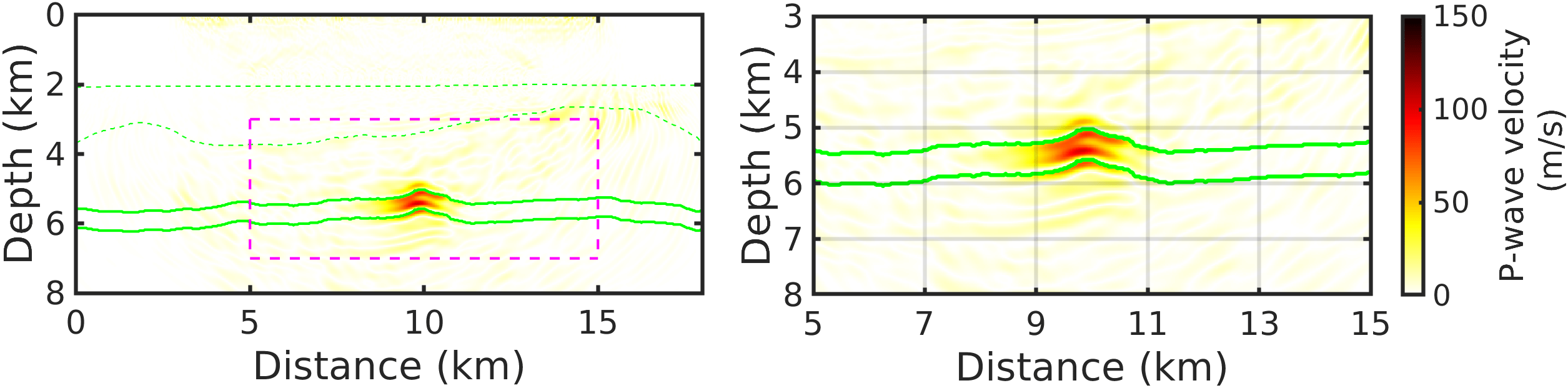}
        }
    \\
    \caption{Time-lapse estimates by the conventional approach (left-hand column) and the receiver-extension strategy (right-hand column) in the combined NR case using \textbf{Bayesian analysis}. Panels (a) and (b) refer to the $BW_1$ model \eqref{eq:delta_m__bw1}, (c) and (d) to the $BW_2$ \eqref{eq:delta_m__bw2}, and (e) and (f) to the $BW_3$ model \eqref{eq:delta_m__bw3}.}    
    \label{fig:resulting_full_NR_Bayes} 
\end{figure*}

Figure \ref{fig:resulting_full_NR_Bayes} shows time-lapse estimates, and provides a view of the area enclosed by the magenta rectangle, for the combined NR case using Bayesian analysis. The left-hand column refers to the estimates derived through the conventional FWI approach, whereas the right-hand column refers to the estimates derived through the receiver-extension strategy. Significant time-lapse noises were effectively mitigated in all Bayesian weighted resulting models, especially those associated with the top of salt body and the deepest region of pre-salt layers. However, although the $BW_2$ model presented a better time-lapse estimate (Fig. \ref{fig:resulting_fullNR_Bayes_mod2_classical}) than the others (Figs. \ref{fig:resulting_fullNR_Bayes_mod1_classical} and \ref{fig:resulting_fullNR_Bayes_mod3_classical}), note that the time-lapse anomalies in all conventional FWI cases (left-hand column of Fig. \ref{fig:resulting_full_NR_Bayes}) are not well delimited and have very low amplitudes of the order of the time-lapse artifacts. In contrast, combining Bayesian analysis with the receiver-extension FWI strategy effectively delineates the time-lapse anomaly, providing a more nuanced and reliable representation of time-lapse changes, as depicted in Figs. \ref{fig:resulting_fullNR_Bayes_mod1_recExt}, \ref{fig:resulting_fullNR_Bayes_mod2_recExt} and \ref{fig:resulting_fullNR_Bayes_mod3_recExt}. The error and similarity measures are summarized in Table \ref{tab:Bayesian_analysis}, revealing that all Bayesian approaches proposed in this work, coupled with the receiver-extension FWI, outperform conventional approaches. Notably, in all cases involving this combination, the error measures in Table \ref{tab:Bayesian_analysis} are smaller than those in Table \ref{tab:combined_NR_issues}. At the same time, the similarity measures in Table \ref{tab:Bayesian_analysis} surpass those in Table \ref{tab:combined_NR_issues}.

\begin{table}[!b]
\centering
\caption{Main statistics comparing the retrieved time-lapse models with the true model for the Bayesian analysis case. \label{tab:Bayesian_analysis}}
\begin{tabular}{*5c}
\toprule
 &  \multicolumn{2}{c}{NRMS} & \multicolumn{2}{c}{R}\\
4D strategy   & C-FWI   & RE-FWI   & C-FWI   & RE-FWI \\
\midrule
$BW_1$  & 1.3685  & 0.5891 & 0.3260  & 0.8368 \\
$BW_2$   &  1.0146 & 0.6653 & 0.3169  & 0.7463\\
$BW_3$   &  0.8154  & \textbf{0.5502} & 0.5012 & \textbf{0.8665} \\
\bottomrule
\end{tabular}\\
\vspace{.1cm} {\footnotesize NRMS, normalized root-mean-square; R, Pearson's coefficient; \\ C-FWI, conventional FWI; RE-FWI, receiver-extension FWI.}
\end{table}


\newpage 

\section{Conclusion \label{sec:conclusion}}

In this work we have evaluated the portability of combining a receiver-extension FWI framework and Bayesian analysis to mitigate time-lapse noises arising from non-repeatability (NR) issues. In the Marmousi case study, we demonstrated that Bayesian-weighted time-lapse models can greatly reduce the time-lapse noises caused by NRs associated with background noise in seismic data. Besides, our findings, using a challenging deep-water Brazilian pre-salt setting, indicate that the receiver relocation strategy can also significantly attenuate time-lapse noises. Moreover, the incorporation of Bayesian analysis, integrated into our proposed weighted time-lapse models, further enhances the time-lapse estimates. Thus, combining Bayesian analysis with the receiver-extension FWI strategy effectively delineates time-lapse anomalies, providing a nuanced and reliable representation of time-lapse changes. 

The numerical experiments have also revealed that the receiver-extension strategy has minimal computational cost and resource requirements, with an additional runtime of less than $3 \%$ compared to the conventional approach. In addition, since the dimensionality of the parameter space is reduced in our context (requiring only two parameters in the Bayesian weighted time-lapse \textit{$\text{BW}_1$} model \eqref{eq:delta_m__bw1} and \textit{$\text{BW}_2$} model \eqref{eq:delta_m__bw2} cases and four parameters in the \textit{$\text{BW}_3$} model \eqref{eq:delta_m__bw3} case), and given the numerical stability of the Gaussian likelihood function, the Bayesian analysis is efficiently performed even with the large number of parameters of the velocity models obtained by FWI. In this sense, the runtime of the Bayesian analysis is minimal, and the computational resources do not require much computational power. For instance, the Bayesian analysis related to the \textit{$\text{BW}_1$} and \textit{$\text{BW}_2$} models required approximately 7.30 minutes of processing time in both study cases, while the \textit{$\text{BW}_3$} case required about 12 minutes, using a computer with an Intel Core i7-3970X processor with 3.50 GHz and 128 GB RAM.

Although the Bayesian weighted time-lapse FWI strategy based on \textit{$\text{BW}_3$} model \eqref{eq:delta_m__bw3}, which generalize the central-difference time-lapse FWI strategy \eqref{eq:delta_m_cd}, generated the best time-lapse estimates, it is worth highlighting that the Bayesian weighted time-lapse \textit{$\text{BW}_2$} model defined in Eq. \eqref{eq:delta_m__bw2} (which combines the parallel and sequential time-lapse FWI strategies) provided similar outcomes by running only 3 FWIs instead of the four inversions as required by the \textit{$\text{BW}_3$} model (and also by the Bayesian weighted time-lapse \textit{$\text{BW}_1$} model \eqref{eq:delta_m__bw1}). The case studies we presented demonstrate the practical applicability of Bayesian analysis in addressing more realistic challenges in time-lapse data analysis. The proposed Bayesian weighted procedure introduces a new workflow for determining time-lapse estimates through statistical analysis of pre-existing models, allowing its application in ongoing 4D projects. 

Finally, it is worth highlighting that both the receiver-extension strategy and Bayesian analysis have inherent limitations. First, the implementation of the receiver-extension strategy may encounter difficulties in dealing with uncertainties in 3D environments where there may be ambiguities in determining the receiver position correction parameters $\Delta \textbf{x}_{s,r}$. Possible improvements in the future could include the implementation of disambiguation constraints. Second, the sensitivity of Bayesian analysis to prior information may limit the accuracy in determining the coefficients used to construct time-lapse models when the prior information is inaccurate or insufficient. As a perspective, we intend to apply the proposed approaches to the analysis of 4D field data in order to investigate the benefits, practicability, and difficulties of integrating Bayesian analysis with receiver-extension FWI in large-scale time-lapse projects. We also aim to explore how Bayesian analysis can add value to advanced time-lapse approaches such as double-difference 4D FWI \cite{Zhang_Huang_2013_Geophysics,Zhang_Huang_2015_Geophysics} and common-model 4D FWI \cite{Hicks_et_al_2016_TIMELAPSEFEI_LeadingEdge} strategies in future research and analysis.

\section*{Acknowledgments}
The authors from Fluminense Federal University (UFF) gratefully acknowledge the support from Shell Brasil Petróleo Ltda through the R\&D project “Refraction seismic for pre-salt reservoirs” (ANP no. 21727-3). The strategic importance of the R\&D levy regulation from the National Agency for Petroleum, Natural Gas and Biofuels (ANP) is also appreciated. We would like to thank Rodrigo S. Stern for the crucial IT support.

\end{document}